\documentclass{nature}
\usepackage[utf8]{inputenc}
\usepackage[T1]{fontenc}  %packages qui permet d'utiliser les caractères du clavier
\usepackage{graphicx}

\usepackage{verbatim}

\usepackage{amssymb}
\usepackage{mathrsfs}
\usepackage{amsfonts}
\usepackage{amsmath}
\usepackage{textcomp}
\usepackage{epsfig}
\usepackage{subcaption}
\usepackage{endnotes}   %permet de faire des notes de bas de page
\let\footnote = \endnote
\usepackage{blindtext}
\usepackage{makeidx} % utile pour créer un index

\usepackage{sidecap}
\usepackage{wrapfig}

\usepackage[table]{xcolor}  % pour colorier les tableaux
\usepackage{caption}
\usepackage{textcomp}
\usepackage{palatino} % font to avoid the ugly default µ

\usepackage{nomencl}
\usepackage{ifthen}
\usepackage{lipsum}

\makenomenclature

\usepackage{hyperref}      %% POUR FAIRE DES LIENS DANS LE TEXTE %%
\hypersetup{
pdfpagemode=UseOutlines,      % UseOutlines, UseThumbs, None, FullScreen : agencement au démarrage
pdfstartview=Fit,             % Fit, FitH, FitB, FitBH : vue de la page au départ (pleine largeur...)
pdffitwindow=true,            % bool: Maximiser
pdfpagelayout=TwoColumnsRight,% SinglePage, TwoColumnsRight/Left, OneColumn : affichage des pages
pdftoolbar=true,              % bool: Affichage de la barre d'outils
pdfmenubar=true,              % bool: Affichage de la barre de menus
bookmarksopen=false,          % bool: Dépliement des signets
bookmarksnumbered=true,       % bool: Numérotation des signets
colorlinks=true,              % bool: Liens colorés
pdfauthor={Ton nom},     % Auteur
pdftitle={Titre PDF},    % Titre
pdfcreator=PDFLaTeX,          % 
pdfproducer=PDFLaTeX,         %
linkcolor=blue,               % Couleur des liens
urlcolor=blue,                %             url
anchorcolor=black,            %         du texte
citecolor=green,              % Couleur de citation 
frenchlinks=true,             % bool: Utiliser des petites majuscules pour les liens, plutôt que de la couleur
pdfborder={0 0 0}             % Ne pas encadrer les liens
}

\usepackage{multicol}    % pour faire plusieurs colonnes de texte

\usepackage{amsmath} % AMS Math Package
\usepackage{amsthm} % Theorem Formatting

\usepackage{multirow}
\usepackage{multicol} % Allows for multiple columns
\usepackage[dvips,letterpaper,margin=0.8in,bottom=0.8in]{geometry}

\usepackage{booktabs}

\newcommand{\ket}[1]{\left| #1 \right>} % for Dirac bras
\let\baraccent=\= % rename builtin command \= to \baraccent
\renewcommand{\=}[1]{\stackrel{#1}{=}} % for putting numbers above =

\theoremstyle{definition}

\theoremstyle{remark}

\newcommand{\ah}[0]{\ensuremath{\hat{a}}}
\newcommand{\ad}[0]{\ensuremath{\hat{a}^\dagger}}

\newcommand{\Ham}[0]{\ensuremath{H}}
\newcommand{\sz}[0]{\ensuremath{\hat{\sigma}_z}}

\newcommand{\ECm}[0]{\ensuremath{E_{C,m}}}
\newcommand{\EC}[0]{\ensuremath{E_\mathrm{C}}}
\newcommand{\ELm}[0]{\ensuremath{E_{L,m}}}
\newcommand{\EL}[0]{\ensuremath{E_{L}}}
\newcommand{\EJ}[0]{\ensuremath{E_\mathrm{J}}}
\newcommand{\Phid}[0]{\ensuremath{\Phi_\mathrm{diff}}}
\newcommand{\nm}[0]{\ensuremath{\hat{n}_m}}
\newcommand{\phim}[0]{\ensuremath{\hat{\varphi}_m}}

\newcommand{\Ll}[0]{\ensuremath{L_l}}
\newcommand{\Cl}[0]{\ensuremath{C_l}}
\newcommand{\Cg}[0]{\ensuremath{C_\mathrm{g}}}

\renewcommand{\figurename}{\textbf{Figure}} 

\usepackage{upgreek}

\usepackage{authblk}

\title{\textbf{Unimon qubit}}

\author[1, *]{Eric Hyypp\"a}
\author[2]{Suman Kundu}
\author[1]{Chun Fai Chan}
\author[2]{Andr\'as Gunyh\'o} %á ó
\author[1]{Juho Hotari}
\author[1]{David Janzso}
\author[1]{Kristinn Juliusson}
\author[2]{Olavi Kiuru}
\author[1]{Janne Kotilahti}
\author[1]{Alessandro Landra}
\author[1]{Wei Liu}
\author[1]{Fabian Marxer}
\author[1]{Akseli M\"akinen}
\author[1]{Jean-Luc Orgiazzi}
\author[1]{Mario Palma}
\author[1]{Mykhailo Savytskyi}
\author[1]{Francesca Tosto}
\author[1]{Jani Tuorila}
\author[2]{Vasilii Vadimov}
\author[1]{Tianyi Li}
\author[1]{Caspar Ockeloen-Korppi}
\author[1, $\dagger$]{Johannes Heinsoo}
\author[1, $\dagger$]{Kuan Yen Tan}
\author[1, $\dagger$]{Juha Hassel}
\author[1,2,3,*, $\dagger$]{Mikko M\"ott\"onen}
\affil[1]{IQM, Keilaranta 19, 02150 Espoo, Finland}
\affil[2]{QCD Labs, QTF Centre of Excellence, Department of Applied Physics, Aalto University, P.O. Box 13500, FIN-00076 Aalto, Finland.}
\affil[3]{VTT Technical Research Centre of Finland Ltd. \& QTF Centre of Excellence, P.O. Box 1000, 02044 VTT, Finland.}
\affil[*]{Corresponding authors. e-mails: eric@meetiqm.com, mikko.mottonen@aalto.fi}
\affil[$\dagger$]{Jointly supervised the work}
\date\today

\makeatletter
\let\saved@includegraphics\includegraphics
\AtBeginDocument{\let\includegraphics\saved@includegraphics}
\renewenvironment*{figure}{\@float{figure}}{\end@float}
\makeatother

\begin{document}

\maketitle %throwing an error but looking ok

%MM NOTE: I added my Zotero bibliography also to this file. Feel free to use it. It is okay to use many bib files but in the end we just need to check that we do not have the references twice in the bibliography. Note that Nature wants the URLs be also in the bibliography in the submitted manuscripts. Those will appear automatically when using Zotero. 

\begin{abstract}%word count: 267
\noindent 
Superconducting qubits\cite{krantz2019quantum,blais_circuit_2021} are one of the most promising candidates to implement quantum computers, with projected applications in physics simulations\cite{yanay2020two}, optimization\cite{farhi_quantum_2014}, machine learning\cite{dunjko_machine_2018}, and chemistry\cite{mcardle_quantum_2020}. The superiority of superconducting quantum computers over any classical device in simulating random but well-determined quantum circuits has already been shown in two independent experiments\cite{arute_quantum_2019,wu_strong_2021} and important steps have been taken in quantum error correction\cite{ofekextending_2016,campagne-ibarcqquantum_2020,chenexponential_2021}. However, the currently wide-spread qubit designs\cite{koch_charge-insensitive_2007,nguyen_high-coherence_2019,pechenezhskiy_superconducting_2020,gyenis_experimental_2021,yan_engineering_2020} do not yet provide high enough performance to enable practical applications or efficient scaling of logical qubits owing to one or several following issues: sensitivity to charge or flux noise leading to decoherence, too weak non-linearity preventing fast operations, undesirably dense excitation spectrum, or complicated design vulnerable to parasitic capacitance. Here, we introduce and demonstrate a superconducting-qubit type, the \emph{unimon}, which combines the desired properties of high non-linearity, full insensitivity to dc charge noise, insensitivity to flux noise, and a simple structure consisting only of a single Josephson junction in a resonator. We measure the qubit frequency, $\omega_{01}/(2\pi)$, and anharmonicity $\alpha$ over the full dc-flux range and observe, in agreement with our quantum models, that the qubit anharmonicity is greatly enhanced at the optimal operation point, yielding, for example, 99.9\% and 99.8\% fidelity for 13-ns single-qubit gates on two qubits with $(\omega_{01},\alpha)=(4.49~\mathrm{GHz}, 434~\mathrm{ MHz})\times 2\pi$ and $(3.55~\mathrm{GHz}, 744~\mathrm{ MHz})\times 2\pi$, respectively. The energy relaxation time $T_1\lesssim 10~\mu\mathrm{s}$ is stable for hours and seems to be limited by dielectric losses. Thus, future improvements of the design, materials, and gate time may promote the unimon to break the 99.99\% fidelity target for efficient quantum error correction and possible quantum advantage with noisy systems.
\end{abstract}

\noindent\textbf{Introduction}

Even though quantum supremacy has already been reached with superconducting qubits in specific computational tasks\cite{arute_quantum_2019,wu_strong_2021}, the current quantum computers still suffer from errors owing to noise. In this so-called noisy intermediate-scale quantum (NISQ) era\cite{preskill_quantum_2018}, the complexity of the implementable quantum computations\cite{nakahara2008quantum} is mostly limited by errors in single- and two-qubit quantum gates. Crudely speaking, the process fidelity of implementing a $d$-deep $n$-qubit logic circuit with gate fidelity $F$ is $F^{dn}$. Thus, to succeed roughly half of the time in a 100-qubit circuit of depth five, one needs at least 99.9\% gate fidelity. In practice, the number of qubits and especially the gate depth required for useful NISQ advantage is likely higher, leading to a fidelity target of 99.99\% for all quantum gates, not yet demonstrated in any superconducting quantum computer.

The effect of gate errors can be reduced to some extent using error mitigation\cite{temme2017error,kandala2019error} or in principle, completely using quantum error correction\cite{gottesman1997stabilizer}. Surface codes\cite{fowler_surface_2012,bonilla2021xzzx} are regarded as some of the most compelling error correction codes for superconducting qubits owing to the two-dimensional topology of the qubit register and their favorable fidelity threshold of roughly 99\% which has been reached with superconducting transmon qubits already in 2014\cite{barends_superconducting_2014}. Despite the recent major developments in implementing distance-two and -three surface codes on superconducting quantum processors\cite{andersen_repeated_2020,marques2021logical,krinner2021realizing,zhao2021realizing}, the gate and readout fidelities of superconducting qubits need to be improved further, preferably above 99.99\%, to enable efficient quantum error correction with a reasonable qubit count.

Currently, most of the superconducting multi-qubit processors utilize transmon qubits\cite{arute_quantum_2019,hertzberg2021laser, andersen_repeated_2020,zhu2021quantum} that can be reproducibly fabricated\cite{hertzberg2021laser} and have coherence times up to several hundred micro\-seconds\cite{place_new_2021,wang2022towards}, leading to record 
average gate fidelities of 99.98\% for single-qubit gates\cite{mckay2017efficient} and 99.8\%--99.9\% for two-qubit gates\cite{sungrealization_2021}. 
The transmon was derived from the charge qubit\cite{nakamura_coherent_1999} by adding a shunt capacitor in parallel with a Josephson junction, with the result of exponentially suppressing the susceptibility of its transition frequency to charge noise. However, the large shunt capacitance results in a relatively low anharmonicity of 200--300 MHz corresponding to only 5\% of the typical qubit frequency\cite{koch_charge-insensitive_2007,barends_coherent_2013}. This limits the speed of quantum gates that can be implemented with transmons since leakage errors to the states beyond the computational subspace need to be suppressed\cite{mckay2017efficient, wood2018quantification}. Similarly, the low anharmonicity also limits the readout speed  of transmon qubits, and a high-power readout tone can even excite the transmon to unconfined states beyond the cosine potential\cite{lescanne_escape_2019}. A higher anharmonicity is preferred to speed up the qubit operations and to allow for higher fidelities limited by the finite coherence time.

Hence, there is an urgent need to find new superconducting qubit types that increase the anharmonicity--coherence-time product. Recently, major progress has been made in the development of fluxonium qubits, one of the most compelling alternatives to transmons thanks to their high anharmonicity and long relaxation and coherence times\cite{manucharyan_fluxonium_2009,bao2021fluxonium,zhang2021universal}. In a fluxonium qubit, a small Josephson junction is shunted by a superinductor implemented by an array of large Josephson junctions\cite{manucharyan_fluxonium_2009,zhang2021universal,nguyen_high-coherence_2019}, a granular aluminum wire\cite{grunhaupt2019granular}, a nanowire with a high kinetic inductance\cite{hazard_nanowire_2019}, or a geometric superinductor\cite{Peruzzo2021Geometric}. The superinductor in the fluxonium ensures that the dephasing and relaxation rates arising from flux noise are reduced, in addition to which all the levels of a fluxonium are fully protected against dephasing arising from low-frequency charge noise. It is possible to add a large shunt capacitor into the fluxonium in order to create a so-called heavy fluxonium\cite{earnestrealization_2018,zhang2021universal}, in which the transition matrix element between the ground state and the first excited state can be suppressed to enhance the relaxation time up to the millisecond regime\cite{earnestrealization_2018}. However, special techniques are required to control, readout, and reset these high-coherence fluxonium qubits due to their low frequency and small transition matrix elements in the vicinity of the half flux quantum operation point\cite{zhang2021universal}. Furthermore, it is not possible to achieve protection against both relaxation and dephasing due to flux noise at a single operation point. Parasitic capacitances in the superinductor may also provide a challenge for the reproducible fabrication of fluxonium qubits and result in parasitic modes.  

By reducing the total inductance of the junction array in the fluxonium, it is possible to implement a plasmonium qubit\cite{liu2021quantum} operated at zero flux or a quarton qubit\cite{yan_engineering_2020} operated at the half-flux-quantum point, both of which have a small size and a high anharmonicity compared with the transmon and a sufficient protection against charge noise in comparison to current coherence times.  On the other hand, an enhancement of the superinductance converts the fluxonium  into a so-called quasicharge qubit\cite{pechenezhskiy_superconducting_2020}, the charge-basis eigenstates of which resemble those of the early charge qubits while retaining the protection against charge noise. Other qubits protected against some sources of relaxation and dephasing include the $0-\pi$ qubit\cite{gyenis_experimental_2021}, bifluxon\cite{kalashnikov2020bifluxon}, and a qubit protected by two-Cooper-pair tunneling\cite{smith2020superconducting}. The $0-\pi$ qubit is protected against both relaxation and dephasing arising from charge and flux noise thanks to its topological features, which unfortunately renders the qubit challenging to operate and its circuit relatively complicated and hence vulnerable to parasitic capacitance. Despite this great progress in fluxonium and protected qubits, they have still not shown broad superiority to the transmons. The race for the new improved mainstream superconducting qubit continues.

\begin{figure}
 \centering
 \includegraphics[width = 0.9\textwidth]{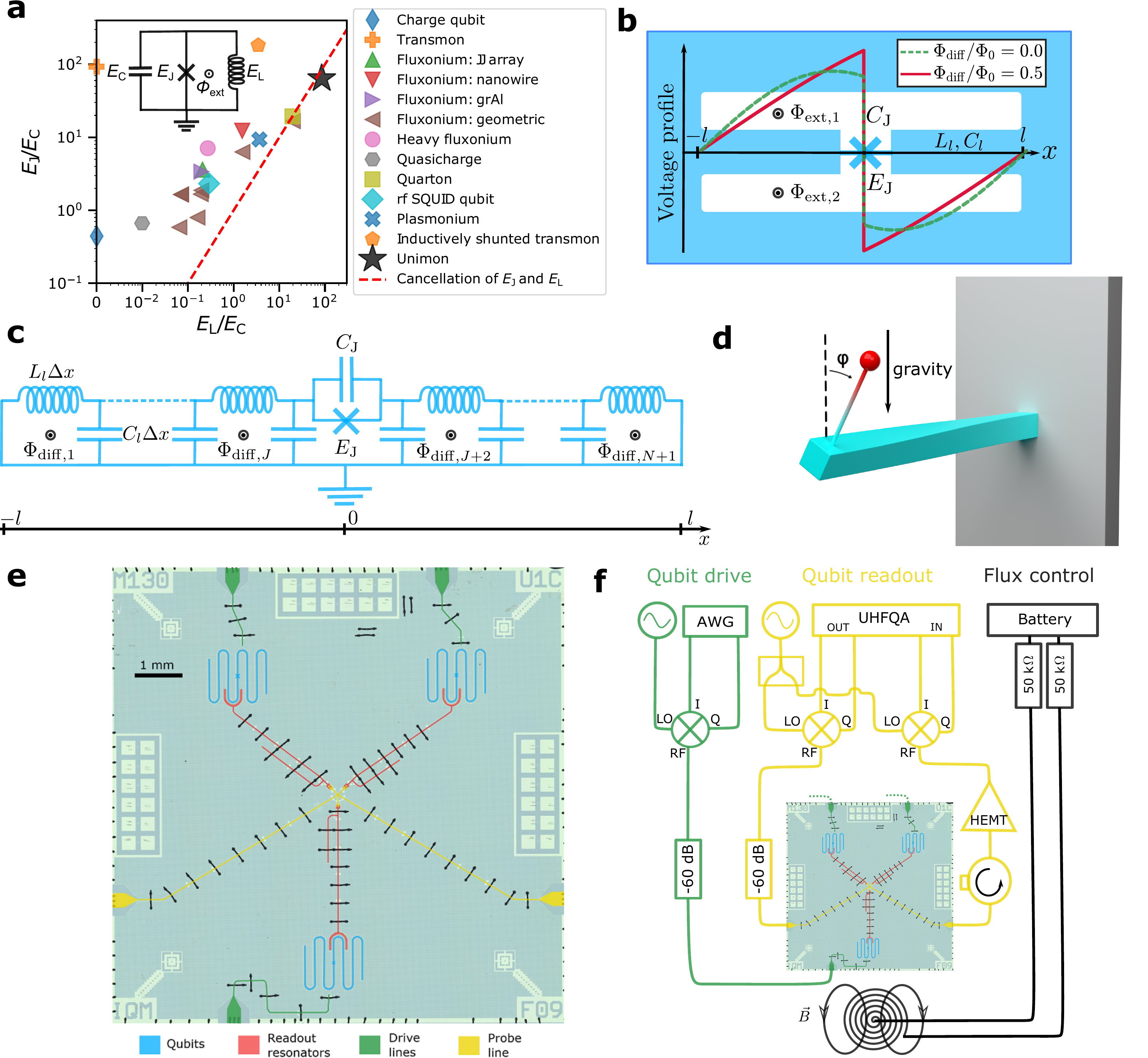}
    \caption{ \label{fig: fig 1}  \textbf{Unimon qubit and its measurement setup.} Caption on the following page.}
\end{figure}

\begin{figure}
 \centering
 \includegraphics[width = 0.95\textwidth]{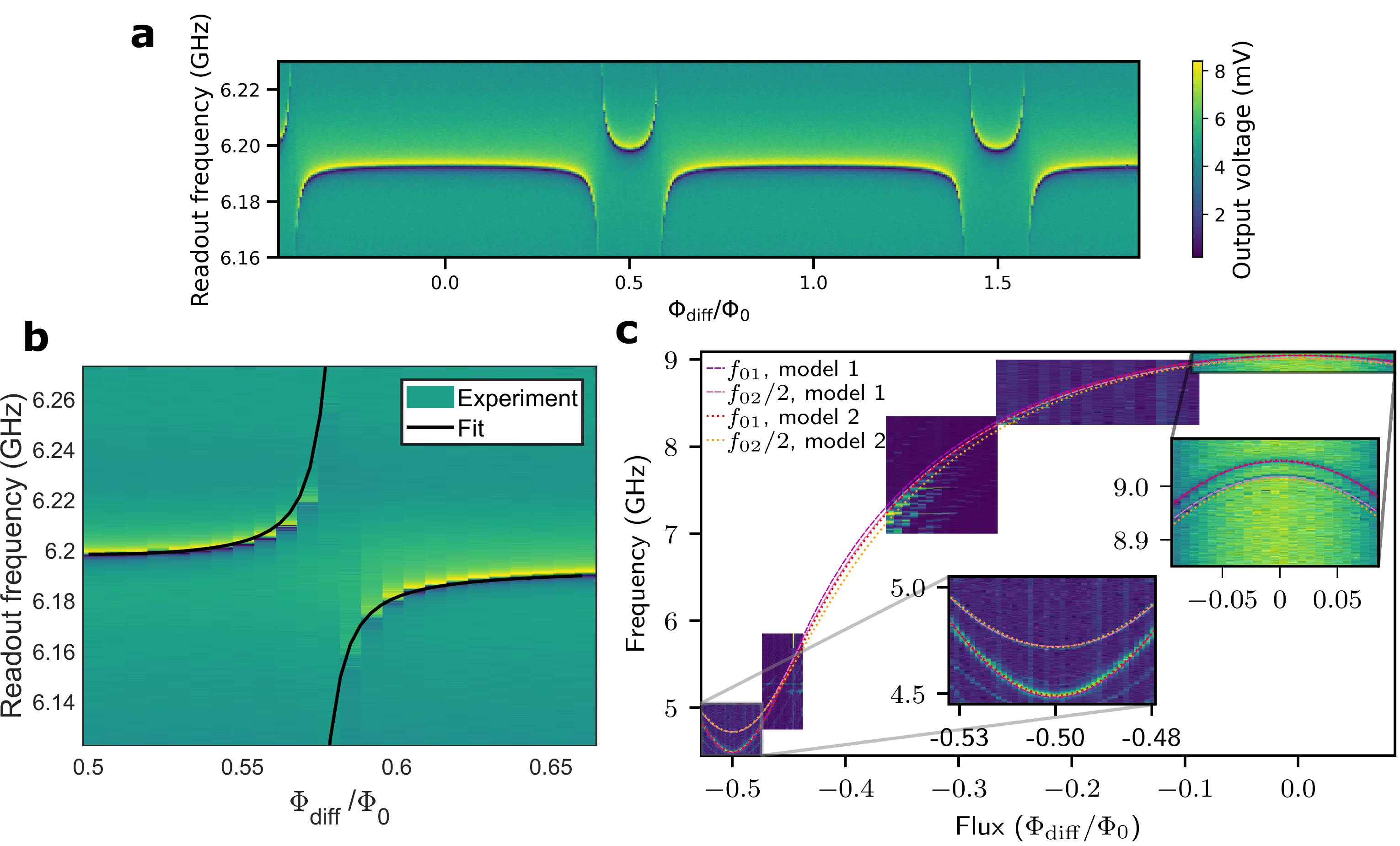}
    \caption{ \label{fig: resonator and qb spectroscopy} \textbf{Resonator and qubit B spectroscopies.} \textbf{a}, Magnitude of the readout signal voltage transmitted through the probe line as a function of  the signal frequency and the flux bias $\Phi_\mathrm{diff}$ of the unimon. \textbf{b}, Magnification at an avoided crossings of \textbf{a}, where a unimon and its readout resonator are close to resonance, together with a fit (solid black line) used to estimate the coupling capacitance $C_\mathrm{g}$ between the qubit and the resonator. The fit is based on diagonalizing equation~\eqref{eq: coupled Hamiltonian final} in Methods. \textbf{c}, Magnitude of the readout signal at a properly chosen readout frequency as a function of the flux bias $\Phid$ and qubit excitation frequency, revealing the spectral lines of the unimon together with global fits to the theoretical model 1 and 2 (Methods). The insets show magnifications at the flux sweet spots, highlighting that at half flux quantum, the unimon frequency is minimized whereas its anharmonicity is maximized.} 
\end{figure}

\begin{figure}
 \centering
 \includegraphics[width = \textwidth]{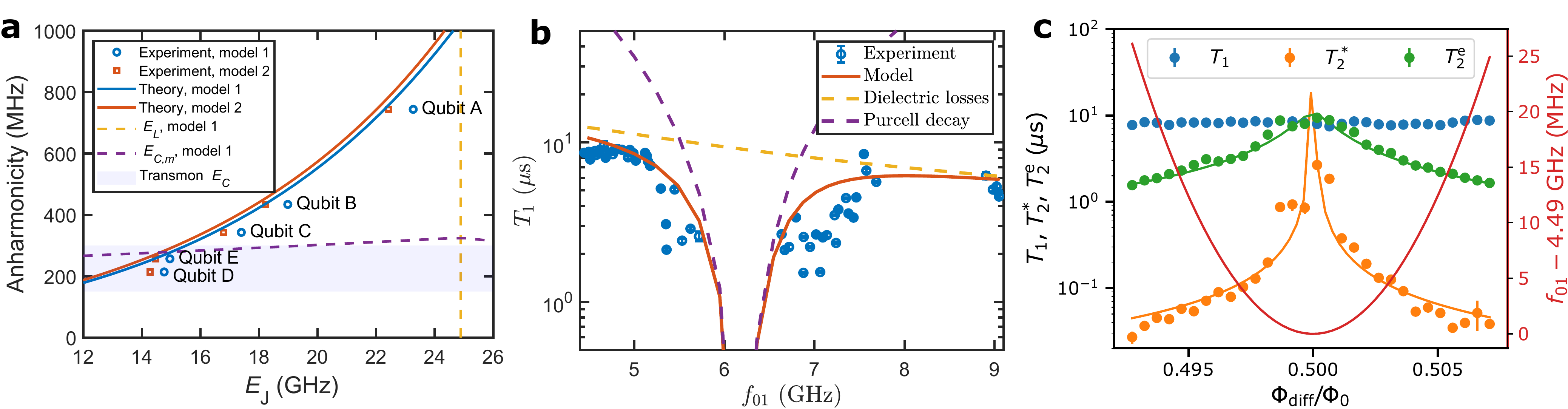}
    \caption{ \label{fig: anharmonicity and coherence times vs flux} \textbf{Measurement of key qubit properties.} \textbf{a}, Measured anharmonicities of five unimon qubits (markers) at flux bias $\Phi_\mathrm{diff}=\Phi_0/2$ as functions of the Josephson energy $\EJ$ estimated from fits to the qubit spectroscopy data similar to Fig.~\ref{fig: resonator and qb spectroscopy}c using the models~1 (blue color) and~2 (orange color) presented in Methods. The solid blue and orange lines show theoretical predictions of the anharmonicity based on models~1 and~2, respectively.  The dashed yellow line illustrates the value of the Josephson energy $\EJ$ that perfectly cancels the linear inductive energy according to model~1. The capacitive energy $\ECm$ of the qubit mode based on model~1 is shown with a dashed purple line, whereas the purple shading visualizes the typical anharmonicity of transmon qubits. 
    \textbf{b}, Experimentally measured mean energy relaxation time $T_1$ of qubit~B (blue circles) as a function of the qubit frequency $f_{01}$ together with a model (solid line) taking into account dielectric losses (dashed yellow line) and the Purcell decay (dashed purple line). The error bars represent the $1\sigma$ uncertainty obtained from 6--30 repetitions of single $T_1$ measurements conducted at each frequency. \textbf{c},  Relaxation time $T_1$, Ramsey coherence time $T_2^*$, and echo coherence time $T_2^\mathrm{e}$ of qubit~B as functions of the flux bias $\Phid$ in the vicinity of the $\Phi_0/2$ flux sweet spot. The error bars represent the $1\sigma$ uncertainty based on 8 repeated experiments. The solid green and orange lines illustrate fits to $T_2^\mathrm{e}$ and $T_2^*$ data based on models regarding the dephasing rate as a linear function of $\partial f_{01}/\partial \Phid$ (Methods and Supplementary Note~III). The solid red line shows the detuning of the qubit frequency from the minimum value of 4.49 GHz.
    }
\end{figure}

\begin{figure}
 \centering
 \includegraphics[width=\textwidth]{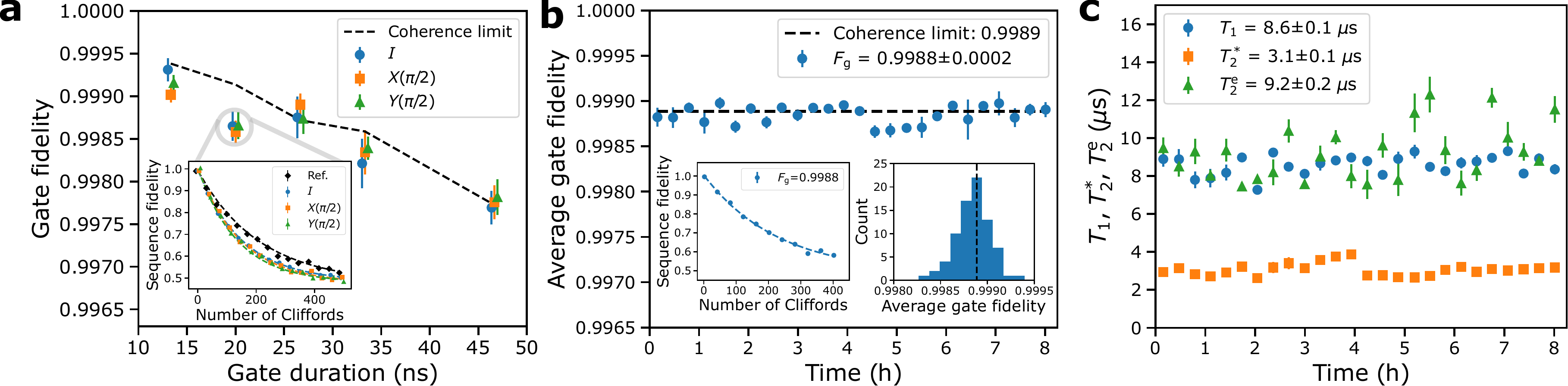}
    \caption{ \label{fig: Gate fidelities and stability} \textbf{Implementation of fast and stable single-qubit gates for qubit B.} \textbf{a}, Average gate fidelity as a function of the gate duration for gates in the set  $\{I, X(\pi/2), Y(\pi/2)\}$. The error bars represent the $1\sigma$ uncertainty based on six interleaved randomized benchmarking experiments. The dashed black line shows the coherence limit computed based on the mean values of the energy relaxation time $T_1$ and the echo coherence time $T_2^\mathrm{e}$. The inset shows  the results of an exemplary interleaved randomized benchmarking experiment corresponding to a gate duration of 20 ns. 
    \textbf{b}, Average gate fidelity measured with randomized benchmarking as a function of time lapsed from the initial gate and parameter calibration. 
    The error bars represent the $1\sigma$ uncertainty %standard deviation 
    from three consecutive randomized benchmarking experiments. The dashed black line shows the coherence limit that has been computed using the mean values of $T_1$ and $T_2^\mathrm{e}$ measured interleaved with the gate fidelity. 
    The left inset illustrates the decay of the sequence fidelity as a function of its length in an exemplary randomized-benchmarking experiment. The right inset shows the histogram of the gate fidelities obtained over the measurement period of eight hours. \textbf{c}, Stability of $T_1$, $T_2^\mathrm{e}$, and the Ramsey coherence time $T_2^*$ over a period of eight hours. The qubit parameters were calibrated only once before the characterization measurements. The error bars represent the $1\sigma$ uncertainty from three
    %standard deviations of 3 
    consecutive measurements.}
\end{figure}

In this work, we introduce and demonstrate a novel superconducting qubit, \emph{the unimon}, that consists of a single Josephson junction shunted by a linear inductor and a capacitor in a largely unexplored parameter regime where the inductive energy is mostly cancelled by the Josephson energy leading to high anharmonicity while being fully resilient against low-frequency charge noise and partially protected from flux noise (Fig.~\ref{fig: fig 1}). We measure the unimon frequency and anharmonicity in a broad range of flux biases and find a very good agreement with first-principles models (Fig.~\ref{fig: resonator and qb spectroscopy}), even for five different qubits (Fig.~\ref{fig: anharmonicity and coherence times vs flux}a). According to our experimental data, the energy relaxation time seems to be limited by dielectric losses (Fig.~\ref{fig: anharmonicity and coherence times vs flux}b), and the coherence time can be protected from flux noise at a flux-insensitive sweet spot (Fig.~\ref{fig: anharmonicity and coherence times vs flux}c). Importantly, we observe that the single-qubit gate fidelity progressively increases with decreasing gate duration, and is stable for hours at 99.9\% for a 13-ns gate duration (Fig.~\ref{fig: Gate fidelities and stability}), demonstrating that the unimon is a promising candidate for a new mainstream superconducting qubit.

\noindent\textbf{Unimon}

In practice, we implement the unimon in a simple superconducting circuit by integrating a single Josephson junction into the center conductor of a superconducting coplanar-waveguide (CPW) resonator grounded at both ends (Fig.~\ref{fig: fig 1}b). There are no charge islands in the circuit, and hence the junction is inductively shunted. In addition to the very recent fluxonium qubit utilizing a geometric superinductance\cite{Peruzzo2021Geometric}, the unimon is the only superconducting qubit with the Josephson junction shunted by a geometric inductance that provides complete protection against low-frequency charge noise. Due to the non-linearity of the Josephson junction, the normal modes of the resonator with a non-zero current across the junction are converted into anharmonic oscillators that can be used as qubits. In this work, we use the lowest anharmonic mode as the qubit since it has the highest anharmonicity.

The frequency of each anharmonic mode can be controlled by applying external fluxes $\Phi_\mathrm{ext,1}$ and $\Phi_\mathrm{ext,2}$ through the two superconducting loops of the resonator structure as illustrated in Fig.~\ref{fig: fig 1}b. The unimon is partially protected against flux noise thanks to its gradiometric structure, which signifies that the superconducting phase across the Josephson junction is dependent on the half difference of the applied external magnetic fluxes $\Phi_\mathrm{diff} = (\Phi_\mathrm{ext,2} - \Phi_\mathrm{ext,1})/2$. Interestingly, the anharmonicity of the unimon is maximized at a flux-insensitive sweet spot, at which the qubit frequency is unaffected by the external flux difference to the first order. This optimal operation point is obtained at $\Phid = \Phi_0/2$ modulo integer flux quanta $\Phi_0 =h/(2e) \approx 2.067 \times 10^{-15}$ Wb, where $h$ is the Planck constant and $e$ is the elementary charge. 

Using the distributed-element circuit model shown in Fig.~\ref{fig: fig 1}c,  the effective Hamiltonian of the qubit mode $m$ can be written (model 1 in Methods) as 
\begin{equation}
    \hat{\Ham}_m = 4 \ECm(\varphi_0) \hat{n}_m^2 + \frac{1}{2} E_{L, m}(\varphi_0) \hat{\varphi}_m^2 + E_L \hat{\varphi}_m \left(\frac{2\pi \Phi_\mathrm{diff}}{\Phi_0} - \varphi_0 \right) - \EJ \cos(\hat{\varphi}_m  -\varphi_0 ), \label{eq: Hamiltonian with external flux}
\end{equation}
where $\varphi_0$ is the Josephson phase of a dc current across the junction,  $\ECm(\varphi_0)$ is the capacitive energy of the qubit mode, $E_{L, m}(\varphi_0)$ is the inductive energy of the qubit mode, $\EL$ is the inductive energy of the dc current,  $\EJ$ is the Josephson energy, and $\nm$ and $\phim$ are the Cooper pair number and phase operators %of the junction 
corresponding to the qubit mode $m$ and satisfying $[\hat{\varphi}_m, \hat{n}_m] = \mathrm{i}$  with $\mathrm{i}$ being the imaginary unit. Note that $\varphi_0$ is treated as a classical variable depending on the flux bias $\Phid$ according to a transcendental equation such that $2\pi\Phid/\Phi_0 - \varphi_0$ is periodic in $\Phid$. (See Extended Data Fig.~\ref{fig: phase basis wave functions, matrix elements etc} for solutions of equation~\eqref{eq: Hamiltonian with external flux}.)

At the sweet spot $\Phid = \Phi_0/2$, the dc phase equals $\varphi_0 = \pi$ and the Hamiltonian of the unimon reduces to 
\begin{equation}
\hat{\Ham}_m = 4 \ECm (\pi) \nm^2 + \frac{1}{2}\ELm(\pi) \phim^2 + \EJ \cos(\phim), \label{eq: sweet spot unimon Hamiltonian}
\end{equation}
where we assume that $\EJ \leq \EL$. Strikingly, this Hamiltonian is exactly analogous to a simple mechanical system visualized in Fig.~\ref{fig: fig 1}d, in which an inverted pendulum is attached to a twisting beam.  
In this analogy, the gravitational potential energy of the pendulum corresponds to the cosine-shaped Josephson potential, the harmonic potential energy associated with the twisting of the beam corresponds to the inductive energy of the unimon, and the moment of inertia of the pendulum is analogous to the capacitance in the unimon. Furthermore, the twist angle $\varphi$ is analogous to the superconducting phase difference $\phim$ of the qubit mode across the Josephson junction. This mechanical analogue provides great intuition to the physics of the unimon. 

In this work, we employ the parameter regime $\EJ \lesssim \ELm(\pi) \approx \EL$ to provide a large anharmonicity without any superinductors. As a result, it is instructive to use the Taylor expansion of the cosine and write the sweet-spot Hamiltonian of the unimon in Eq.~\eqref{eq: sweet spot unimon Hamiltonian} as
    \begin{equation}
        \hat{\Ham}_m = 4 \ECm (\pi) \nm^2 + \frac{\ELm(\pi) - \EJ}{2} \phim^2 + \frac{\EJ}{24} \phim^4 + \mathcal{O}(\phim^6).
\end{equation}
The quadratic term proportional to $\propto (\ELm(\pi) - \EJ)$ is mostly cancelled in the unimon regime, which emphasizes the high-order terms in the potential energy and hence increases the anharmonicity of the qubit. This cancellation bears resemblance to the quarton qubit\cite{yan_engineering_2020} with the distinctive difference that the quadratic inductive energy of a quarton qubit is only an approximation for the actual potential energy function of a short Josephson junction array, as a result of which the quarton circuit is not fully protected against low-frequency charge noise unlike the unimon.

To experimentally demonstrate the unimon qubit, we design and fabricate samples, each of which consists of three unimon qubits as illustrated in Fig.~\ref{fig: fig 1}e. We use niobium as the superconducting material apart from the Josephson junctions, in which the superconducting leads are fabricated using aluminum (see Sample Fabrication in Methods). The CPW structure of the unimon is designed for characteristic impedance $Z = 100$~$\Omega$ to reduce the total capacitance of the unimon in comparison to a standard 50-$\Omega$ resonator. Each qubit is capacitively coupled to an individual drive line that enables single-qubit rotations in a similar manner as for conventional transmon qubits by applying attenuated microwave pulses along the drive line as illustrated in the simplified schematic of the experimental setup in Fig.~\ref{fig: fig 1}f (see also Extended Data Fig.~\ref{fig: experimental setup}). All experiments are carried out at 10-mK base temperature of a pulse-tube-cooled dilution refrigerator. Furthermore, each qubit is capacitively coupled to a readout resonator using a U-shaped capacitor in order to enable dispersive qubit state measurements\cite{blais_cavity_2004,siddiqi2006dispersive} similar to those conventionally used with transmon qubits\cite{krinner2021realizing}. The frequency of the qubits is tuned by applying a current through an external coil attached to the sample holder such that one flux quantum $\Phi_0$ approximately corresponds to 10~$\mu$A.

\noindent\textbf{Results}

We experimentally study five unimon qubits, A--E, on two different chips. In all of the qubits, the geometry of the CPW resonator is similar, but the qubits have different Josephson energies $\EJ$ corresponding to different amounts of cancellation $\propto (\ELm(\pi) - \EJ)$ of the quadratic potential energy terms. Furthermore, the coupling capacitance between a qubit and its readout resonator has been designed to be different on the two chips. We present the main measured properties for all of the five qubits in Extended Data Tables~\ref{tab: basic qubit properties} and~\ref{tab: results of characterization}. Design targets of the parameter values are provided in Extended Data Table~\ref{tab: designed parameters}. The results discussed below are obtained from qubit~B unless otherwise stated.    

In Fig.~\ref{fig: resonator and qb spectroscopy}a, we show the microwave response of the readout resonator as a function of the flux bias $\Phi_\mathrm{diff}$ through the unimon loops. We observe that the frequency of the readout resonator changes periodically, as expected, since a change of flux by a flux quantum has no observable effect on the full circuit Hamiltonian in equation~\eqref{eq: Hamiltonian with external flux}. Furthermore, the frequency of the readout resonator exhibits an avoided crossing where the first transition frequency of the bare qubit $f_{01} = \omega_{01}/(2\pi)$ crosses the bare resonator frequency. By fitting our theoretical model of the coupled unimon-resonator system (see Methods and Supplementary Note~II) to the experimental data of the avoided crossing shown in Fig.~\ref{fig: resonator and qb spectroscopy}b, we estimate that the coupling capacitance between the qubit and the readout resonator is $\Cg =10.0$~fF in good agreement with the design value of 10.4~fF obtained from our classical electromagnetic simulations.

Figure~\ref{fig: resonator and qb spectroscopy}(c) shows the results of a two-tone experiment to map the qubit frequency spectrum (Methods). We observe that the single-photon transition between the ground state $|0\rangle$ and the first excited state $|1\rangle$ has a minimum frequency of $f_{01} = 4.488$~GHz at $\Phi_\mathrm{diff}/\Phi_0 = -0.5$ and a maximum frequency of $f_{01} = 9.05$~GHz at $\Phi_\mathrm{diff} = 0$. The two-photon transition $|0\rangle \leftrightarrow |2\rangle$ is also clearly visible, which allows us to verify that the anharmonicity $\alpha/(2\pi) = f_{12} - f_{01}$ of the qubit is enhanced at the sweet spot $\Phi_\mathrm{diff}/\Phi_0 = -0.5$ to $\alpha/(2\pi) = 434$~MHz. (See Extended Data Fig.~\ref{fig: ef rabis} for an alternative agreeing way to measure the anharmonicity.)

Figure~\ref{fig: resonator and qb spectroscopy}(c) presents fits to the experimental transition frequencies $f_{01}$ and $f_{02}/2$ based on two theoretical models of the circuit Hamiltonian, the first of which corresponds to Eq.~\eqref{eq: Hamiltonian with external flux} (model 1 in Methods) and the second of which is based on a path integral approach that does not require the dc phase $\varphi_0$ to be treated as a classical variable (model 2 in Methods). The fits agree very well with the experimental transition frequencies, especially near the sweet spots $\Phid = 0$ and $\Phid/\Phi_0 = -0.5$. Importantly, this good agreement with the models and the qubit frequency and anharmonicity is obtained with only three fitting parameters in a broad range of flux biases, and hence confirms our interpretation of the unimon physics (Fig.~\ref{fig: fig 1}d) and justifies the use of the models for reliable predictions of promising parameter regimes. According to the fits of model 1 (model 2), the capacitance and inductance per unit length of the unimon have a value of $\Cl = 87.1$ pF/m ($\Cl =79.8$ pF/m) and $\Ll = 0.821$ $\mu$H/m ($\Ll =0.893$ $\mu$H/m), respectively, in good agreement with the design values of  $\Cl = 83$ pF/m and $\Ll = 0.83$ $\mu$H/m. 

The measured sweet-spot anharmonicities of the five qubits are shown in Fig.~\ref{fig: anharmonicity and coherence times vs flux}a as functions of the Josephson energy $\EJ$ that is estimated by fitting the models 1 and 2 to the qubit spectroscopy data as in Figure~\ref{fig: resonator and qb spectroscopy}(c). The measured anharmonicities are slightly lower, but very close to the values predicted by the two theoretical models.
The qubits A and B exhibit the highest anharmonicities of  $\alpha/(2\pi) = 744$~MHz and  $434$~MHz, respectively, as a result of the largest cancellation between the inductive energy $\ELm$ and the Josephson energy $\EJ$. Importantly, the anharmonicity of the qubits A and B is significantly higher than that of typical transmon qubits, 200--300~MHz\cite{barends_coherent_2013}. Furthermore, the measured anharmonicities greatly exceed the capacitive energy $\ECm$ of the qubit mode unlike for transmons. 

To study the mechanisms determining the energy relaxation time $T_1$ of the unimon, we measure $T_1$ as a function of the qubit frequency as shown in Fig.~\ref{fig: anharmonicity and coherence times vs flux}b (see also Extended Data Figs.~\ref{fig: readout} and~\ref{fig: T1 example and range}). At the $\Phid = \Phi_0/2$ sweet spot, we find $T_1 \approx 8.6$~$\mu$s, whereas $T_1 \approx 4.6$ $\mu$s at $\Phid = 0$. Between these flux sweet spots, the relaxation time attains a minimum in a frequency range close to the frequency of the readout resonator $f_\mathrm{r} = 6.198$~GHz.  This behaviour of $T_1$ can be reasonably explained by dielectric losses with an effective quality factor of $Q_C \approx 1.7 \times 10^5$ and Purcell decay through the readout resonator (See Methods and Supplementary Note~III).  This suggests the qubit energy relaxation to be dominated by dielectric losses at $\Phid = \Phi_0/2$. The estimated quality factor of this first unimon qubit is 3--7 times higher than for the other geometric-superinductance qubit\cite{peruzzo_surpassing_2020}, but 2--3 times lower than in fluxoniums\cite{zhang2021universal,nguyen_high-coherence_2019} and an order of magnitude lower than in state-of-the-art transmons\cite{place_new_2021}. Improvements to design, materials, and fabrication processes are expected to reduce the dielectric losses in future unimon qubits compared with the very first samples presented here. 

To characterize the sensitivity of the qubit to flux noise, we measure the Ramsey coherence time $T_2^*$ and the echo coherence time $T_2^\mathrm{e}$ with a single echo $\pi$-pulse (Extended Data Fig.~\ref{fig: T2 and T2 echo examples}) as a function of the flux bias $\Phid$. Figure~\ref{fig: anharmonicity and coherence times vs flux}(c) shows that $T_2^*$ and $T_2^\mathrm{e}$ are both maximized at $\Phid = \Phi_0/2$, reaching 3.1~$\mu$s and 9.2~$\mu$s, respectively. Away from the sweet spot, the Ramsey coherence time $T_2^*$ degrades quickly, but the echo coherence time $T_2^\mathrm{e}$ stays above 1 $\mu$s even if the qubit frequency is tuned from the sweet spot by over 30 MHz. Assuming that the flux noise is described by a $1/f$ noise model $S_{\Phid}(\omega) = 2\pi A_{\Phid}^2/\omega$, we estimate a flux noise density of $A_{\Phid}/\sqrt{\textrm{Hz}} = 15.0$ $\mu\Phi_0/\sqrt{\textrm{Hz}}$ based on the flux dependence of $T_2^\mathrm{e}$ (Methods). The estimated flux noise density is an order of magnitude greater than in state-of-the-art SQUIDS\cite{braumuller2020characterizing}, but an order of magnitude lower than reported for all previous geometric-superinductance qubits\cite{Peruzzo2021Geometric}.

At $\Phid = 0$ in contrast, we measure a Ramsey coherence time of $T_2^* = 6.8$~$\mu$s and a $T_1$-limited echo coherence time of $T_2^\mathrm{e} = 9.9$~$\mu$s. The dephasing rate is lower here than at $\Phid = -\Phi_0/2$ since the qubit frequency is less sensitive to the external flux difference due to lower $|\partial \omega_{01}/\partial \Phid |$. Note that the anharmonicity of the qubit at $\Phid = 0$ is only $\alpha/(2\pi) = -58$~MHz, and hence this operation point is not of great interest for implementations of high-fidelity quantum logic.

 Next, we demonstrate that the high anharmonicity of the unimon and its protection against charge and flux noise enable us to implement fast high-fidelity single-qubit gates. To this end, we calibrate single-qubit gates of duration $t_\mathrm{g} \in [13.3, 46.6]$~ns using microwave pulses parametrized according to the derivative removal by adiabatic gate (DRAG) framework\cite{motzoi_simple_2009,chow_optimized_2010}. To characterize the average fidelity of gates in the set $\{I, X(\pi/2), Y(\pi/2) \}$, we utilize interleaved randomized benchmarking\cite{magesan2012efficient} (Methods). 
 Figure~\ref{fig: Gate fidelities and stability}(a) shows that we reach a practically coherence-limited fidelity of 99.9\% for $I$, $X(\pi/2)$, and $Y(\pi/2)$ gates at 13.3-ns duration. Our electronics limit the shortest gate pulses to 13.3~ns although the anharmonicity should allow for high-fidelity gates down to 5-ns duration corresponding to a gate fidelity of 99.97\% with the reported coherence properties.  
 
 To study the long-term stability of the gate fidelity, we first calibrate 20-ns single-qubit gates and then conduct repetitive measurements of the average gate fidelity using standard randomized benchmarking\cite{magesan_scalable_2011,magesan_characterizing_2012} without any recalibration between repetitions. % of the qubit or the gate parameters.
 Figure~\ref{fig: Gate fidelities and stability}(b) indicates that the measured gate fidelity is stable over the full period of eight hours with an average fidelity of $99.88 \pm 0.02$\%, practically coinciding with the coherence limit of 99.89\%. This stability can be attributed to the relaxation time $T_1$ and the coherence times $T_2^*$ and $T_2^\mathrm{e}$ staying practically constant in time as illustrated in Fig. ~\ref{fig: Gate fidelities and stability}c.

\noindent\textbf{Conclusions and outlook}

In conclusion, we introduced and demonstrated the unimon qubit that has a relatively high anharmonicity while requiring only a single Josephson junction without any superinductors, and being protected against both low-frequency charge noise and flux noise. The geometric inductance of the unimon has the potential for higher predictability and reproducibility than the junction-array-based superinductors in conventional fluxoniums or in quartons. Thus, the unimon constitutes a promising candidate for achieving single-qubit gate fidelities beyond 99.99\% in superconducting qubits with the help of the following future improvements: (i) redesign of the geometry to minimize dielectric losses\cite{lahtinen_effects_2020} currently dominating the energy relaxation, (ii) use of recently found low-loss materials\cite{place_new_2021}, and (iii) reduction of the gate duration to values well below 10~ns allowed even by the anharmonicities achieved here.
Future unimon research is also needed to study and minimize the various on-chip cross talks, implement two-qubit gates, and to scale up to many-qubit processors. To further reduce the sensitivity of the unimon to flux noise and to scale up the qubit count, it is likely beneficial to reduce the footprint of a single unimon qubit using, e.g., a superconductor with a high kinetic inductance in the coplanar-waveguide resonator. 
The anharmonicity of the unimon at flux bias $\Phid = \Phi_0/2$ has an opposite sign to that of the transmon, which may be helpful to suppress the unwanted residual $ZZ$ interaction with two-qubit-gate schemes that utilize qubits with opposite-sign anharmoncities\cite{yan_tunable_2018,zhao2020high}. The distributed-element nature of the unimon provides further opportunities for implementing a high connectivity and distant couplings in multi-qubit processors. In the future, we also aim to study the utilization of other modes of the unimon circuit, for example, for additional qubits and qubit readout.

\section*{References}

\bibliographystyle{naturemag}
%\bibliography{bibliography,references_Mikko_Zotero}{}
\bibliography{ms}{}

\begin{thebibliography}{10}
\expandafter\ifx\csname url\endcsname\relax
  \def\url#1{\texttt{#1}}\fi
\expandafter\ifx\csname urlprefix\endcsname\relax\def\urlprefix{URL }\fi
\providecommand{\bibinfo}[2]{#2}
\providecommand{\eprint}[2][]{\url{#2}}

\bibitem{krantz2019quantum}
\bibinfo{author}{Krantz, P.} \emph{et~al.}
\newblock \bibinfo{title}{A quantum engineer's guide to superconducting
  qubits}.
\newblock \emph{\bibinfo{journal}{Applied Physics Reviews}}
  \textbf{\bibinfo{volume}{6}}, \bibinfo{pages}{021318} (\bibinfo{year}{2019}).

\bibitem{blais_circuit_2021}
\bibinfo{author}{Blais, A.}, \bibinfo{author}{Grimsmo, A.~L.},
  \bibinfo{author}{Girvin, S.} \& \bibinfo{author}{Wallraff, A.}
\newblock \bibinfo{title}{Circuit quantum electrodynamics}.
\newblock \emph{\bibinfo{journal}{Reviews of Modern Physics}}
  \textbf{\bibinfo{volume}{93}}, \bibinfo{pages}{025005}
  (\bibinfo{year}{2021}).
\newblock
  \urlprefix\url{https://link.aps.org/doi/10.1103/RevModPhys.93.025005}.
\newblock \bibinfo{note}{Publisher: American Physical Society}.

\bibitem{yanay2020two}
\bibinfo{author}{Yanay, Y.}, \bibinfo{author}{Braum{\"u}ller, J.},
  \bibinfo{author}{Gustavsson, S.}, \bibinfo{author}{Oliver, W.~D.} \&
  \bibinfo{author}{Tahan, C.}
\newblock \bibinfo{title}{Two-dimensional hard-core bose--hubbard model with
  superconducting qubits}.
\newblock \emph{\bibinfo{journal}{npj Quantum Information}}
  \textbf{\bibinfo{volume}{6}}, \bibinfo{pages}{1--12} (\bibinfo{year}{2020}).

\bibitem{farhi_quantum_2014}
\bibinfo{author}{Farhi, E.}, \bibinfo{author}{Goldstone, J.} \&
  \bibinfo{author}{Gutmann, S.}
\newblock \bibinfo{title}{A {Quantum} {Approximate} {Optimization}
  {Algorithm}}.
\newblock \emph{\bibinfo{journal}{arXiv:1411.4028 [quant-ph]}}
  (\bibinfo{year}{2014}).
\newblock \urlprefix\url{http://arxiv.org/abs/1411.4028}.
\newblock \bibinfo{note}{ArXiv: 1411.4028}.

\bibitem{dunjko_machine_2018}
\bibinfo{author}{Dunjko, V.} \& \bibinfo{author}{Briegel, H.~J.}
\newblock \bibinfo{title}{Machine learning \& artificial intelligence in the
  quantum domain: a review of recent progress}.
\newblock \emph{\bibinfo{journal}{Reports on Progress in Physics}}
  \textbf{\bibinfo{volume}{81}}, \bibinfo{pages}{074001}
  (\bibinfo{year}{2018}).
\newblock \urlprefix\url{https://doi.org/10.1088/1361-6633/aab406}.
\newblock \bibinfo{note}{Publisher: IOP Publishing}.

\bibitem{mcardle_quantum_2020}
\bibinfo{author}{McArdle, S.}, \bibinfo{author}{Endo, S.},
  \bibinfo{author}{Aspuru-Guzik, A.}, \bibinfo{author}{Benjamin, S.~C.} \&
  \bibinfo{author}{Yuan, X.}
\newblock \bibinfo{title}{Quantum computational chemistry}.
\newblock \emph{\bibinfo{journal}{Reviews of Modern Physics}}
  \textbf{\bibinfo{volume}{92}}, \bibinfo{pages}{015003}
  (\bibinfo{year}{2020}).
\newblock
  \urlprefix\url{https://link.aps.org/doi/10.1103/RevModPhys.92.015003}.

\bibitem{arute_quantum_2019}
\bibinfo{author}{Arute, F.} \emph{et~al.}
\newblock \bibinfo{title}{Quantum supremacy using a programmable
  superconducting processor}.
\newblock \emph{\bibinfo{journal}{Nature}} \textbf{\bibinfo{volume}{574}},
  \bibinfo{pages}{505--510} (\bibinfo{year}{2019}).
\newblock \urlprefix\url{https://www.nature.com/articles/s41586-019-1666-5}.

\bibitem{wu_strong_2021}
\bibinfo{author}{Wu, Y.} \emph{et~al.}
\newblock \bibinfo{title}{Strong quantum computational advantage using a
  superconducting quantum processor}.
\newblock \emph{\bibinfo{journal}{arXiv:2106.14734 [quant-ph]}}
  (\bibinfo{year}{2021}).
\newblock \urlprefix\url{http://arxiv.org/abs/2106.14734}.
\newblock \bibinfo{note}{ArXiv: 2106.14734}.

\bibitem{ofekextending_2016}
\bibinfo{author}{Ofek, N.} \emph{et~al.}
\newblock \bibinfo{title}{Extending the lifetime of a quantum bit with error
  correction in superconducting circuits}.
\newblock \emph{\bibinfo{journal}{Nature}} \textbf{\bibinfo{volume}{536}}
  (\bibinfo{year}{2016}).
\newblock \urlprefix\url{https://www.nature.com/articles/nature18949}.

\bibitem{campagne-ibarcqquantum_2020}
\bibinfo{author}{Campagne-Ibarcq, P.} \emph{et~al.}
\newblock \bibinfo{title}{Quantum error correction of a qubit encoded in grid
  states of an oscillator}.
\newblock \emph{\bibinfo{journal}{Nature}} \textbf{\bibinfo{volume}{584}}
  (\bibinfo{year}{2020}).
\newblock \urlprefix\url{https://www.nature.com/articles/s41586-020-2603-3}.

\bibitem{chenexponential_2021}
\bibinfo{author}{Chen, Z.} \emph{et~al.}
\newblock \bibinfo{title}{Exponential suppression of bit or phase errors with
  cyclic error correction}.
\newblock \emph{\bibinfo{journal}{Nature}} \textbf{\bibinfo{volume}{595}}
  (\bibinfo{year}{2021}).
\newblock \urlprefix\url{https://www.nature.com/articles/s41586-021-03588-y}.

\bibitem{koch_charge-insensitive_2007}
\bibinfo{author}{Koch, J.} \emph{et~al.}
\newblock \bibinfo{title}{Charge-insensitive qubit design derived from the
  {Cooper} pair box}.
\newblock \emph{\bibinfo{journal}{Physical Review A}}
  \textbf{\bibinfo{volume}{76}}, \bibinfo{pages}{042319}
  (\bibinfo{year}{2007}).
\newblock \urlprefix\url{https://link.aps.org/doi/10.1103/PhysRevA.76.042319}.

\bibitem{nguyen_high-coherence_2019}
\bibinfo{author}{Nguyen, L.~B.} \emph{et~al.}
\newblock \bibinfo{title}{High-{Coherence} {Fluxonium} {Qubit}}.
\newblock \emph{\bibinfo{journal}{Physical Review X}}
  \textbf{\bibinfo{volume}{9}}, \bibinfo{pages}{041041} (\bibinfo{year}{2019}).
\newblock \urlprefix\url{https://link.aps.org/doi/10.1103/PhysRevX.9.041041}.
\newblock \bibinfo{note}{Publisher: American Physical Society}.

\bibitem{pechenezhskiy_superconducting_2020}
\bibinfo{author}{Pechenezhskiy, I.~V.}, \bibinfo{author}{Mencia, R.~A.},
  \bibinfo{author}{Nguyen, L.~B.}, \bibinfo{author}{Lin, Y.-H.} \&
  \bibinfo{author}{Manucharyan, V.~E.}
\newblock \bibinfo{title}{The superconducting quasicharge qubit}.
\newblock \emph{\bibinfo{journal}{Nature}} \textbf{\bibinfo{volume}{585}},
  \bibinfo{pages}{368--371} (\bibinfo{year}{2020}).
\newblock \urlprefix\url{https://www.nature.com/articles/s41586-020-2687-9}.
\newblock \bibinfo{note}{Bandiera\_abtest: a Cg\_type: Nature Research Journals
  Number: 7825 Primary\_atype: Research Publisher: Nature Publishing Group
  Subject\_term: Quantum information;Qubits;Superconducting devices
  Subject\_term\_id: quantum-information;qubits;superconducting-devices}.

\bibitem{gyenis_experimental_2021}
\bibinfo{author}{Gyenis, A.} \emph{et~al.}
\newblock \bibinfo{title}{Experimental {Realization} of a {Protected}
  {Superconducting} {Circuit} {Derived} from the
  \$0\$--\${\textbackslash}ensuremath\{{\textbackslash}pi\}\$ {Qubit}}.
\newblock \emph{\bibinfo{journal}{PRX Quantum}} \textbf{\bibinfo{volume}{2}},
  \bibinfo{pages}{010339} (\bibinfo{year}{2021}).
\newblock \urlprefix\url{https://link.aps.org/doi/10.1103/PRXQuantum.2.010339}.

\bibitem{yan_engineering_2020}
\bibinfo{author}{Yan, F.} \emph{et~al.}
\newblock \bibinfo{title}{Engineering {Framework} for {Optimizing}
  {Superconducting} {Qubit} {Designs}}.
\newblock \emph{\bibinfo{journal}{arXiv:2006.04130 [quant-ph]}}
  (\bibinfo{year}{2020}).
\newblock \urlprefix\url{http://arxiv.org/abs/2006.04130}.
\newblock \bibinfo{note}{ArXiv: 2006.04130}.

\bibitem{preskill_quantum_2018}
\bibinfo{author}{Preskill, J.}
\newblock \bibinfo{title}{Quantum {Computing} in the {NISQ} era and beyond}.
\newblock \emph{\bibinfo{journal}{Quantum}} \textbf{\bibinfo{volume}{2}},
  \bibinfo{pages}{79} (\bibinfo{year}{2018}).
\newblock \urlprefix\url{https://quantum-journal.org/papers/q-2018-08-06-79/}.

\bibitem{nakahara2008quantum}
\bibinfo{author}{Nakahara, M.}
\newblock \emph{\bibinfo{title}{Quantum computing: from linear algebra to
  physical realizations}} (\bibinfo{publisher}{CRC press},
  \bibinfo{year}{2008}).

\bibitem{temme2017error}
\bibinfo{author}{Temme, K.}, \bibinfo{author}{Bravyi, S.} \&
  \bibinfo{author}{Gambetta, J.~M.}
\newblock \bibinfo{title}{Error mitigation for short-depth quantum circuits}.
\newblock \emph{\bibinfo{journal}{Physical review letters}}
  \textbf{\bibinfo{volume}{119}}, \bibinfo{pages}{180509}
  (\bibinfo{year}{2017}).

\bibitem{kandala2019error}
\bibinfo{author}{Kandala, A.} \emph{et~al.}
\newblock \bibinfo{title}{Error mitigation extends the computational reach of a
  noisy quantum processor}.
\newblock \emph{\bibinfo{journal}{Nature}} \textbf{\bibinfo{volume}{567}},
  \bibinfo{pages}{491--495} (\bibinfo{year}{2019}).

\bibitem{gottesman1997stabilizer}
\bibinfo{author}{Gottesman, D.}
\newblock \emph{\bibinfo{title}{Stabilizer codes and quantum error
  correction}}.
\newblock Ph.D. thesis, \bibinfo{school}{California Institute of Technology}
  (\bibinfo{year}{1997}).

\bibitem{fowler_surface_2012}
\bibinfo{author}{Fowler, A.~G.}, \bibinfo{author}{Mariantoni, M.},
  \bibinfo{author}{Martinis, J.~M.} \& \bibinfo{author}{Cleland, A.~N.}
\newblock \bibinfo{title}{Surface codes: {Towards} practical large-scale
  quantum computation}.
\newblock \emph{\bibinfo{journal}{Physical Review A}}
  \textbf{\bibinfo{volume}{86}}, \bibinfo{pages}{032324}
  (\bibinfo{year}{2012}).
\newblock \urlprefix\url{https://link.aps.org/doi/10.1103/PhysRevA.86.032324}.

\bibitem{bonilla2021xzzx}
\bibinfo{author}{Bonilla~Ataides, J.~P.}, \bibinfo{author}{Tuckett, D.~K.},
  \bibinfo{author}{Bartlett, S.~D.}, \bibinfo{author}{Flammia, S.~T.} \&
  \bibinfo{author}{Brown, B.~J.}
\newblock \bibinfo{title}{The xzzx surface code}.
\newblock \emph{\bibinfo{journal}{Nature communications}}
  \textbf{\bibinfo{volume}{12}}, \bibinfo{pages}{1--12} (\bibinfo{year}{2021}).

\bibitem{barends_superconducting_2014}
\bibinfo{author}{Barends, R.} \emph{et~al.}
\newblock \bibinfo{title}{Superconducting quantum circuits at the surface code
  threshold for fault tolerance}.
\newblock \emph{\bibinfo{journal}{Nature}} \textbf{\bibinfo{volume}{508}},
  \bibinfo{pages}{500--503} (\bibinfo{year}{2014}).
\newblock \urlprefix\url{https://www.nature.com/articles/nature13171}.

\bibitem{andersen_repeated_2020}
\bibinfo{author}{Andersen, C.~K.} \emph{et~al.}
\newblock \bibinfo{title}{Repeated quantum error detection in a surface code}.
\newblock \emph{\bibinfo{journal}{Nature Physics}}
  \textbf{\bibinfo{volume}{16}}, \bibinfo{pages}{875--880}
  (\bibinfo{year}{2020}).
\newblock \urlprefix\url{https://www.nature.com/articles/s41567-020-0920-y}.

\bibitem{marques2021logical}
\bibinfo{author}{Marques, J.} \emph{et~al.}
\newblock \bibinfo{title}{Logical-qubit operations in an error-detecting
  surface code}.
\newblock \emph{\bibinfo{journal}{Nature Physics}} \bibinfo{pages}{1--7}
  (\bibinfo{year}{2021}).

\bibitem{krinner2021realizing}
\bibinfo{author}{Krinner, S.} \emph{et~al.}
\newblock \bibinfo{title}{Realizing repeated quantum error correction in a
  distance-three surface code}.
\newblock \emph{\bibinfo{journal}{arXiv preprint arXiv:2112.03708}}
  (\bibinfo{year}{2021}).

\bibitem{zhao2021realizing}
\bibinfo{author}{Zhao, Y.} \emph{et~al.}
\newblock \bibinfo{title}{Realizing an error-correcting surface code with
  superconducting qubits}.
\newblock \emph{\bibinfo{journal}{arXiv preprint arXiv:2112.13505}}
  (\bibinfo{year}{2021}).

\bibitem{hertzberg2021laser}
\bibinfo{author}{Hertzberg, J.~B.} \emph{et~al.}
\newblock \bibinfo{title}{Laser-annealing josephson junctions for yielding
  scaled-up superconducting quantum processors}.
\newblock \emph{\bibinfo{journal}{npj Quantum Information}}
  \textbf{\bibinfo{volume}{7}}, \bibinfo{pages}{1--8} (\bibinfo{year}{2021}).

\bibitem{zhu2021quantum}
\bibinfo{author}{Zhu, Q.} \emph{et~al.}
\newblock \bibinfo{title}{Quantum computational advantage via 60-qubit 24-cycle
  random circuit sampling}.
\newblock \emph{\bibinfo{journal}{Science Bulletin}}  (\bibinfo{year}{2021}).

\bibitem{place_new_2021}
\bibinfo{author}{Place, A. P.~M.} \emph{et~al.}
\newblock \bibinfo{title}{New material platform for superconducting transmon
  qubits with coherence times exceeding 0.3 milliseconds}.
\newblock \emph{\bibinfo{journal}{Nature Communications}}
  \textbf{\bibinfo{volume}{12}}, \bibinfo{pages}{1779} (\bibinfo{year}{2021}).
\newblock \urlprefix\url{https://www.nature.com/articles/s41467-021-22030-5}.

\bibitem{wang2022towards}
\bibinfo{author}{Wang, C.} \emph{et~al.}
\newblock \bibinfo{title}{Towards practical quantum computers: transmon qubit
  with a lifetime approaching 0.5 milliseconds}.
\newblock \emph{\bibinfo{journal}{npj Quantum Information}}
  \textbf{\bibinfo{volume}{8}}, \bibinfo{pages}{1--6} (\bibinfo{year}{2022}).

\bibitem{mckay2017efficient}
\bibinfo{author}{McKay, D.~C.}, \bibinfo{author}{Wood, C.~J.},
  \bibinfo{author}{Sheldon, S.}, \bibinfo{author}{Chow, J.~M.} \&
  \bibinfo{author}{Gambetta, J.~M.}
\newblock \bibinfo{title}{Efficient z gates for quantum computing}.
\newblock \emph{\bibinfo{journal}{Physical Review A}}
  \textbf{\bibinfo{volume}{96}}, \bibinfo{pages}{022330}
  (\bibinfo{year}{2017}).

\bibitem{sungrealization_2021}
\bibinfo{author}{Sung, Y.} \emph{et~al.}
\newblock \bibinfo{title}{Realization of {High}-{Fidelity} {CZ} and {ZZ}-{Free}
  {iSWAP} {Gates} with a {Tunable} {Coupler}}.
\newblock \emph{\bibinfo{journal}{Physical Review X}}
  \textbf{\bibinfo{volume}{11}}, \bibinfo{pages}{021058}
  (\bibinfo{year}{2021}).
\newblock \urlprefix\url{https://link.aps.org/doi/10.1103/PhysRevX.11.021058}.
\newblock \bibinfo{note}{Publisher: American Physical Society}.

\bibitem{nakamura_coherent_1999}
\bibinfo{author}{Nakamura, Y.}, \bibinfo{author}{Pashkin, Y.~A.} \&
  \bibinfo{author}{Tsai, J.~S.}
\newblock \bibinfo{title}{Coherent control of macroscopic quantum states in a
  single-{Cooper}-pair box}.
\newblock \emph{\bibinfo{journal}{Nature}} \textbf{\bibinfo{volume}{398}},
  \bibinfo{pages}{786--788} (\bibinfo{year}{1999}).
\newblock \urlprefix\url{https://www.nature.com/articles/19718}.

\bibitem{barends_coherent_2013}
\bibinfo{author}{Barends, R.} \emph{et~al.}
\newblock \bibinfo{title}{Coherent {Josephson} {Qubit} {Suitable} for
  {Scalable} {Quantum} {Integrated} {Circuits}}.
\newblock \emph{\bibinfo{journal}{Physical Review Letters}}
  \textbf{\bibinfo{volume}{111}}, \bibinfo{pages}{080502}
  (\bibinfo{year}{2013}).
\newblock
  \urlprefix\url{https://link.aps.org/doi/10.1103/PhysRevLett.111.080502}.

\bibitem{wood2018quantification}
\bibinfo{author}{Wood, C.~J.} \& \bibinfo{author}{Gambetta, J.~M.}
\newblock \bibinfo{title}{Quantification and characterization of leakage
  errors}.
\newblock \emph{\bibinfo{journal}{Physical Review A}}
  \textbf{\bibinfo{volume}{97}}, \bibinfo{pages}{032306}
  (\bibinfo{year}{2018}).

\bibitem{lescanne_escape_2019}
\bibinfo{author}{Lescanne, R.} \emph{et~al.}
\newblock \bibinfo{title}{Escape of a {Driven} {Quantum} {Josephson} {Circuit}
  into {Unconfined} {States}}.
\newblock \emph{\bibinfo{journal}{Physical Review Applied}}
  \textbf{\bibinfo{volume}{11}}, \bibinfo{pages}{014030}
  (\bibinfo{year}{2019}).
\newblock
  \urlprefix\url{https://link.aps.org/doi/10.1103/PhysRevApplied.11.014030}.
\newblock \bibinfo{note}{Publisher: American Physical Society}.

\bibitem{manucharyan_fluxonium_2009}
\bibinfo{author}{Manucharyan, V.~E.}, \bibinfo{author}{Koch, J.},
  \bibinfo{author}{Glazman, L.~I.} \& \bibinfo{author}{Devoret, M.~H.}
\newblock \bibinfo{title}{Fluxonium: {Single} {Cooper}-{Pair} {Circuit} {Free}
  of {Charge} {Offsets}}.
\newblock \emph{\bibinfo{journal}{Science}} \textbf{\bibinfo{volume}{326}},
  \bibinfo{pages}{113--116} (\bibinfo{year}{2009}).
\newblock \urlprefix\url{https://science.sciencemag.org/content/326/5949/113}.

\bibitem{bao2021fluxonium}
\bibinfo{author}{Bao, F.} \emph{et~al.}
\newblock \bibinfo{title}{Fluxonium: an alternative qubit platform for
  high-fidelity operations}.
\newblock \emph{\bibinfo{journal}{arXiv preprint arXiv:2111.13504}}
  (\bibinfo{year}{2021}).

\bibitem{zhang2021universal}
\bibinfo{author}{Zhang, H.} \emph{et~al.}
\newblock \bibinfo{title}{Universal fast-flux control of a coherent,
  low-frequency qubit}.
\newblock \emph{\bibinfo{journal}{Physical Review X}}
  \textbf{\bibinfo{volume}{11}}, \bibinfo{pages}{011010}
  (\bibinfo{year}{2021}).

\bibitem{grunhaupt2019granular}
\bibinfo{author}{Gr{\"u}nhaupt, L.} \emph{et~al.}
\newblock \bibinfo{title}{Granular aluminium as a superconducting material for
  high-impedance quantum circuits}.
\newblock \emph{\bibinfo{journal}{Nature materials}}
  \textbf{\bibinfo{volume}{18}}, \bibinfo{pages}{816--819}
  (\bibinfo{year}{2019}).

\bibitem{hazard_nanowire_2019}
\bibinfo{author}{Hazard, T.} \emph{et~al.}
\newblock \bibinfo{title}{Nanowire {Superinductance} {Fluxonium} {Qubit}}.
\newblock \emph{\bibinfo{journal}{Physical Review Letters}}
  \textbf{\bibinfo{volume}{122}}, \bibinfo{pages}{010504}
  (\bibinfo{year}{2019}).
\newblock
  \urlprefix\url{https://link.aps.org/doi/10.1103/PhysRevLett.122.010504}.
\newblock \bibinfo{note}{Publisher: American Physical Society}.

\bibitem{Peruzzo2021Geometric}
\bibinfo{author}{Peruzzo, M.} \emph{et~al.}
\newblock \bibinfo{title}{Geometric superinductance qubits: Controlling phase
  delocalization across a single josephson junction}.
\newblock \emph{\bibinfo{journal}{PRX Quantum}} \textbf{\bibinfo{volume}{2}},
  \bibinfo{pages}{040341} (\bibinfo{year}{2021}).
\newblock \urlprefix\url{https://link.aps.org/doi/10.1103/PRXQuantum.2.040341}.

\bibitem{earnestrealization_2018}
\bibinfo{author}{Earnest, N.} \emph{et~al.}
\newblock \bibinfo{title}{Realization of a {Lambda}-{System} with {Metastable}
  {States} of a {Capacitively} {Shunted} {Fluxonium}}.
\newblock \emph{\bibinfo{journal}{Physical Review Letters}}
  \textbf{\bibinfo{volume}{120}}, \bibinfo{pages}{150504}
  (\bibinfo{year}{2018}).
\newblock
  \urlprefix\url{https://link.aps.org/doi/10.1103/PhysRevLett.120.150504}.

\bibitem{liu2021quantum}
\bibinfo{author}{Liu, F.-M.} \emph{et~al.}
\newblock \bibinfo{title}{Quantum design for advanced qubits}.
\newblock \emph{\bibinfo{journal}{arXiv preprint arXiv:2109.00994}}
  (\bibinfo{year}{2021}).

\bibitem{kalashnikov2020bifluxon}
\bibinfo{author}{Kalashnikov, K.} \emph{et~al.}
\newblock \bibinfo{title}{Bifluxon: Fluxon-parity-protected superconducting
  qubit}.
\newblock \emph{\bibinfo{journal}{PRX Quantum}} \textbf{\bibinfo{volume}{1}},
  \bibinfo{pages}{010307} (\bibinfo{year}{2020}).

\bibitem{smith2020superconducting}
\bibinfo{author}{Smith, W.}, \bibinfo{author}{Kou, A.}, \bibinfo{author}{Xiao,
  X.}, \bibinfo{author}{Vool, U.} \& \bibinfo{author}{Devoret, M.}
\newblock \bibinfo{title}{Superconducting circuit protected by two-cooper-pair
  tunneling}.
\newblock \emph{\bibinfo{journal}{npj Quantum Information}}
  \textbf{\bibinfo{volume}{6}}, \bibinfo{pages}{1--9} (\bibinfo{year}{2020}).

\bibitem{blais_cavity_2004}
\bibinfo{author}{Blais, A.}, \bibinfo{author}{Huang, R.-S.},
  \bibinfo{author}{Wallraff, A.}, \bibinfo{author}{Girvin, S.~M.} \&
  \bibinfo{author}{Schoelkopf, R.~J.}
\newblock \bibinfo{title}{Cavity quantum electrodynamics for superconducting
  electrical circuits: {An} architecture for quantum computation}.
\newblock \emph{\bibinfo{journal}{Physical Review A}}
  \textbf{\bibinfo{volume}{69}}, \bibinfo{pages}{062320}
  (\bibinfo{year}{2004}).
\newblock \urlprefix\url{https://link.aps.org/doi/10.1103/PhysRevA.69.062320}.

\bibitem{siddiqi2006dispersive}
\bibinfo{author}{Siddiqi, I.} \emph{et~al.}
\newblock \bibinfo{title}{Dispersive measurements of superconducting qubit
  coherence with a fast latching readout}.
\newblock \emph{\bibinfo{journal}{Physical Review B}}
  \textbf{\bibinfo{volume}{73}}, \bibinfo{pages}{054510}
  (\bibinfo{year}{2006}).

\bibitem{peruzzo_surpassing_2020}
\bibinfo{author}{Peruzzo, M.}, \bibinfo{author}{Trioni, A.},
  \bibinfo{author}{Hassani, F.}, \bibinfo{author}{Zemlicka, M.} \&
  \bibinfo{author}{Fink, J.~M.}
\newblock \bibinfo{title}{Surpassing the {Resistance} {Quantum} with a
  {Geometric} {Superinductor}}.
\newblock \emph{\bibinfo{journal}{Physical Review Applied}}
  \textbf{\bibinfo{volume}{14}}, \bibinfo{pages}{044055}
  (\bibinfo{year}{2020}).
\newblock
  \urlprefix\url{https://link.aps.org/doi/10.1103/PhysRevApplied.14.044055}.
\newblock \bibinfo{note}{Publisher: American Physical Society}.

\bibitem{braumuller2020characterizing}
\bibinfo{author}{Braum{\"u}ller, J.} \emph{et~al.}
\newblock \bibinfo{title}{Characterizing and optimizing qubit coherence based
  on squid geometry}.
\newblock \emph{\bibinfo{journal}{Physical Review Applied}}
  \textbf{\bibinfo{volume}{13}}, \bibinfo{pages}{054079}
  (\bibinfo{year}{2020}).

\bibitem{motzoi_simple_2009}
\bibinfo{author}{Motzoi, F.}, \bibinfo{author}{Gambetta, J.~M.},
  \bibinfo{author}{Rebentrost, P.} \& \bibinfo{author}{Wilhelm, F.~K.}
\newblock \bibinfo{title}{Simple {Pulses} for {Elimination} of {Leakage} in
  {Weakly} {Nonlinear} {Qubits}}.
\newblock \emph{\bibinfo{journal}{Physical Review Letters}}
  \textbf{\bibinfo{volume}{103}}, \bibinfo{pages}{110501}
  (\bibinfo{year}{2009}).
\newblock
  \urlprefix\url{https://link.aps.org/doi/10.1103/PhysRevLett.103.110501}.
\newblock \bibinfo{note}{Publisher: American Physical Society}.

\bibitem{chow_optimized_2010}
\bibinfo{author}{Chow, J.~M.} \emph{et~al.}
\newblock \bibinfo{title}{Optimized driving of superconducting artificial atoms
  for improved single-qubit gates}.
\newblock \emph{\bibinfo{journal}{Physical Review A}}
  \textbf{\bibinfo{volume}{82}}, \bibinfo{pages}{040305}
  (\bibinfo{year}{2010}).
\newblock \urlprefix\url{https://link.aps.org/doi/10.1103/PhysRevA.82.040305}.
\newblock \bibinfo{note}{Publisher: American Physical Society}.

\bibitem{magesan2012efficient}
\bibinfo{author}{Magesan, E.} \emph{et~al.}
\newblock \bibinfo{title}{Efficient measurement of quantum gate error by
  interleaved randomized benchmarking}.
\newblock \emph{\bibinfo{journal}{Physical review letters}}
  \textbf{\bibinfo{volume}{109}}, \bibinfo{pages}{080505}
  (\bibinfo{year}{2012}).

\bibitem{magesan_scalable_2011}
\bibinfo{author}{Magesan, E.}, \bibinfo{author}{Gambetta, J.~M.} \&
  \bibinfo{author}{Emerson, J.}
\newblock \bibinfo{title}{Scalable and {Robust} {Randomized} {Benchmarking} of
  {Quantum} {Processes}}.
\newblock \emph{\bibinfo{journal}{Physical Review Letters}}
  \textbf{\bibinfo{volume}{106}}, \bibinfo{pages}{180504}
  (\bibinfo{year}{2011}).
\newblock
  \urlprefix\url{https://link.aps.org/doi/10.1103/PhysRevLett.106.180504}.
\newblock \bibinfo{note}{Publisher: American Physical Society}.

\bibitem{magesan_characterizing_2012}
\bibinfo{author}{Magesan, E.}, \bibinfo{author}{Gambetta, J.~M.} \&
  \bibinfo{author}{Emerson, J.}
\newblock \bibinfo{title}{Characterizing quantum gates via randomized
  benchmarking}.
\newblock \emph{\bibinfo{journal}{Physical Review A}}
  \textbf{\bibinfo{volume}{85}}, \bibinfo{pages}{042311}
  (\bibinfo{year}{2012}).
\newblock \urlprefix\url{https://link.aps.org/doi/10.1103/PhysRevA.85.042311}.

\bibitem{lahtinen_effects_2020}
\bibinfo{author}{Lahtinen, V.} \& \bibinfo{author}{Möttönen, M.}
\newblock \bibinfo{title}{Effects of device geometry and material properties on
  dielectric losses in superconducting coplanar-waveguide resonators}.
\newblock \emph{\bibinfo{journal}{Journal of Physics: Condensed Matter}}
  \textbf{\bibinfo{volume}{32}}, \bibinfo{pages}{405702}
  (\bibinfo{year}{2020}).
\newblock \urlprefix\url{https://doi.org/10.1088/1361-648x/ab98c8}.
\newblock \bibinfo{note}{Publisher: IOP Publishing}.

\bibitem{yan_tunable_2018}
\bibinfo{author}{Yan, F.} \emph{et~al.}
\newblock \bibinfo{title}{Tunable {Coupling} {Scheme} for {Implementing}
  {High}-{Fidelity} {Two}-{Qubit} {Gates}}.
\newblock \emph{\bibinfo{journal}{Physical Review Applied}}
  \textbf{\bibinfo{volume}{10}}, \bibinfo{pages}{054062}
  (\bibinfo{year}{2018}).
\newblock
  \urlprefix\url{https://link.aps.org/doi/10.1103/PhysRevApplied.10.054062}.
\newblock \bibinfo{note}{Publisher: American Physical Society}.

\bibitem{zhao2020high}
\bibinfo{author}{Zhao, P.} \emph{et~al.}
\newblock \bibinfo{title}{High-contrast z z interaction using superconducting
  qubits with opposite-sign anharmonicity}.
\newblock \emph{\bibinfo{journal}{Physical Review Letters}}
  \textbf{\bibinfo{volume}{125}}, \bibinfo{pages}{200503}
  (\bibinfo{year}{2020}).

\bibitem{bourassa2012josephson}
\bibinfo{author}{Bourassa, J.}, \bibinfo{author}{Beaudoin, F.},
  \bibinfo{author}{Gambetta, J.~M.} \& \bibinfo{author}{Blais, A.}
\newblock \bibinfo{title}{Josephson-junction-embedded transmission-line
  resonators: From kerr medium to in-line transmon}.
\newblock \emph{\bibinfo{journal}{Physical Review A}}
  \textbf{\bibinfo{volume}{86}}, \bibinfo{pages}{013814}
  (\bibinfo{year}{2012}).

\bibitem{vool2017introduction}
\bibinfo{author}{Vool, U.} \& \bibinfo{author}{Devoret, M.}
\newblock \bibinfo{title}{Introduction to quantum electromagnetic circuits}.
\newblock \emph{\bibinfo{journal}{International Journal of Circuit Theory and
  Applications}} \textbf{\bibinfo{volume}{45}}, \bibinfo{pages}{897--934}
  (\bibinfo{year}{2017}).

\bibitem{kqcircuits}
\bibinfo{author}{Heinsoo, J.} \emph{et~al.}
\newblock \bibinfo{title}{{KQC}ircuits} (\bibinfo{year}{2021}).
\newblock \urlprefix\url{https://github.com/iqm-finland/KQCircuits}.

\bibitem{klayout}
\bibinfo{author}{K{\"o}fferlein, M.}
\newblock \bibinfo{title}{{KL}ayout} (\bibinfo{year}{2021}).
\newblock \urlprefix\url{https://www.klayout.de/}.

\bibitem{simons2004coplanar}
\bibinfo{author}{Simons, R.~N.}
\newblock \emph{\bibinfo{title}{Coplanar waveguide circuits, components, and
  systems}} (\bibinfo{publisher}{John Wiley \& Sons}, \bibinfo{year}{2004}).

\bibitem{schoelkopf2003qubits}
\bibinfo{author}{Schoelkopf, R.}, \bibinfo{author}{Clerk, A.},
  \bibinfo{author}{Girvin, S.}, \bibinfo{author}{Lehnert, K.} \&
  \bibinfo{author}{Devoret, M.}
\newblock \bibinfo{title}{Qubits as spectrometers of quantum noise}.
\newblock In \emph{\bibinfo{booktitle}{Quantum noise in mesoscopic physics}},
  \bibinfo{pages}{175--203} (\bibinfo{publisher}{Springer},
  \bibinfo{year}{2003}).

\bibitem{bylander2011noise}
\bibinfo{author}{Bylander, J.} \emph{et~al.}
\newblock \bibinfo{title}{Noise spectroscopy through dynamical decoupling with
  a superconducting flux qubit}.
\newblock \emph{\bibinfo{journal}{Nature Physics}}
  \textbf{\bibinfo{volume}{7}}, \bibinfo{pages}{565--570}
  (\bibinfo{year}{2011}).

\bibitem{yan_flux_2016}
\bibinfo{author}{Yan, F.} \emph{et~al.}
\newblock \bibinfo{title}{The flux qubit revisited to enhance coherence and
  reproducibility}.
\newblock \emph{\bibinfo{journal}{Nature Communications}}
  \textbf{\bibinfo{volume}{7}}, \bibinfo{pages}{12964} (\bibinfo{year}{2016}).
\newblock \urlprefix\url{https://www.nature.com/articles/ncomms12964}.

\bibitem{ithier2005decoherence}
\bibinfo{author}{Ithier, G.} \emph{et~al.}
\newblock \bibinfo{title}{Decoherence in a superconducting quantum bit
  circuit}.
\newblock \emph{\bibinfo{journal}{Physical Review B}}
  \textbf{\bibinfo{volume}{72}}, \bibinfo{pages}{134519}
  (\bibinfo{year}{2005}).

\bibitem{reed2013entanglement}
\bibinfo{author}{Reed, M.}
\newblock \emph{\bibinfo{title}{Entanglement and quantum error correction with
  superconducting qubits}}.
\newblock Ph.D. thesis, \bibinfo{school}{Yale University}
  (\bibinfo{year}{2013}).

\bibitem{nielsen2002simple}
\bibinfo{author}{Nielsen, M.~A.}
\newblock \bibinfo{title}{A simple formula for the average gate fidelity of a
  quantum dynamical operation}.
\newblock \emph{\bibinfo{journal}{Physics Letters A}}
  \textbf{\bibinfo{volume}{303}}, \bibinfo{pages}{249--252}
  (\bibinfo{year}{2002}).

\bibitem{epstein_investigating_2014}
\bibinfo{author}{Epstein, J.~M.}, \bibinfo{author}{Cross, A.~W.},
  \bibinfo{author}{Magesan, E.} \& \bibinfo{author}{Gambetta, J.~M.}
\newblock \bibinfo{title}{Investigating the limits of randomized benchmarking
  protocols}.
\newblock \emph{\bibinfo{journal}{Physical Review A}}
  \textbf{\bibinfo{volume}{89}}, \bibinfo{pages}{062321}
  (\bibinfo{year}{2014}).
\newblock \urlprefix\url{https://link.aps.org/doi/10.1103/PhysRevA.89.062321}.

\bibitem{nakamura1999coherent}
\bibinfo{author}{Nakamura, Y.}, \bibinfo{author}{Pashkin, Y.~A.} \&
  \bibinfo{author}{Tsai, J.}
\newblock \bibinfo{title}{Coherent control of macroscopic quantum states in a
  single-cooper-pair box}.
\newblock \emph{\bibinfo{journal}{nature}} \textbf{\bibinfo{volume}{398}},
  \bibinfo{pages}{786--788} (\bibinfo{year}{1999}).

\bibitem{barends2013coherent}
\bibinfo{author}{Barends, R.} \emph{et~al.}
\newblock \bibinfo{title}{Coherent josephson qubit suitable for scalable
  quantum integrated circuits}.
\newblock \emph{\bibinfo{journal}{Physical review letters}}
  \textbf{\bibinfo{volume}{111}}, \bibinfo{pages}{080502}
  (\bibinfo{year}{2013}).

\bibitem{manucharyan2009fluxonium}
\bibinfo{author}{Manucharyan, V.~E.}, \bibinfo{author}{Koch, J.},
  \bibinfo{author}{Glazman, L.~I.} \& \bibinfo{author}{Devoret, M.~H.}
\newblock \bibinfo{title}{Fluxonium: Single cooper-pair circuit free of charge
  offsets}.
\newblock \emph{\bibinfo{journal}{Science}} \textbf{\bibinfo{volume}{326}},
  \bibinfo{pages}{113--116} (\bibinfo{year}{2009}).

\bibitem{hazard2019nanowire}
\bibinfo{author}{Hazard, T.} \emph{et~al.}
\newblock \bibinfo{title}{Nanowire superinductance fluxonium qubit}.
\newblock \emph{\bibinfo{journal}{Physical review letters}}
  \textbf{\bibinfo{volume}{122}}, \bibinfo{pages}{010504}
  (\bibinfo{year}{2019}).

\bibitem{pechenezhskiy2020superconducting}
\bibinfo{author}{Pechenezhskiy, I.~V.}, \bibinfo{author}{Mencia, R.~A.},
  \bibinfo{author}{Nguyen, L.~B.}, \bibinfo{author}{Lin, Y.-H.} \&
  \bibinfo{author}{Manucharyan, V.~E.}
\newblock \bibinfo{title}{The superconducting quasicharge qubit}.
\newblock \emph{\bibinfo{journal}{Nature}} \textbf{\bibinfo{volume}{585}},
  \bibinfo{pages}{368--371} (\bibinfo{year}{2020}).

\bibitem{yan2020engineering}
\bibinfo{author}{Yan, F.} \emph{et~al.}
\newblock \bibinfo{title}{Engineering framework for optimizing superconducting
  qubit designs}.
\newblock \emph{\bibinfo{journal}{arXiv preprint arXiv:2006.04130}}
  (\bibinfo{year}{2020}).

\bibitem{peltonen2018hybrid}
\bibinfo{author}{Peltonen, J.} \emph{et~al.}
\newblock \bibinfo{title}{Hybrid rf squid qubit based on high kinetic
  inductance}.
\newblock \emph{\bibinfo{journal}{Scientific reports}}
  \textbf{\bibinfo{volume}{8}}, \bibinfo{pages}{1--8} (\bibinfo{year}{2018}).

\bibitem{hassani2022superconducting}
\bibinfo{author}{Hassani, F.} \emph{et~al.}
\newblock \bibinfo{title}{A superconducting qubit with noise-insensitive
  plasmon levels and decay-protected fluxon states}.
\newblock \emph{\bibinfo{journal}{arXiv preprint arXiv:2202.13917}}
  (\bibinfo{year}{2022}).

\end{thebibliography}


\begin{thebibliography}{10}
\expandafter\ifx\csname url\endcsname\relax
  \def\url#1{\texttt{#1}}\fi
\expandafter\ifx\csname urlprefix\endcsname\relax\def\urlprefix{URL }\fi
\providecommand{\bibinfo}[2]{#2}
\providecommand{\eprint}[2][]{\url{#2}}

\bibitem{deaver1961experimental}
\bibinfo{author}{Deaver~Jr, B.~S.} \& \bibinfo{author}{Fairbank, W.~M.}
\newblock \bibinfo{title}{Experimental evidence for quantized flux in
  superconducting cylinders}.
\newblock \emph{\bibinfo{journal}{Physical Review Letters}}
  \textbf{\bibinfo{volume}{7}}, \bibinfo{pages}{43} (\bibinfo{year}{1961}).

\bibitem{doll1961experimental}
\bibinfo{author}{Doll, R.} \& \bibinfo{author}{N{\"a}bauer, M.}
\newblock \bibinfo{title}{Experimental proof of magnetic flux quantization in a
  superconducting ring}.
\newblock \emph{\bibinfo{journal}{Physical Review Letters}}
  \textbf{\bibinfo{volume}{7}}, \bibinfo{pages}{51} (\bibinfo{year}{1961}).

\bibitem{bourassa2012josephson}
\bibinfo{author}{Bourassa, J.}, \bibinfo{author}{Beaudoin, F.},
  \bibinfo{author}{Gambetta, J.~M.} \& \bibinfo{author}{Blais, A.}
\newblock \bibinfo{title}{Josephson-junction-embedded transmission-line
  resonators: From kerr medium to in-line transmon}.
\newblock \emph{\bibinfo{journal}{Physical Review A}}
  \textbf{\bibinfo{volume}{86}}, \bibinfo{pages}{013814}
  (\bibinfo{year}{2012}).

\bibitem{blais2021circuit}
\bibinfo{author}{Blais, A.}, \bibinfo{author}{Grimsmo, A.~L.},
  \bibinfo{author}{Girvin, S.} \& \bibinfo{author}{Wallraff, A.}
\newblock \bibinfo{title}{Circuit quantum electrodynamics}.
\newblock \emph{\bibinfo{journal}{Reviews of Modern Physics}}
  \textbf{\bibinfo{volume}{93}}, \bibinfo{pages}{025005}
  (\bibinfo{year}{2021}).

\bibitem{schoelkopf2003qubits}
\bibinfo{author}{Schoelkopf, R.}, \bibinfo{author}{Clerk, A.},
  \bibinfo{author}{Girvin, S.}, \bibinfo{author}{Lehnert, K.} \&
  \bibinfo{author}{Devoret, M.}
\newblock \bibinfo{title}{Qubits as spectrometers of quantum noise}.
\newblock In \emph{\bibinfo{booktitle}{Quantum noise in mesoscopic physics}},
  \bibinfo{pages}{175--203} (\bibinfo{publisher}{Springer},
  \bibinfo{year}{2003}).

\bibitem{bylander2011noise}
\bibinfo{author}{Bylander, J.} \emph{et~al.}
\newblock \bibinfo{title}{Noise spectroscopy through dynamical decoupling with
  a superconducting flux qubit}.
\newblock \emph{\bibinfo{journal}{Nature Physics}}
  \textbf{\bibinfo{volume}{7}}, \bibinfo{pages}{565--570}
  (\bibinfo{year}{2011}).

\bibitem{hazard2019nanowire}
\bibinfo{author}{Hazard, T.} \emph{et~al.}
\newblock \bibinfo{title}{Nanowire superinductance fluxonium qubit}.
\newblock \emph{\bibinfo{journal}{Physical review letters}}
  \textbf{\bibinfo{volume}{122}}, \bibinfo{pages}{010504}
  (\bibinfo{year}{2019}).

\bibitem{zhang2021universal}
\bibinfo{author}{Zhang, H.} \emph{et~al.}
\newblock \bibinfo{title}{Universal fast-flux control of a coherent,
  low-frequency qubit}.
\newblock \emph{\bibinfo{journal}{Physical Review X}}
  \textbf{\bibinfo{volume}{11}}, \bibinfo{pages}{011010}
  (\bibinfo{year}{2021}).

\bibitem{vool2017introduction}
\bibinfo{author}{Vool, U.} \& \bibinfo{author}{Devoret, M.}
\newblock \bibinfo{title}{Introduction to quantum electromagnetic circuits}.
\newblock \emph{\bibinfo{journal}{International Journal of Circuit Theory and
  Applications}} \textbf{\bibinfo{volume}{45}}, \bibinfo{pages}{897--934}
  (\bibinfo{year}{2017}).

\bibitem{koch2007charge}
\bibinfo{author}{Koch, J.} \emph{et~al.}
\newblock \bibinfo{title}{Charge-insensitive qubit design derived from the
  cooper pair box}.
\newblock \emph{\bibinfo{journal}{Physical Review A}}
  \textbf{\bibinfo{volume}{76}}, \bibinfo{pages}{042319}
  (\bibinfo{year}{2007}).

\end{thebibliography}

\section*{Methods}
\label{sec: Methods}

\noindent \textbf{Hamiltonian based on a model of coupled normal modes (model 1)} \\
Here, we provide a brief summary of the theoretical model 1 that is used to derive a Hamiltonian for the unimon qubit, starting from the  normal modes of the distributed-element circuit illustrated in Fig.~\ref{fig: fig 1}c.  A complete derivation is provided in Supplementary Methods~I. In this theoretical model, we extend the approach of Ref.\cite{bourassa2012josephson} to  model phase-biased Josephson junctions in distributed-element resonators in the presence of an external magnetic flux.

In the discretized circuit of Fig.~\ref{fig: fig 1}c, the Josephson junction is located at $x_\mathrm{J}$ and we model the CPW resonator of length $2l$  using $N$ inductances $\Ll \Delta x$ and $N$ capacitances $\Cl \Delta x$ with $\Delta x = 2l/N$. 
Based on this circuit model, we construct the classical Lagrangian of the system using the node fluxes $\Psi_i = \int_{-\infty}^t V_i(t') \, \mathrm{d}t'$ as the coordinates with $V_i$ denoting the voltage across the $i$th capacitor\cite{vool2017introduction}. From the Lagrangian, we derive the classical equations of motion for the node fluxes and take the continuum limit resulting in a continuous node-flux function $\psi(x)= \int_{-\infty}^t V(x,t') \, \mathrm{d}t'$. Under the assumption of a sufficiently homogeneous magnetic field, the flux at the center conductor $\psi(x)$ is described with the wave equation $\ddot{\psi} = v_\mathrm{p}^2\partial_{xx} \psi$, where the phase velocity is given by $v_\mathrm{p} = 1/\sqrt{\Ll \Cl}$, where $\Ll$ and $\Cl$ denote the inductance and capacitance per unit length.
Furthermore, we obtain a set of boundary conditions corresponding to the grounding of the CPW at its end points and the current continuity across the junction.  

In the regime of small oscillations about a minimum of the potential energy, the classical flux $ \psi(x)$ can be decomposed into a sum of a dc component and oscillating normal modes. Using this decomposition and linearizing the junction in the vicinity of the dc operation point, we derive the classical normal-mode frequencies $\{\omega_m/(2\pi) \}_{m=0}^\infty$ and dimensionless flux-envelope functions $\{u_m(x) \}_{m=0}^\infty$. Subsequently, we invoke a single-mode approximation, in which the flux $\psi(x)$ is expressed as $\psi(x,t) = \Phi_0 \varphi_0 u_0(x)/(2\pi) + \psi_m(t) u_m(x)$, where $m$ is the mode index corresponding to the qubit, $\varphi_0$ is the dc Josephson phase, and $\psi_m(t)$ describes the temporal evolution of the flux for the qubit mode $m$. The dc phase $\varphi_0$ is controlled by the flux bias $\Phid$ as $\varphi_0 + 2l \Ll \sin(\varphi_0)/ L_\mathrm{J} = 2\pi \Phid/\Phi_0$, where $L_\mathrm{J}  = \Phi_0^2/(2\pi)^2/\EJ$ is the effective Josephson inductance.  

Finally, we quantize the classical Hamiltonian under the single-mode approximation and obtain
\begin{equation}
    \hat{\Ham}_m = 4 \ECm(\varphi_0) \hat{n}_m^2 + \frac{1}{2} E_{L, m}(\varphi_0) \hat{\varphi}_m^2 + E_L \hat{\varphi}_m \left(\frac{2\pi \Phi_\mathrm{diff}}{\Phi_0} - \varphi_0 \right) - \EJ \cos(\hat{\varphi}_m  -\varphi_0 ), \label{eq: single mode Hamiltonian with external flux}
\end{equation}
where $\nm$ and $\phim$ are the charge and phase operators %of the junction 
corresponding to the qubit mode and satisfying $[\hat{\varphi}_m, \hat{n}_m] = \mathrm{i}$, $\EL = \Phi_0^2/(2\pi)^2/(2l \Ll)$ is the inductive energy of the dc component, and the capacitive energy $\ECm(\varphi_0)$ and the inductive energy $E_{L, m}(\varphi_0)$ of the qubit mode $m$ are functions of $\Phid$ and circuit parameters according to Eqs. (S27), (S30-S34), (S37-S38), and (S40) in Supplementary Methods~I. %checked 24.3. \textcolor{red}{Check before submission}. %Supplementary equation numbers need to be checked before submission!!

The phase-basis wave functions and the potential energy  based on the Hamiltonian in equation~\eqref{eq: single mode Hamiltonian with external flux} are illustrated in Extended Data Figs.~\ref{fig: phase basis wave functions, matrix elements etc}a,~b for the parameter values of the qubit B. In Extended Data Figs.~\ref{fig: phase basis wave functions, matrix elements etc}c--e, we further show the characteristic energy scales of the unimon ($\ECm$, $\ELm$, $\EL$), the charge matrix elements $\langle i | \nm | j \rangle $ and the phase matrix elements $\langle i | \phim | j \rangle $ as functions of $\Phid$, where we denote the $k$-photon state of mode $m$ by $\ket{k}$.

In our qubit samples, each unimon is coupled to a readout resonator via a capacitance $C_\mathrm{g}$ at a location $x_\mathrm{g}$ to allow measurements of the qubit state. As derived in Supplementary Methods~II, the Hamiltonian of the coupled resonator-unimon system is given  by %EH: No RWA here
\begin{equation}
    \hat{\Ham} = \hbar \omega_\mathrm{r} \ad_\mathrm{r} \ah_\mathrm{r} + \sum_j \hbar \omega_j |j\rangle \langle j| +  \hbar \sum_{i,j} \left( g_{ij} |i \rangle \langle j | \ad_\mathrm{r} +  g_{ij}^* |j \rangle \langle i | \ah_\mathrm{r} \right),  \label{eq: coupled Hamiltonian final}
\end{equation}
where $f_\mathrm{r} = \omega_\mathrm{r}/(2\pi)$ is the resonator frequency, $\ah_\mathrm{r}$ is the annihilation operator of the readout resonator, $\{\hbar \omega_j \}$ and $\{|j\rangle \}$ are the eigenenergies and eigenstates of the bare unimon qubit, and the coupling strengths $g_{ij}$ are given by 
\begin{equation}
     g_{ij} \approx 4 \omega_\mathrm{r} \frac{C_\mathrm{g} u_m(x_\mathrm{g})\Delta u_m}{2\Cl l + C_\mathrm{J} + C_\mathrm{g} u_m(x_\mathrm{g})^2} \sqrt{\frac{ Z_\mathrm{tr}}{R_K}} \langle i |\mathrm{i} \hat{n}_m | j\rangle,
\end{equation}
where $\Delta u_m = u_m(x_\mathrm{J}^+) - u_m(x_\mathrm{J}^-)$, $C_\mathrm{J}$ is the junction capacitance, $Z_\mathrm{tr}$ is the characteristic impedance of the resonator, and $R_K = h/e^2$ is the von Klitzing constant. Assuming that $|\omega_1 - \omega_0 - \omega_r| \gg |g_{01}|$, we invoke the dispersive approximation allowing us to simplify equation~\eqref{eq: coupled Hamiltonian final} as (see Supplementary Methods~II)
\begin{equation}
    \hat{\Ham}_\mathrm{disp} \approx  \hbar \omega_\mathrm{r}' \ad_\mathrm{r} \ah_\mathrm{r} - \frac{\hbar \omega_{01}'}{2} \sz - \hbar \chi  \ad_\mathrm{r} \ah_\mathrm{r}  \sz, \label{eq: disp Hamiltonian}
\end{equation}
where $\omega_\mathrm{r}'$ and $\omega_{01}'$ are the renormalized resonator and qubit frequencies, $\sz = |0\rangle \langle 0 | -  |1\rangle \langle 1|$, and the dispersive shift $\chi$ is approximately given by  %EH: updated sign of \sz 9.3.2022
\begin{equation}
    \chi =  \frac{|g_{01}|^2}{\omega_1 - \omega_0 - \omega_\mathrm{r}} - \frac{1}{2} \frac{|g_{12}|^2}{\omega_2 - \omega_1 - \omega_\mathrm{r}}. \label{eq: dispersive shift}
\end{equation}
Although the dispersive approximation involves a minor transformation of the qubit and resonator operators, we have for simplicity used identical symbols for the transformed and original operators. 

\noindent \textbf{Hamiltonian  based on a path integral approach (model 2)} \\
%414 words, 380 without Eq. (copied raw text to word)
Here, we summarize our alternative theoretical approach for evaluation of the unimon spectrum. The unimon consists of a non-linear element (the Josephson junction) embedded into a linear non-dissipative environment (the $\lambda / 2$ resonator) as shown in Fig.~\ref{fig: fig 1}. This environment can be integrated out by the means of a path-integral formalism resulting in an effective action for a single variable, the flux difference $\psi_-$ across the junction. This action appears to be both non-Gaussian and non-local in imaginary time, and hence extremely challenging to integrate it analytically. In order to obtain the low-frequency spectrum of the unimon, we approximate the non-local part of the action by coupling the $\psi_-$ degree of freedom to~$M$ auxiliary linear modes, each described by a flux coordinate $\chi_k$, $k = 1,\ldots, M$. As described in detail in the Supplementary Note~I, the effective Hamiltonian of the unimon in this model reads as
\begin{equation}
    \hat H_M = \frac{\hat Q_-^2}{2C} + \frac{\hat\psi_-^2}{2 L_\psi} - E_\mathrm J \cos\left[\frac{2\pi}{\Phi_0}\left(\hat\psi_- + \Phi_\mathrm{diff}\right)\right] + \sum\limits_{k=1}^M\left[ \frac{\hat q_k^2}{2C} + \frac{C \hat\chi_k^2}{2}\left(\frac{\pi k v_\mathrm p}{ 2 l}\right)^2+ \alpha_k \hat\chi_k \hat\psi_-\right],
\end{equation}
where $[\hat \chi_k,\hat q_m]=\textrm{i}\hbar\delta_{km}$, $[\hat\psi_-,\hat Q_-]=\textrm{i}\hbar$, and all other single-operator commutators are zero, and the parameters $C$, $L_\psi$, and $\alpha_k$ are determined by Eqs.~(S76), (S77), and~(S78). In the limit $M\to \infty$, this approximation becomes exact. We restrict our analysis to the lowest auxiliary mode which gives a non-vanishing contribution to the unimon spectrum. Note that if the unimon is symmetric ($x_\mathrm J = 0$), the coupling of the Josephson junction to the first mode of the resonator vanishes, i.e., $\alpha_1 = 0$, and hence we need to consider the case $M = 2$. This approximation defines our model~2 which appears accurate enough for the quantitative analysis of the experimental data.

In addition to the technicalities related to the derivation of the models, the main difference between models~1 and~2 lies within the different employed approximations. In model~1, we take the linear part of the unimon into account exactly after linearizing the circuit at the minimum of the potential given by the dc phase, but we apply the single-mode approximation. Model~2 does not require us to solve the dc phase, and consequently we can conveniently work also in the regime $E_\mathrm J > E_L$ which is problematic for model~1 owing to multiple solutions for the dc phase. The price we pay for this advantage is that we consider the linear part of the problem to some extent approximately and that we need to solve a multidimensional Schr\"odinger equation.

\noindent \textbf{Design of the qubit samples} \\
%404 words including symbols. 
The samples are designed using KQCircuits\cite{kqcircuits} software which is built to work with the open-source computer-automated-design program KLayout\cite{klayout}. The designs are code-generated and parametrized for convenient adjustments during the design process. As illustrated in Fig.~\ref{fig: fig 1}e, each of the qubit chips comprise three unimon qubits which are capacitively coupled to individual readout resonators via U-shaped capacitors. All readout resonators are coupled with finger capacitors to the probe line using a single waveguide splitter. For multiplexed readout, the frequencies of the readout resonators are designed to be separated by 300~MHz. All of the unimons have the Josephson junction at the mid-point of the waveguide and are capacitively coupled to individual drive lines.

We present the design values of the main characteristic properties for all of the measured five qubits in Extended Data Table~\ref{tab: designed parameters}. To obtain the geometries of the qubit circuits that yield the desired physical properties, first, the dimensions of the center conductor of the qubit are chosen in an effort to obtain the characteristic impedance of $Z=\sqrt{L_l/C_l}=100$~$\Omega$. Here, the capacitance per unit length is $C_l=2\epsilon_0 (\epsilon_r-1)r_1+C_{\mathrm{air}}$ and the inductance per unit length is $L_l=1/(C_{\mathrm{air}}c^2)$, where $\epsilon_0$ is the vacuum electric permittivity, $\epsilon_r=11.45$ is the relative dielectric constant of the substrate, $r_1=K(r_2^2)/K(1-r_2^2)$, where $K$ denotes the complete elliptic integral of the first kind, $r_2=\tanh[\pi a/(4\eta)]/\tanh[\pi b/(4\eta)]$, $a$ is the width of the center conductor of the qubit, $\eta$ is the thickness of the substrate, $b$ is the total width of the qubit waveguide, $C_{\mathrm{air}}=2\epsilon_0(r_1+r_3)$, where $r_3=K(r_4^2)/K(1-r_4^2)$,  $r_4=a/b$, and $c$ is the speed of light\cite{simons2004coplanar}. Second, a series of finite-element simulations is executed on Ansys Q3D Extractor software to obtain the dimensions of the U-shaped capacitors with the target values of approximately $10$~fF for the coupling capacitances $C_\mathrm{g}$ between the readout resonators and the qubits. Third, the dispersive shift of the qubit is approximated based on Eq.~\eqref{eq: dispersive shift} as $\chi=\alpha |g_{01}|^2/[\Delta(\Delta+\alpha)]$, where $\alpha/(2\pi)=500$~MHz is a rough estimate for the anharmonicity of the unimon, $|g_{01}|/(2\pi)$ is the targeted  coupling strength between the qubit and its readout resonator, and $\Delta=2\pi(f_{01}-f_\mathrm{r})$. Finally, we adjust the length of the readout resonator and the capacitance $C_\kappa$ between the resonator and the probe line in order to obtain a resonator linewidth of $\kappa \approx \chi$ and a resonator frequency of $f_\mathrm{r}$. To this end, we carry out the microwave modelling of the device netlist, from which we obtain estimates for the resonant modes and their respective linewidths.

\noindent \textbf{Sample fabrication} \\
%236 words
The qubit devices were fabricated at the facilities of OtaNano Micronova cleanroom.  First, we sputter a 200-nm-thick layer of highly pure Nb on a high-resistivity ($\rho$  > 10 k$\Omega$cm) non-oxidised undoped $n$-type (100) 6-inch silicon wafer. Then, the coplanar waveguide is defined in a mask aligner using photo resist. After development, the Nb film is etched with a reactive ion etching (RIE) system. After etching, the resist residuals are cleaned in ultrasonic bath with acetone and isopropyl alcohol (IPA), and dried with a nitrogen gun.  Subsequently, the 6-inch wafer is cleaved into 3 $\times$ 3 cm$^{2}$ dices by Disco DAD3220, including nine chips in total. Each chip is 1 $\times$ 1 cm$^{2}$.

The tunnel junctions are patterned by a 100-keV EPBG5000pES electron beam lithography (EBL) system with a bilayer of methyl methacrylate/poly methyl methacrylate (MMA/PMMA) resist on a single chip. This is followed by a development in a solution of Methyl isobutyl ketone (MIBK) and IPA (1:3) for 20 s, Methyl Glycol for 20 s, and IPA for 20 s. The resist residues are cleaned with oxygen descum for 15 s. The two-angle shadow evaporation technique is applied to form the SIS junctions in an electron beam evaporator.  Before evaporation, the native oxides are removed by Ar ion milling. Aluminum is deposited at a rate of 5 Å/s. After lift off in acetone, each chip is cleaved by Disco DAD3220, then packaged and bonded with Al wires.

\noindent \textbf{Measurement setup}\\
%325 words. This is all very standard, 
For the experimental characterization, the packaged qubit devices are cooled down to a temperature of 10 mK using a commercial dilution refrigerator. The packaged samples are shielded by nested mu-metal and Aluminum shields. The ports of the sample holder are connected to room temperature electronics according to the schematic diagram shown in Extended Data Fig.~\ref{fig: experimental setup}. 

To implement the microwave signals for driving the qubits, we up-convert in-phase ($I$) and quadrature-phase ($Q$) waveforms generated by an arbitrary waveform generator with the help of an $IQ$ mixer and a local oscillator signal. The generated microwave signal is passed through a room temperature dc block and 60 dB of attenuation within the cryostat before reaching the sample. 

For the qubit-state readout, we use an ultrahigh-frequency quantum analyzer (UHFQA) by Zurich Instruments. Using the UHFQA, we create an intermediate-frequency voltage signal that is up-converted to the frequency of the readout resonator with an $IQ$ mixer and a local oscillator. The obtained microwave signal is passed through 60 dB of attenuation within the cryostat before entering the probe line. The output readout signal passes through two microwave isolators and a cryogenic high-electron-mobility transistor (HEMT) for amplification. At room temperature, the output signal is further amplified using a series of amplifiers and down-converted back to an intermediate frequency. In the UHFQA, the down-converted voltage signal is digitized and numerically converted to the base band. Due to the qubit-state-dependent dispersive shift of the readout resonator [see equation~\eqref{eq: disp Hamiltonian}], the measured output voltage is also dependent on the qubit state. To enable convenient calibration of the $IQ$ mixer used for the qubit drive, the setup also includes a room temperature switch enabling us to alternatively down-convert and measure the up-converted drive signal. 

To control the external flux difference, we use an external coil connected to a dc voltage source via two 50-k$\Omega$ resistors and a series of low-pass filters at room temperature and at the 100-mK stage of the cryostat.

\noindent \textbf{Measurement and analysis of qubit frequency and anharmonicity} \\
%310 words
To measure the frequencies of the one-photon $|0\rangle \leftrightarrow |1\rangle$ transition and the two-photon $|0\rangle \leftrightarrow |2\rangle$ transition, we use a standard two-tone qubit spectroscopy experiment illustrated in Fig.~\ref{fig: resonator and qb spectroscopy}c. In the experiment, we apply a continuous microwave signal to the drive line of the qubit while applying a readout signal through the probe line of the sample. At the sweet spot $\Phid = \Phi_0/2$, we further measure the $|1\rangle \leftrightarrow |2\rangle$ transition frequency with an ef-Rabi experiment (see Extended Data Fig.~\ref{fig: ef rabis}) in order to verify the anharmonicities shown in Fig.~\ref{fig: anharmonicity and coherence times vs flux}a and summarized in Extended Data Table~\ref{tab: basic qubit properties}. In the ef-Rabi experiment, the qubit is first prepared to the excited state with a $\pi$-pulse followed by another pulse with a varying amplitude and a varying frequency around the estimated  $|1\rangle \leftrightarrow |2\rangle$ transition. After the drive pulses, a readout pulse is applied and an oscillating output voltage is observed as a result of Rabi oscillations between the states $|1\rangle$ and $|2\rangle$. 

To estimate the circuit parameters presented in Extended Data Table~\ref{tab: basic qubit properties}, we use the following approach. First, we fit the theoretical Hamiltonian in equation~\eqref{eq: single mode Hamiltonian with external flux} to the experimental transition frequencies of qubit B in order to estimate $\Ll$, $\Cl$, and $\EJ$. Subsequently, the coupling capacitance $C_\mathrm{g}$ of qubit B is estimated by fitting equation~\eqref{eq: coupled Hamiltonian final} to the data of the avoided crossing in Fig.~\ref{fig: resonator and qb spectroscopy}b. For the other qubits, it is assumed that $\Ll$ and $\Cl$ are equal to those of qubit B due to an identical geometry of the CPW. For these qubits, the Josephson energy $\EJ$ is first approximately fitted based on the measured $|0\rangle \leftrightarrow |1\rangle$ transition followed by an estimation of $C_\mathrm{g}$ using data of an avoided unimon--resonator crossing.

\noindent \textbf{Characterization for readout} \\
%225 words. Eric: this text resembles results a bit. 
To characterize the device for qubit readout, we measure the dispersive shift $\chi/(2\pi)$ for all of the qubits. This is achieved using an experiment, in which the output readout signal is measured as a function of the signal frequency after preparing the qubit either to its ground or first excited state. In Extended Data Fig.~\ref{fig: readout}a, the measured dispersive shifts are compared against theoretical predictions computed with equation~\eqref{eq: dispersive shift} based on the fitted circuit parameters, and the measured qubit frequency $\omega_{01}/(2\pi)$ and anharmonicity $\alpha/(2\pi)$. The good agreement between the experiment and the theory validates the dispersive approximation in equation~\eqref{eq: disp Hamiltonian}. 

We further measure the single-shot readout fidelity for qubit E with $\chi/(2\pi) = 4.1$~MHz. This is achieved by alternately preparing the qubit to the ground state and to the first excited state followed by a state measurement with a 1.6-$\mu$s-long readout pulse. The output readout voltage is obtained as an unweighted average of the voltage during a 1.6-$\mu$s-long integration window. This experiment is repeated 2000 times. Using an optimized threshold voltage, we extract a readout fidelity $[P(|0\rangle | |0\rangle) + P(|1\rangle | |1\rangle)]/2$ of 89.0\% as shown in Extended Data Figs.~\ref{fig: readout}b,~c. The readout error is dominated by qubit relaxation during the readout pulse. Note that the measured fidelity is reached without a quantum-limited amplifier suggesting that high-fidelity single-shot readout is possible with the unimon.

\noindent \textbf{Measurement and analysis of energy relaxation time} \\
%244 words all (226 without Eq)
To measure the energy relaxation time $T_1$, an initial $\pi$-pulse is applied to the ground-state-initialized qubit followed by a varying delay and a subsequent measurement of the qubit population. We use a single exponential function for fitting the qubit population, which is supported by the experimental data of qubit B shown in Extended Data Fig.~\ref{fig: T1 example and range}a. Thus, there is no evidence of quasiparticle-induced losses that result in a double-exponential decay. 

For qubit B, the relaxation time is characterized across $\Phid/\Phi_0 \in [0.0, 0.5]$ in order to determine the mechanisms limiting $T_1$. As detailed in Supplementary Methods~III, we model the relaxation rate $\Gamma_1 = 1/T_1$ due to a noise source~$\lambda$ as\cite{schoelkopf2003qubits}
\begin{equation}
    \Gamma_1 = \frac{|\langle 0 | \partial \hat{\Ham}_m/ \partial \lambda |1\rangle |^2}{\hbar^2} S_{\lambda}(\omega_{01}), \label{eq: general loss rate}
\end{equation}
where  $S_{\lambda}(\omega_{01})$ is the symmetrized noise spectral density of the variable $\lambda$ at the qubit angular frequency $\omega_{01}$. In Extended Data Fig.~\ref{fig: T1 example and range}b, we compare the frequency dependence of the measured relaxation rate to the theoretical models based on Ohmic flux noise, $1/f$ flux noise, dielectric losses, inductive losses, radiative losses, and Purcell decay through the resonator by scaling the theoretical predictions to coincide with the experimental data at $\Phid = \Phi_0/2$. As illustrated in Fig.~\ref{fig: anharmonicity and coherence times vs flux}b, the experimental data is most accurately explained by a model including Purcell decay and dielectric losses with an effective dielectric quality factor of $Q_C = 1.7 \times 10^5$.

\noindent \textbf{Measurement and analysis of coherence time} \\
%420 words all, 379 without Eqs. 
The coherence time of the qubits is characterized using standard Ramsey and Hahn echo measurements\cite{krantz2019quantum}. At the sweet spots, we estimate the Ramsey coherence time $T_2^*$ by fitting an exponentially decaying sinusoidal function to the measured qubit population, whereas we obtain the echo coherence time $T_2^\mathrm{e}$ using an exponential fit. As illustrated in Extended Data Figs.~\ref{fig: T2 and T2 echo examples}a--c, these models agree well with the experimental data of qubit B at the flux-insensitive sweet spots yielding $T_2^* = 3.1$ $\mu$s and $T_2^\mathrm{e} = 8.9$ $\mu$s for $\Phid = \Phi_0/2$, and $T_2^* = 6.4$ $\mu$s and $T_2^* = 9.5$ $\mu$s for $\Phid = 0$. 

To study the sensitivity of the qubits to flux noise, we conduct Ramsey and Hahn echo measurements as a function of the external flux bias in the vicinity of $\Phid = \Phi_0/2$ (see Fig.~\ref{fig: anharmonicity and coherence times vs flux}c). In superconducting qubits, flux noise is often accurately described by $1/f$ noise\cite{bylander2011noise,yan_flux_2016}
\begin{equation}
S_{\Phid}(\omega) = \int_{-\infty}^\infty \mathrm{d}t \exp(-\mathrm{i} \omega t) \langle\Phid(0) \Phid(t) \rangle  = 2\pi \frac{A_{\Phid}^2}{\omega},
\end{equation}
where $A_{\Phid}/\sqrt{\mathrm{Hz}}$ is the flux noise density at 1 Hz. The $1/f$-noise gives rise to a Gaussian decay in the echo experiment\cite{ithier2005decoherence,braumuller2020characterizing}, due to which we model the Hahn echo decay with a product of Gaussian and exponential functions, $\propto \mathrm{exp}(-\Gamma_{\varphi, \Phi}^\mathrm{e}t^2 - \Gamma_{\varphi, 0}^\mathrm{e}t)$, as illustrated in Extended Data Fig. \ref{fig: T2 and T2 echo examples}d. The corresponding $T_2^\mathrm{e}$ is evaluated as the $1/\mathrm{e}$ decay time given by\cite{zhang2021universal}
\begin{equation}
    T_2^\mathrm{e} = \frac{\sqrt{4(\Gamma_{\varphi, \Phi}^\mathrm{e})^2 + (\Gamma_{\varphi, 0}^\mathrm{e})^2} -\Gamma_{\varphi, 0}^\mathrm{e}}{2(\Gamma_{\varphi, \Phi}^\mathrm{e})^2}.
\end{equation}
Under the assumption of $1/f$-noise, the Gaussian dephasing rate $\Gamma_{\varphi, \Phi}^\mathrm{e}$ obtained from an echo measurement is related to the flux noise density as\cite{ithier2005decoherence, braumuller2020characterizing}
    \begin{equation}
        \Gamma_{\varphi, \Phi}^\mathrm{e} = \sqrt{\ln 2} A_{\Phid}\bigg|\frac{\partial \omega_{01}}{\partial \Phid }\bigg| + \Gamma_{\varphi, \mathrm{x}}^\mathrm{e},
    \end{equation}
where $\Gamma_{\varphi, x}^\mathrm{e}$ is a small residual Gaussian decay rate at the sweet spot. For each of the qubits, we estimate the parameter $A_{\Phid}$ in Extended Data Table~\ref{tab: results of characterization} by a linear least-squares fit to $(|\partial \omega_{01}/\partial \Phid|, \Gamma_{\varphi, \Phi}^\mathrm{e})$ data, where $\partial \omega_{01}/\partial \Phid$ is estimated by fitting a parabola $\omega_{01} = \tilde{a} \Phid^2 + \tilde{b} \Phid + \tilde{c}$ to the measured $\omega_{01}$ near the sweet spot and then evaluating $\partial \omega_{01}/\partial \Phid = 2\tilde{a}\Phid + \tilde{b}$. 

For Ramsey experiments, we use an exponential decay model also away from the sweet spot to constrain the number of fitting parameters. The theoretical fit shown in Fig.~\ref{fig: anharmonicity and coherence times vs flux}c is based on a simple model of the form $1/T_2^* = a' |\partial \omega_{01}/\partial \Phid| + b'$.

\noindent \textbf{Implementation and benchmark of high-fidelity single-qubit gates}\\
%372 words all, 353 without equations
To implement fast high-fidelity single-qubit gates, we use the derivative removal by adiabatic gate (DRAG) framework\cite{motzoi_simple_2009}. Thus, we parametrize the microwave pulses implementing the gates  $V_\mathrm{rf}(t) = I_\mathrm{qb}(t) \cos(\omega_\mathrm{d} t + \theta) + Q_\mathrm{qb}(t) \sin(\omega_\mathrm{d} t + \theta)$  as 
\begin{align}
    I_\mathrm{qb}(t) &= A \mathrm{exp} \left[\frac{-(t- t_\mathrm{g}/2)^2}{2\sigma^2} \right],~t \in [0, t_\mathrm{g}],\\
    Q_\mathrm{qb}(t) &= \beta I_\mathrm{qb}'(t),~t \in [0, t_\mathrm{g}],
\end{align}
where $\omega_\mathrm{d}/(2\pi)$ is the drive frequency, $\theta$ determines the rotation axis of the gate, $A$ and $\beta$ are amplitudes of $I_\mathrm{qb}$ and $Q_\mathrm{qb}$, respectively, $t_\mathrm{g}$ is the gate duration, and $\sigma=t_\mathrm{g}/4$ is the standard deviation of the Gaussian. The drive frequency $\omega_\mathrm{d}/(2\pi)$ is set to the qubit frequency $\omega_{01}/(2\pi)$ measured in a Ramsey experiment. The amplitude $A$ of the Gaussian pulse is determined using error amplification by applying repeated $\pi$ pulses with varying amplitudes $A$ after an initial $\pi/2$ pulse. The amplitude $\beta$ of the derivative component is chosen to minimize the difference of qubit populations measured after gate sequences $(X(\pi), Y(\pi/2))$ and $(Y(\pi), X(\pi/2))$\cite{reed2013entanglement}.

To characterize the accuracy of the calibrated single-qubit gates, we use the definition of average gate fidelity\cite{nielsen2002simple}. To measure the average gate fidelity, we use standard and interleaved randomized benchmarking (RB) protocols\cite{magesan_scalable_2011,magesan_characterizing_2012, magesan2012efficient}. In the standard RB protocol, we apply random sequences of Clifford gates appended with a final inverting gate and estimate the average fidelity of gates in the Clifford group $F_\mathrm{Cl}$ based on the decay rate of the ground state probability as a function of the sequence length. We decompose the Clifford gates based on Table~1 in Ref.\cite{epstein_investigating_2014} using the native gate set $\{I, X(\pm \pi/2), Y(\pm \pi/2), X(\pi), Y(\pi) \}$ such that each Clifford gate contains on average 1.875 native gates. The average fidelity per a single native gate is estimated as $F_\mathrm{g} = 1 - (1 - F_\mathrm{Cl})/1.875$. To estimate the average gate fidelity of individual gates in the set $\{I, X(\pi/2), Y(\pi/2) \}$, we utilize the interleaved RB protocol, in which the average gate fidelity is measured by comparing the decay rates for sequences with and without the gate of interest interleaved after each random Clifford gate. 

The theoretical coherence limit for the gate fidelity is computed based on the measured $T_1$ and $T_2^\mathrm{e}$ as $    F_\mathrm{g,lim} = 1/6\times\left(3 + \exp(-t_\mathrm{g}/T_1) + 2\exp(-t_\mathrm{g}/T_2^\mathrm{e}) \right)$\cite{bao2021fluxonium}.

\section*{Acknowledgements}
%MM: I updated this now.
We have received funding from the European Research Council under Consolidator Grant No. 681311 (QUESS), European Commission through H2020 program projects QMiCS (grant agreement 820505, Quantum Flagship), the Academy of Finland through its Centers of Excellence Program (project Nos. 312300, and 336810), and Research Impact Foundation. We acknowledge the provision of facilities and technical support by Aalto University at OtaNano – Micronova Nanofabrication Center and LTL infrastructure which is part of European Microkelvin Platform (EMP, No. 824109 EU Horizon 2020). We thank the whole staff at IQM and QCD Labs for their support. 
Especially, we acknowledge the help with the experimental setup from Roope Kokkoniemi, code and software support from Joni Ikonen, Tuukka Hiltunen, Shan Jolin, Miikka Koistinen,  Jari Rosti, Vasilii Sevriuk, and Natalia Vorobeva, and useful discussions with Brian Tarasinski.

\section*{Author contributions}
%Both Juha Hassel, Juho Hotari and Johannes Heinsoo are J.H. -> Juha Hassel J.HA., Juho Hotari J.HO., Johannes J.HE.

The concept of the unimon qubit was conceived by E.H. and M.M. The theoretical model 1 was developed by E.H. with theory support from M.M., J.T., and J.HA. The theoretical model 2 was developed by V.V. The qubit samples were designed by A.L., A.M., and C.O.-K. with support from S.K. J.K. and D.J. had a significant role in developing the KQCircuits software used for designing the unimon qubit devices. W.L. and T.L. designed Josephson Junctions. W.L. fabricated the qubit devices and benchmarked the room temperature resistance. M.P. helped on sample lift-off process. J.HO packaged the device.
E.H. conducted the qubit measurements at IQM with support from F.M., C.F.C., and J.-L.O. regarding the experimental setup and from F.T., M.S., and K.J. regarding the measurement code. S.K. conducted the qubit measurements at QCD with support from A.G. E.H. analyzed the measurement data with support from S.K., V.V. and O.K. The manuscript was written by E.H. and M.M with support from V.V., A.M., W.L., and S.K. All authors commented on the manuscript. Different aspects of the work were supervised by J.HE., C.O.-K., T.L., J.HA, and K.Y.T. M.M supervised the work in all respects.

\section*{Additional information}
\textbf{Supplementary information} is available as a part of the arxiv submission. \\
\textbf{Correspondence and requests for materials} should be addressed to E.H. and M. M. \\
%\textbf{Peer review information} <insert text>\\
%\textbf{Reprints and permissions information} is available at http://www.nature.com/reprints.

\clearpage

\clearpage
%%%%%%%%%%%%%%%%Figure 1

\clearpage
\textbf{Figure 1: Unimon qubit and its measurement setup.} \textbf{a}, Map of superconducting-qubit types based on their energy scales: Josephson energy $\EJ$ and inductive energy $E_\textrm{L}$ compared with the charging energy $\EC$. The map includes those superconducting qubits that can be described with the circuit model shown in the inset. The proposed unimon qubit corresponds to the vicinity of the red dashed line leading to the cancellation of the linear inductive energy by the quadratic contribution of the Josephson energy  at half flux quantum $\Phi_0/2$. The black star denotes the unimons realized in this work and the other experimental data points are from Refs.\cite{nakamura1999coherent,barends2013coherent,manucharyan2009fluxonium, hazard2019nanowire,grunhaupt2019granular,Peruzzo2021Geometric,zhang2021universal,pechenezhskiy2020superconducting,yan2020engineering,peltonen2018hybrid,liu2021quantum,hassani2022superconducting}. \textbf{b}, Schematic illustration of the unimon circuit consisting of a Josephson junction ($\EJ$, $C_\mathrm{J}$) in a grounded coplanar-waveguide (CPW) resonator that has a length of $2l$ and an inductance and capacitance per unit length of $\Ll$ and $\Cl$, respectively.  The voltage envelope functions of the qubit mode are also illustrated at external-flux biases $\Phi_\mathrm{diff} = (\Phi_\mathrm{ext,2} - \Phi_\mathrm{ext,1})/2=0.0$ (dashed line) and $\Phi_\mathrm{diff}=\Phi_0/2$ (solid line). \textbf{c}, Distributed-element circuit model of the unimon qubit, in which the CPW is modeled with $N$ inductors $\Ll \Delta x$ and capacitors  $\Cl \Delta x$ with $\Delta x = 2l /N$ denoting the length scale of discretization. \textbf{d}, Schematic illustration of a mechanical inverted pendulum system, the Hamiltonian of which is identical to that of the lumped-element unimon circuit in \textbf{a}. In this analogy, the gravitational potential energy corresponds to the Josephson potential, the harmonic potential energy of the twisting beam corresponds to the inductive energy, the moment of inertial corresponds to the capacitance of the unimon, and the angle of the zero twist position corresponds to the flux bias $\Phid$.  \textbf{e}, False-color microscope image of a silicon chip containing three unimon qubits (blue) together with their readout resonators (red), drive lines (green), and a joint probe line (yellow). \textbf{f}, Simplified schematic of the experimental setup used to measure the unimon qubits at 10 mK (see also Extended Data Fig.~\ref{fig: experimental setup}).

%Figure 2
\clearpage

%Figure 3
\clearpage

%Figure 4
\clearpage

\renewcommand{\figurename}{\textbf{Extended Data Fig.}} 
\setcounter{figure}{0} 

\renewcommand{\tablename}{\textbf{Extended Data Table}} 
\setcounter{table}{0}

\clearpage
\begin{figure}
 \centering
 \includegraphics[width = \textwidth]{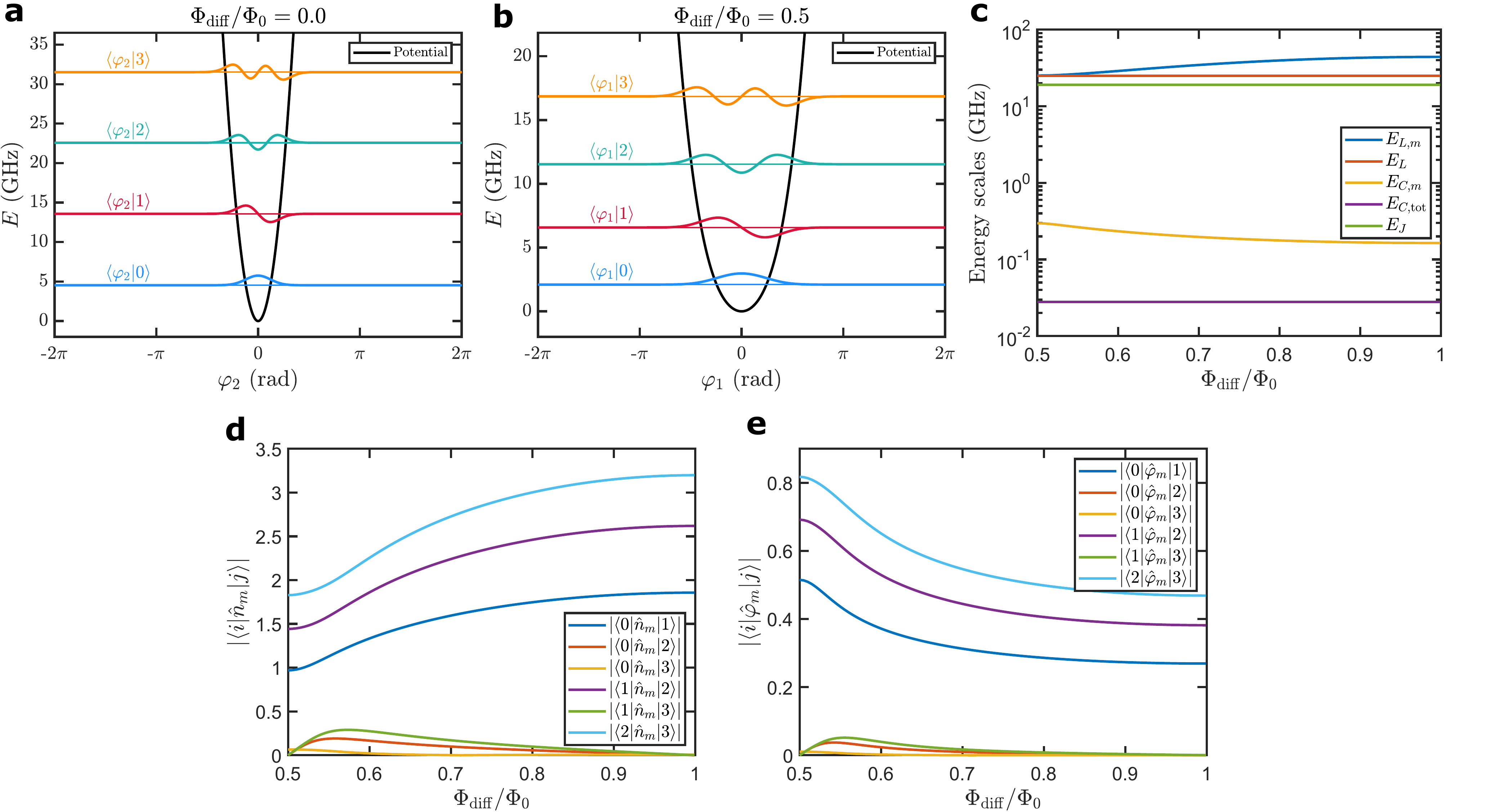}
    \caption{ \label{fig: phase basis wave functions, matrix elements etc} \textbf{Potential-energy function, energy scales, and matrix elements of of qubit~B based on model~1.} \textbf{a}, \textbf{b}, Potential energy of the unimon based on equation~\eqref{eq: single mode Hamiltonian with external flux} in Methods as a function of the phase variable $\varphi_m$ of mode $m$ together with the four lowest eigenenergies and corresponding phase-basis wave functions at flux biases of $\Phi_\mathrm{diff} = 0$ (\textbf{a}) and $\Phi_\mathrm{diff} = \Phi_0/2$ (\textbf{b}). Note that the second mode of the circuit is used as the qubit at $\Phi_\mathrm{diff} = 0$ and the first mode at $\Phi_\mathrm{diff} = \Phi_0/2$. \textbf{c}, Energy scales $\ECm$, $\ELm$, $\EL$, and $\EJ$ of the qubit as functions of $\Phid$. Here, $\ECm$ and $\ELm$ are the capacitive and inductive energies of the qubit mode, $\EL$ is the inductive energy corresponding to a dc current in the center conductor of the qubit, and $\EJ$ is the Josephson energy.  We also show an effective charging energy $E_{C, \mathrm{tot}} = e^2/(4 \Cl l)$ computed based on the total capacitance $2 \Cl l$ of the transmission line of the unimon. \textbf{d}, Off-diagonal matrix elements of the Cooper pair number operator $\hat{n}_m$ for the four lowest-energy states of the qubit mode $m$ as functions of $\Phid$. \textbf{e}, As \textbf{d} but for the phase operator $\hat{\varphi}_m$. In all panels, the results have been obtained by using the theoretical model~1 in Methods and the measured parameter values of qubit~B reported in Extended Data Table~\ref{tab: basic qubit properties}.}
\end{figure}

\begin{figure}
 \centering
 \includegraphics[width=0.8\textwidth]{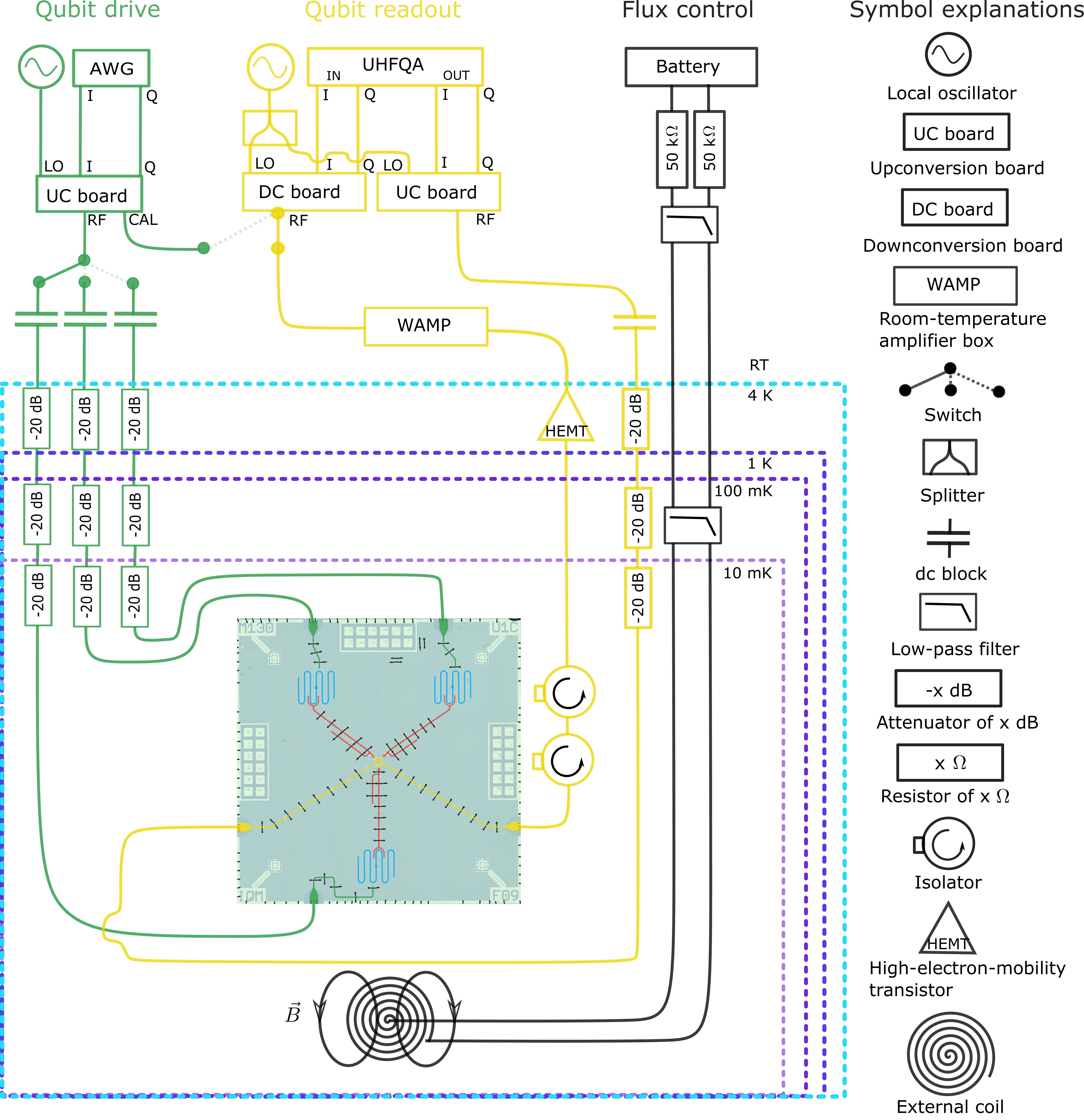}
    \caption{ \label{fig: experimental setup} \textbf{Experimental setup.} Schematic experimental setup used in the measurements of the unimon qubits of this work. Electronics and components related to qubit drive, qubit readout, and flux control are denoted with green, yellow, and black colors, respectively. The dashed rectangles illustrate boundaries between the different temperature stages of the cryostat. Brief explanations of the symbols are provided in the rightmost column of the figure. Note that the qubit readout does not utilize any parametric amplifier.}
\end{figure}
%Note mu-metal and aluminum shields are not depicted

\clearpage
\begin{figure}
 \centering
 \includegraphics[width = 0.9\textwidth]{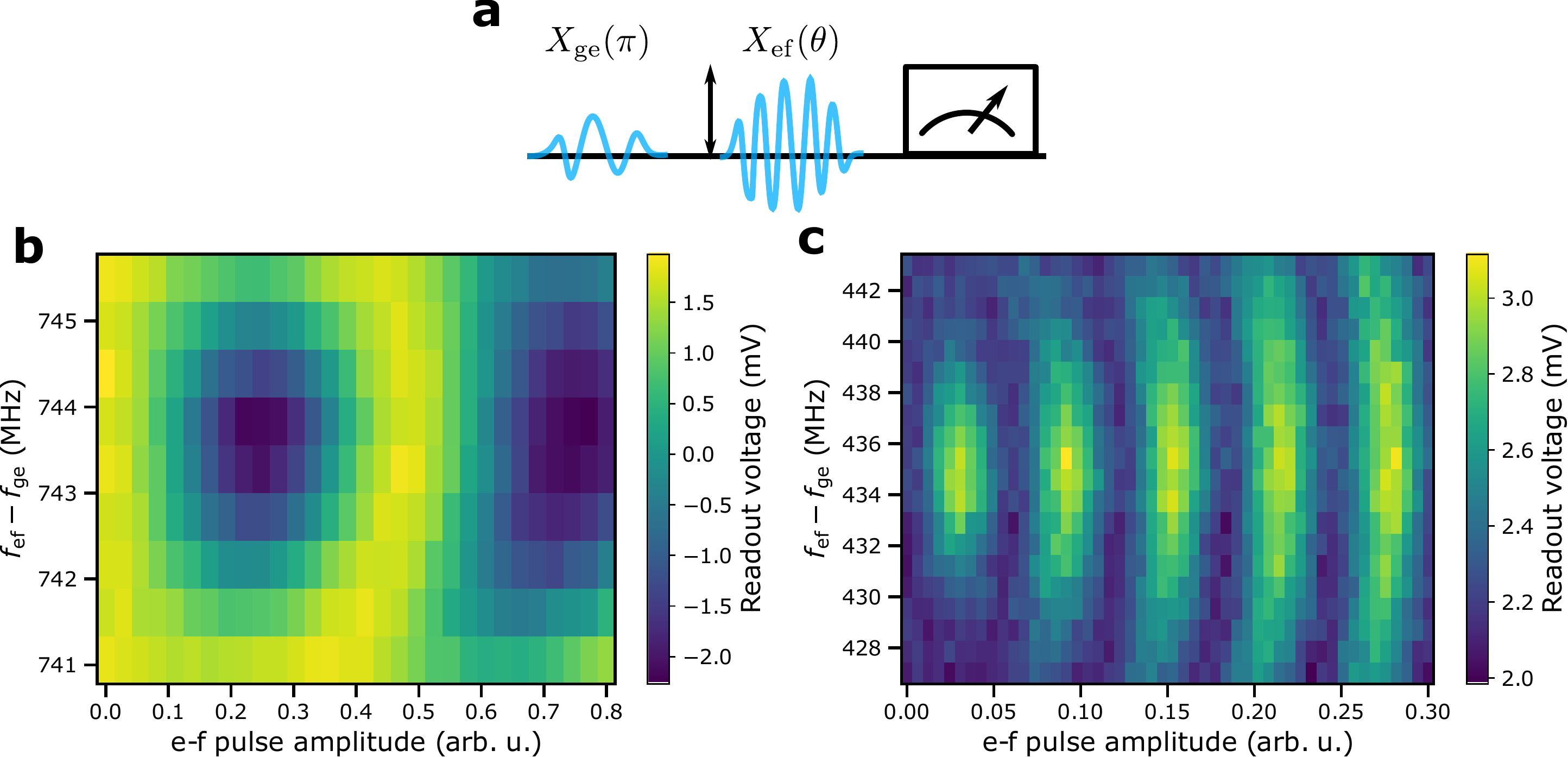}
    \caption{ \label{fig: ef rabis} \textbf{Estimation of qubit anharmonicity using Rabi oscillations.} \textbf{a} Measurement sequence of an ef-Rabi experiment. In the experiment, the qubit is first prepared in the excited state with a $\pi$-pulse at the frequency $f_\mathrm{ge}$ matching with the qubit frequency $f_{01}$ followed by another pulse with a fixed duration but a varying frequency $f_\mathrm{ef}$ and a varying amplitude. After applying the pulses, the output readout voltage is measured. \textbf{b},\textbf{c} Output readout voltage as a function of the amplitude of the ef-pulse and the frequency difference $f_\mathrm{ef} - f_\mathrm{ge}$ in an ef-Rabi experiment conducted for qubits A (\textbf{b}) and B (\textbf{c}). The resulting observed Rabi oscillations between the first and the second excited states of the unimon comfirm that the anharmonicities $\alpha$ of qubits~A and~B are approximately $2\pi\times 744$~MHz and $2\pi\times 434$~MHz, respectively, as measured using a two-tone qubit spectroscopy such as that shown in Fig.~\ref{fig: resonator and qb spectroscopy}c. }
\end{figure}

%EH: No TWPA used for the readout 
\clearpage
\begin{figure}
 \centering
 \includegraphics[width = \textwidth]{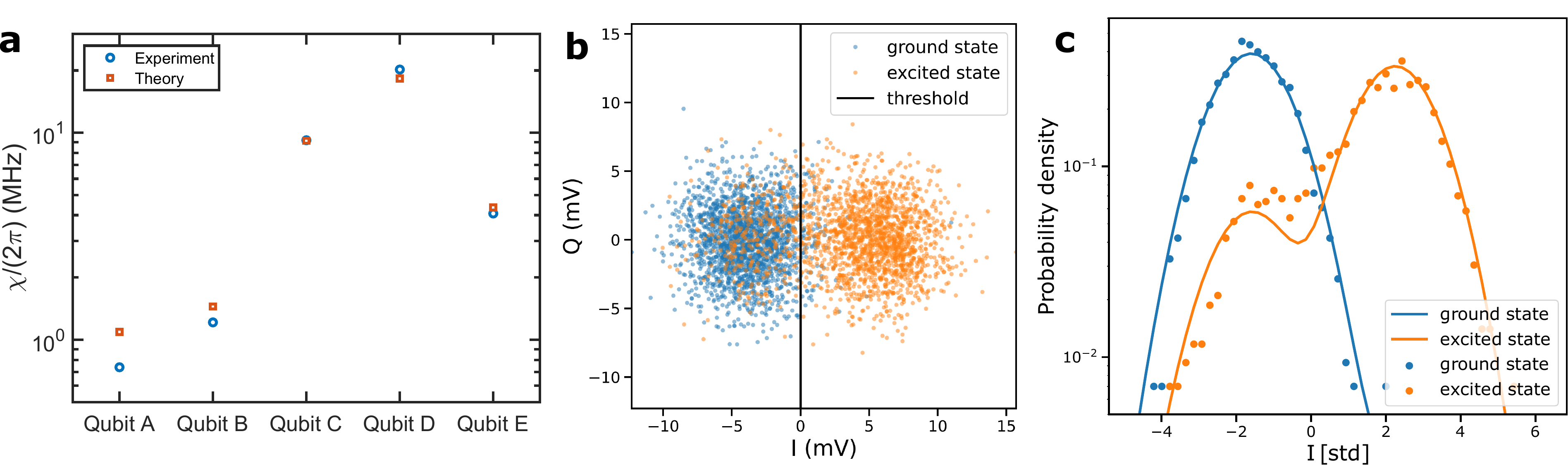}
    \caption{ \label{fig: readout} \textbf{Dispersive shift and single-shot qubit readout}  \textbf{a}, Measured dispersive shift $\chi$ for different unimon qubits at a flux bias of $\Phid = \Phi_0/2$ (blue circles) together with theoretical predictions (orange squares) computed using equation~\eqref{eq: dispersive shift} in Methods based on the measured qubit frequency, measured anharmonicity, and the fitted coupling capacitance $C_\mathrm{g}$, all of which are reported in Extended Data Table~\ref{tab: basic qubit properties}.  %EH: updated to use parameter values without Cg
    \textbf{b}, Readout voltages in the $IQ$ plane for a single-shot readout experiment conducted for qubit E. The qubit is prepared either to the ground state (blue dots) or to the excited state (orange dots)  followed by a single-shot state measurement implemented using a readout pulse with a duration of 1.6 $\mu$s.  The state preparation was repeated 2000 times for both the ground state and the excited state. The solid black line illustrates the optimal single-shot classification boundary corresponding to a readout fidelity of 89.0\%. The classification errors are 5.0\% and 17.0\% when preparing the qubit to the ground state and to the excited state, respectively. \textbf{c}, Probability density distributions for the single-shot voltage corresponding to the qubit prepared to the ground state (blue dots) or to the excited state (orange dots). The single-shot voltages have been projected along a line perpendicular to the classification boundary. The solid lines denote fits to the measured probability densities based on a model involving a sum of two Gaussian functions.}
\end{figure}

\clearpage
\begin{figure}
 \centering
 \includegraphics[width = 0.8\textwidth]{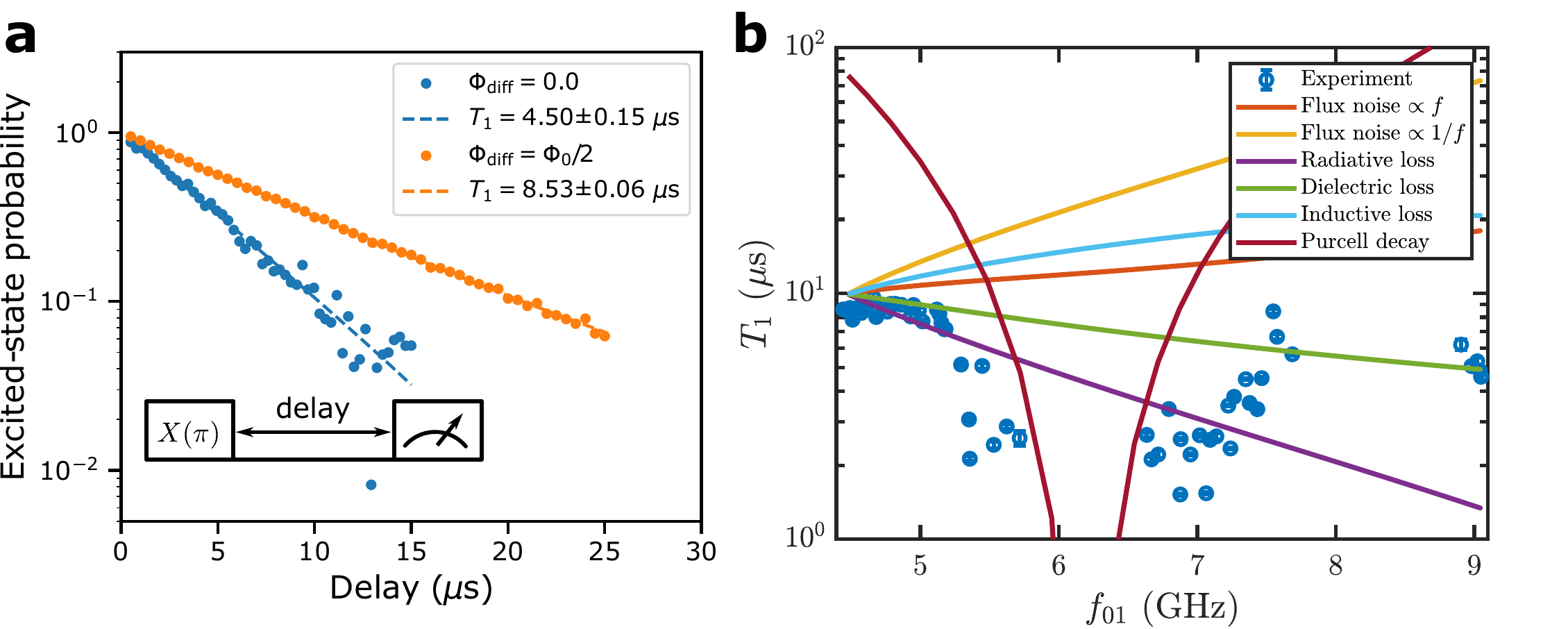}
    \caption{ \label{fig: T1 example and range} \textbf{Characterization of the energy relaxation time for qubit B.} \textbf{a}, Excited-state probability as a function of the length of the delay between the state preparation and readout (see inset) at the sweet spots $\Phi_\mathrm{diff} = 0$ (blue color) and $\Phi_\mathrm{diff} = \Phi_0/2$ (orange color). The slopes of the exponential fits (dashed lines) yield energy relaxation times $T_1=4.5 \pm 0.2$ $\mu$s and $T_1=8.5 \pm 0.1$~$\mu$s at $\Phi_\mathrm{diff} = 0$ and~$\Phi_0/2$, respectively. \textbf{b}, Measured mean $T_1$  as a function of the qubit frequency (blue circles) together with theoretical predictions of $T_1$ based on different indicated loss mechanisms including Ohmic flux noise, $1/f$ flux noise, radiative losses, dielectric losses, inductive losses, and Purcell decay through the resonator.  The error bars provide the $1\sigma$ uncertainty. The theoretical predictions (apart from the Purcell decay) have been scaled to coincide with the experimental data at $\Phi_\mathrm{diff} = \Phi_0/2$.}%  %EH: updated to use parameter values without Cg  }
\end{figure}

\clearpage
\begin{figure}
 \centering
 \includegraphics[width=0.85\textwidth]{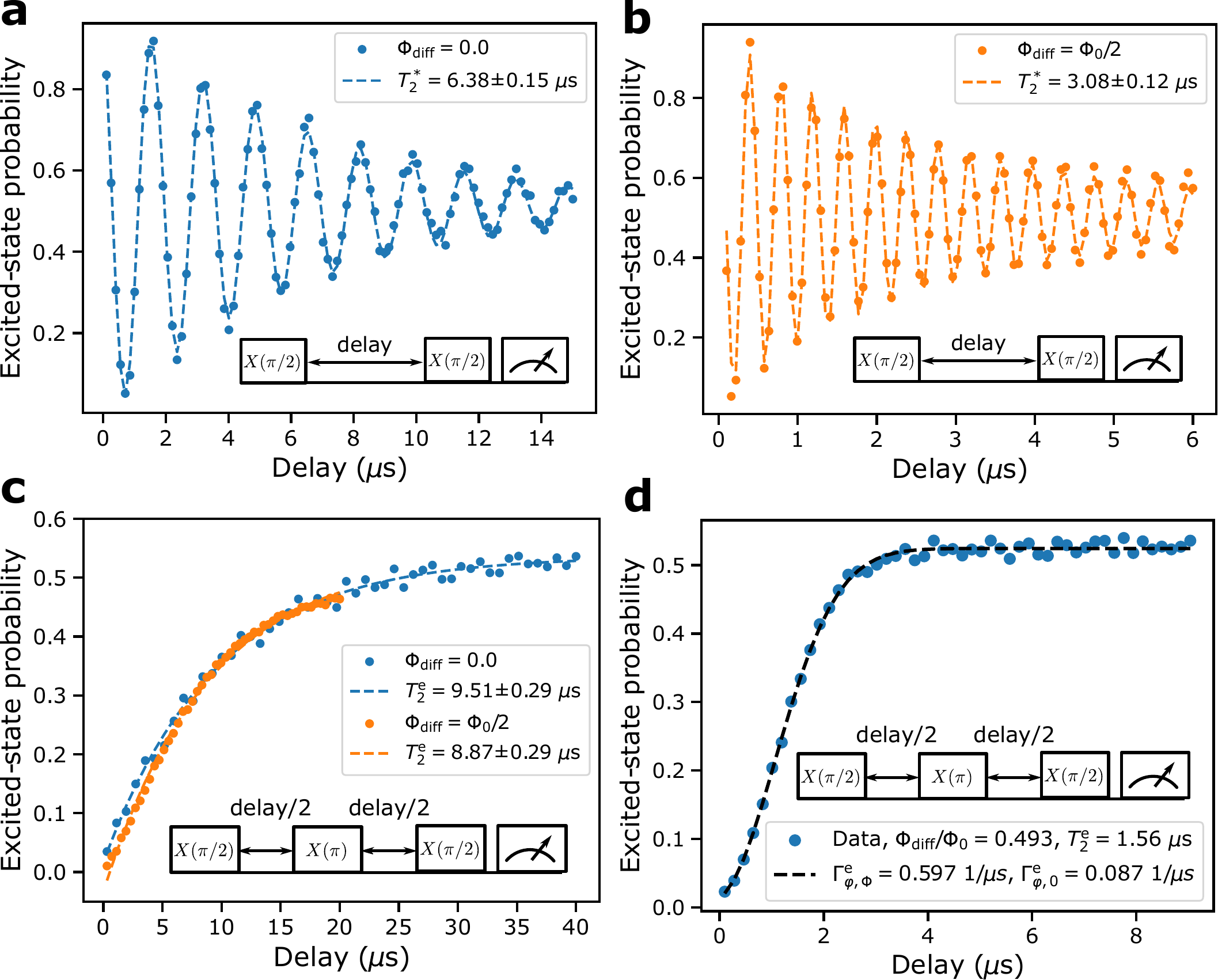}
    \caption{ \label{fig: T2 and T2 echo examples} \small{\textbf{Characterization of the coherence time for qubit B.} \textbf{a}, Excited-state probability as a function of the length of the delay between the $X(\pi/2)$ pulses of a Ramsey measurement (see inset) conducted at the sweet spot $\Phi_\mathrm{diff} = 0$ using a detuning of 0.6 MHz between the drive frequency and the qubit frequency. The markers show the experimental data and the dash line provides a corresponding least-squares fit of an exponentially decaying sinusoidal function. 
    \textbf{b}, As \textbf{a} but for $\Phid = \Phi_0/2$ and a detuning of 2.5 MHz between the drive frequency and the qubit frequency. \textbf{c}, Measured excited-state probability (markers) as a function of the length of the delay between the $X(\pi/2)$ pulses in a $T_2$ echo measurement (see inset) conducted at the sweet spots $\Phi_\mathrm{diff} = 0$ (blue color) and $\Phi_\mathrm{diff} = \Phi_0/2$ (orange color). The dash lines provide exponential fits to the data. \textbf{d} Same as \textbf{c} but for a flux bias of $\Phid /\Phi_0 = 0.493$. Here, the dashed line shows a model corresponding to a product of a Gaussian and an exponential with decay rates $\Gamma_{\varphi, \Phi}^\mathrm{e} = 0.60$ 1/$\mu$s and $\Gamma_{\varphi, 0}^\mathrm{e} = 0.09$ 1/$\mu$s, respectively.}
    }
    % The results in all of the panels have been obtained with the qubit B. }
\end{figure}

%EH: For the table, I use the following mapping 
%U1C, qb2 -> A
%U1C, qb3 -> B
%U1D, qb1 -> C
%U1D, qb2 -> D
%U1D, qb3 -> E
%Data is from 
\clearpage
\begin{table}
\centering
\caption{\label{tab: basic qubit properties} \textbf{Characteristic parameter values for the five measured unimon qubits.} For each of the characterized qubits, the data includes the total length $2l$ of the center conductor in the qubit, the fitted %MM: from which theory? EH: added below
inductance $L_l$ and capacitance $C_l$ per unit length of the center conductor, the fitted Josephson energy $E_\mathrm{J}$, %MM: from which theory?
the inductive and capacitive energies $E_{L, m}$ and $\ECm$, the measured qubit frequency $\omega_{01}/(2\pi)$, the measured anharmonicity $\alpha/(2\pi)$, the location $x_\mathrm{g}$ of the unimon--resonator coupling point with respect to the junction, the measured %MM: Changed to measured (from fitted) since this fit is so simple that essentially to me it feels like that it is measured
coupling strength $|g_{01}|/(2\pi)$ between the qubit and its readout resonator providing the corresponding coupling capacitance $C_\mathrm{g}$, the measured dispersive shift $\chi/(2\pi)$, the measured frequency of the readout resonator $f_\mathrm{r}$, and the measured linewidth of the readout resonator $\kappa/(2\pi)$.  All the fitted values were estimated using theory model 1 (Methods). The values of the flux-dependent quantities $E_{L, m}$, $\ECm$, $\omega_{01}/(2\pi)$, $\alpha/(2\pi)$, $g_{01}/(2\pi)$, $\chi/(2\pi)$, and $f_\mathrm{r}$ are reported at $\Phi_\mathrm{diff}= \Phi_0/2$. The asterisks denote the fact that the inductance $L_l$ and capacitance $C_l$ per unit length are estimated by fitting the theoretical model to the spectrum of qubit B and that equal values are used for the other qubits due to an identical cross section of the co-planar waveguide in all qubits. The inductive and capacitive energies $E_{L, m}$ and $\ECm$ are computed using the theoretical model 1 (Methods) which is also used to obtain the measured values of $E_\mathrm{J}$, $L_l$ and $C_l$.
}
\vspace{0.2 cm}
\resizebox{\textwidth}{!}{
\begin{tabular}{lrrrrrrrrrrrrrr}
\toprule
Qubit & $2l$ & $L_l$ & $C_l$ & $E_\mathrm{J}/h$ & $E_{L, m}/h$ & $\ECm/h$ & $\omega_{01}/(2\pi)$ & $\alpha/(2\pi)$ & $x_\mathrm{g}$ &  $C_\mathrm{g}$ & $|g_{01}|/(2\pi)$ & $\chi/(2\pi)$ & $f_r$ & $\kappa/(2\pi)$   \\
 & (mm) & ($\mu$H/m) & (pF/m) & (GHz) & (GHz) & (GHz) & (GHz) & (MHz) & (mm) & (fF) &  (MHz) & (MHz) & (GHz) &  (MHz)   \\
\hline
\textbf{A} & 8.0 & 0.821$^*$ & 87.1$^*$ & 23.3 & 24.9 & 0.318 & 3.547 & 744 & 0.596 & 9 & 53.5 & 0.74 & 5.826 & 0.43 \\
\textbf{B} &  8.0 & 0.821 & 87.1 & 19.0 & 25.2 & 0.297 & 4.488 & 434 & 0.596 & 10 & 70.0 & 1.2 & 6.198 & 1.24\\
\textbf{C} & 8.0 & 0.821$^*$ & 87.1$^*$ & 17.4 & 25.3 & 0.290 & 4.781 & 343 &  0.596 & 12.5 & 79.7 & 9.20 & 5.522 & 9.2\\
\textbf{D} & 8.0 & 0.821$^*$ & 87.1$^*$ & 14.8 & 25.7 & 0.278 & 5.257 & 214 &  0.596&  12.5 & 85.7 & 20.2 & 5.699 &  10.0\\
\textbf{E} & 8.0 & 0.821$^*$ & 87.1$^*$ & 15.0 & 25.7 & 0.279 & 5.224 & 257 & 0.596 &  12.5 & 92.3 & 4.1 & 6.156 & 1.8 \\
\bottomrule
\end{tabular}}
\end{table}

\clearpage
\begin{table}
\centering
\caption{\label{tab: results of characterization} \textbf{Results of coherence characterization and randomized-benchmarking (RB) experiments for the five unimon qubits.} The measured energy relaxation time $T_1$, the measured Ramsey coherence time $T_2^*$, %the fitted flux noise density $A_\phi^*$ based on Ramsey coherence times, 
the measured echo coherence time $T_2^\mathrm{e}$, the flux noise density parameter $A_{\Phid}$ at 1~Hz estimated from the echo coherence time, the average fidelity per microwave gate from standard RB experiments, and the gate duration used in the RB experiments.  All of the values are measured in the immediate vicinity of the sweet spot $\Phi_\mathrm{diff} = \Phi_0/2$.}
\vspace{0.2 cm}
\resizebox{0.65\textwidth}{!}{
\begin{tabular}{@{\extracolsep{4pt}} lrrrrrr @{}}
\toprule
 & \multicolumn{6}{c}{\textbf{$\Phid = \Phi_0/2$}} \\
 \cline{2-7} 
Qubit & $T_1$ & $T_2^*$ %& $A_\phi^*$ 
& $T_2^\mathrm{e}$ & $A_{\Phid}$ & RB fidelity & Gate duration \\
 & ($\mu$s) & ($\mu$s) %& ($\mu \Phi_0$) 
 & ($\mu$s) & ($\mu \Phi_0$) &  (\%) & (ns) \\
\hline
\textbf{A} & 7.2 & 1.9 %& 317 &  &7.2
& 6.8 & 6.1 & 99.81 & 13.3\\ % & 8.6 old flux noise
\textbf{B} & 8.6 & 3.1 %& 788 & &16.6
& 9.2 & 15.0 & 99.90 & 13.3 \\ % & 19.9 old flux noise
\textbf{C} & 5.8 & 2.3 %& 731 & & 9.2
& 9.3 & 11.2  & 99.54 & 40 \\ %11.1 old flux noise
\textbf{D} & 3.9 & 2.3 %& 760 & &7.3
& 7.0 & 11.1 & 99.63 & 20 \\ %8.8 öld flux noise
\textbf{E} &  5.6 & 2.5 %& 750 & & 12.4
& 11.4 & 14.3 & 99.86 & 13.3 \\ %14.9 old flux noise
\bottomrule
\end{tabular}}
\end{table}

\clearpage
\begin{table}
\centering
\caption{\label{tab: designed parameters} \textbf{Design values of the characteristic parameters of the five unimon qubits.} The design values listed in the table include the inductance $L_l$ and capacitance $C_l$ per unit length of the center conductor of the qubit, the characteristic impedance $Z$ of the transmission line in the qubit, the coupling capacitance $C_\mathrm{g}$ and coupling strength $|g_{01}|/(2\pi)$ between the qubit and its readout resonator, the coupling capacitance $C_\mathrm{d}$ between the qubit and its drive line, the dispersive shift $\chi/(2\pi)$ of the qubit, the frequency $f_r$ of the readout resonator, the coupling capacitance $C_\kappa$ between the readout resonator and the readout transmission line, and the linewidth of the readout resonator $\kappa/(2\pi)$. Note that the coupling strength $|g_{01}|/(2\pi)$ is estimated by a simulation, in which the resonator frequency is tuned on resonance with a linearized unimon circuit in order to measure $|g_{01}|/(2\pi)$ from the frequency separation of the resonances in the formed avoided crossing. }
\vspace{0.2 cm}
\resizebox{0.8\textwidth}{!}{
\begin{tabular}{lrrrrrrrrrr}
\toprule
Qubit & $L_l$ & $C_l$ & $Z$ & $C_\mathrm{g}$ & $|g_{01}|/(2\pi)$ & $C_\mathrm{d}$ & $\chi/(2\pi)$ & $f_r$ & $C_\kappa$ & $\kappa/(2\pi)$   \\
 & ($\mu$H/m) & (pF/m) & ($\Omega$) & (fF) & (MHz) & (fF) & (MHz) & (GHz) &  (fF) & (MHz)   \\
\hline
\textbf{A} & 0.83 & 83 & 100 & 10.4 & 53.2 & 0.083 & 0.89 & 6.0 & 16.8 & 0.89 \\
\textbf{B} & 0.83 & 83 & 100 & 10.4 & 53.2 & 0.083 & 0.58 & 6.3 & 12.8 & 0.58 \\
\textbf{C} & 0.83 & 83 & 100 & 14.0 & 69.4 & 0.083 & 2.48  & 5.7 & 29.6 & 2.48 \\
\textbf{D} & 0.83 & 83 & 100 & 14.0 & 69.4 & 0.083 & 1.44 & 6.0 & 28.1 & 1.44 \\
\textbf{E} & 0.83 & 83 & 100 & 14.0 & 69.4 & 0.083 & 0.94 & 6.3 & 21.5 & 0.94 \\
\bottomrule
\end{tabular}}
\end{table}

\pagebreak

\end{document}

%\begin{center}
%    \LARGE{Supplementary Materials for \\ \textbf{Unimon qubit}}
%\end{center}
\maketitle
\thispagestyle{empty}
\clearpage
\pagenumbering{arabic}

\renewcommand{\theequation}{S\arabic{equation}}
\renewcommand{\thefigure}{\textbf{S\arabic{figure}}}
\renewcommand{\thetable}{\textbf{S\arabic{table}}}
% \addto\captionsenglish{\renewcommand{\figurename}{\textbf{Supplementary Figure}}}
% \renewcommand{\theequation}{S\arabic{equation}}
\renewcommand{\figurename}{\textbf{Supplementary Fig.}} 
\setcounter{figure}{0} 
\renewcommand{\tablename}{\textbf{Supplementary Table}} 
\setcounter{table}{0} 

\section*{Supplementary Methods I: Derivation for the Hamiltonian of the unimon qubit using a distributed-element circuit model}
In this section, we provide two different derivations for the Hamiltonian of the unimon qubit, both starting from the distributed-element circuit model. These theoretical models correspond to models~1 and~2 in the main text. We begin with model~1 that is based on expressing the Hamiltonian in the basis of interacting classical normal modes that are obtained by linearizing the circuit in the vicinity of its potential energy minimum. Subsequently, we proceed to model~2 which utilizes the path integral formalism to analytically eliminate exactly solvable linear parts of the Lagrangian and to approximate the remaining action with the help of a truncated polynomial.

\begin{figure}[tb]
	\centering	
	\includegraphics[width = 0.35\textwidth]{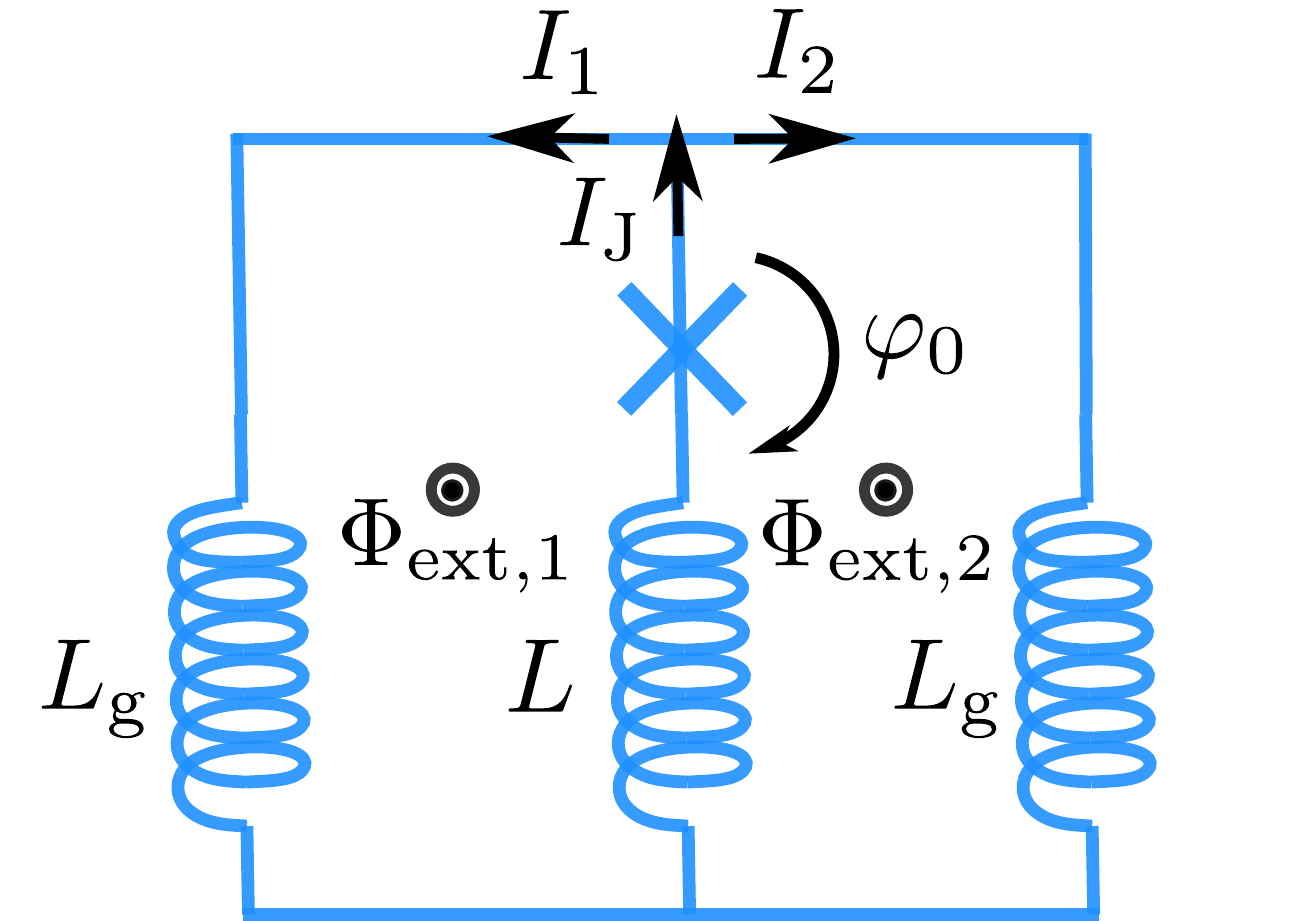}
    \caption{ \label{fig: schematic with flux} \textbf{Direct-current circuit model of the unimon.} Circuit model used for deriving the flux quantization condition of the unimon circuit in the dc regime. Here, $L$ denotes the lumped-element inductance of the center conductor of the CPW resonator, $L_\mathrm{g}$ denotes the inductance of the right and left arms of the outer superconducting loop, $I_\mathrm{J}$ denotes the dc current through the Josephson junction embedded in the center conductor of the CPW resonator, and $I_1$ and $I_2$ denote the dc current in the left and right branch of the outer superconducting loop, respectively. Furthermore, $\Phi_\mathrm{ext, 1}$ and $\Phi_\mathrm{ext, 2}$ denote the external dc magnetic flux through the left and right superconducting loop, respectively, and $\varphi_0$ denotes the dc superconducting phase difference across the Josephson junction due to the external magnetic flux. Note that the used inductor symbols do not show the handedness of the inductors which is defined by equations~\eqref{eq: Phi 1} and~\eqref{eq: Phi 2}.}
\end{figure}

\noindent \textbf{Gradiometricity and classical treatment of the dc supercurrent} \\
First, we study the dc response of the unimon circuit in the presence of an external magnetic flux in order to justify why the circuit is gradiometric, and hence it is only sensitive to the external flux difference. To this end, we first consider the lumped-element circuit model for the currents and fluxes shown in Fig.~\ref{fig: schematic with flux}. The capacitance can be neglected in the case of dc currents. In our model, we denote the total flux through the left (right) loop of the grounded coplanar waveguide (CPW) structure of the unimon by $\Phi_1$ ($\Phi_2$). The loop flux $\Phi_1$ is given by a sum of the external flux and the flux generated by the induced supercurrents as
\begin{equation}
    \Phi_1 = \Phi_{\mathrm{ext},1} + L_\mathrm{g} I_1 + L I_\mathrm{J}, \label{eq: Phi 1}
\end{equation}
where $\Phi_{\mathrm{ext},1}$ is the external dc magnetic flux through the left loop, $L_\mathrm{g}$ is the geometric inductance of the left branch of the ground loop, $L$ is the total inductance of the center conductor, and the currents $I_1$ and $I_\mathrm{J}$ are defined as in Fig.~\ref{fig: schematic with flux}. Similarly, the total dc flux through the right loop $\Phi_2$ is given by
\begin{equation}
    \Phi_2 = \Phi_{\mathrm{ext},2} -L_\mathrm{g} I_2 - L I_\mathrm{J}, \label{eq: Phi 2}
\end{equation}
where $\Phi_{\mathrm{ext},2}$ is the external dc magnetic flux through the right loop,  and $I_2$ is the current in the right branch. Note that we assume that the left and right branch of the ground loop are symmetric. 

Importantly, we model the outer loop as an ideal superconducting loop with a geometric inductance of $L_\mathrm{g}$ in both arms of the loop. Due to the phenomenon of flux quantization\cite{deaver1961experimental,doll1961experimental}, the total dc flux $\Phi_1 + \Phi_2$ must be quantized in units of the flux quantum 
\begin{equation}
    \Phi_1 + \Phi_2 = \Phi_{\mathrm{ext}, 1} + \Phi_{\mathrm{ext}, 2} + L_\mathrm{g}(I_1 - I_2) = n \Phi_0 = 0, \label{eq: flux sum}
\end{equation} 
where $\Phi_0 = h/(2e) \approx 2.067 \times 10^{-15}$ Wb is the flux quantum, and we have set $n=0$ in the last step. Naturally, this implies that the fluxes through the left and right loop must satisfy $\Phi_1 = - \Phi_2$. By considering the flux quantization through the left loop, we obtain an equation that relates the dc phase across the Josephson junction to $\Phi_1$ as
\begin{equation}
    \Phi_1 + \frac{\Phi_0}{2\pi} \varphi_0 = 0 \Rightarrow \varphi_0 = - \frac{2\pi}{\Phi_0} \Phi_1, \label{eq: phase junction flux 1}
\end{equation}
where we have denoted the dc Josephson phase across the Josephson junction with the symbol $\varphi_0$. Note that an identical equation is obtained by considering the right superconducting loop.  

By combining the information in equations~\eqref{eq: Phi 1}--\eqref{eq: phase junction flux 1} and using the Kirchhoff current law $I_1 + I_2 = I_\mathrm{J}$, we find that the dc phase across the Josephson junction reads
\begin{equation}
    \varphi_0 =  \frac{2\pi}{\Phi_0}[\Phi_\mathrm{diff} - (L + L_\mathrm{g}/2) I_\mathrm{J}],\label{eq: flux quantization 1}
\end{equation}
where we have additionally introduced the shorthand notation $ \Phi_\mathrm{diff} = (\Phi_{\mathrm{ext}, 2} - \Phi_{\mathrm{ext}, 1})/2$ for the half difference of the external fluxes through the right and the left loop. Importantly, the dc current across the Josephson junction is given by the dc Josephson relation $I_\mathrm{J} = I_\mathrm{c} \sin(\varphi_0)$, where $I_\mathrm{c}$ is the critical current of the junction. Consequently, the dc phase across the Josephson junction is obtained as a solution of the transcendental equation
\begin{equation}
    \varphi_0 + \frac{L_\mathrm{CPW}}{L_\mathrm{J}} \sin( \varphi_0) = 2\pi  \frac{\Phi_\mathrm{diff}}{\Phi_0}, \label{eq: flux quantization 3}
\end{equation}
where we have used the expression for the Josephson inductance $L_\mathrm{J} = \Phi_0/(2\pi I_\mathrm{c})$ and defined the shorthand notation $L_\mathrm{CPW} = L + L_\mathrm{g}/2$. Importantly, the Josephson phase across the junction is controlled by the difference of the fluxes through the two loops since the circuit contains a gradiometric loop. Since the Josephson phase is only affected by the difference of the external fluxes, the circuit and the corresponding qubit are protected against flux noise which is homogeneous over the length scale of the transverse width of the CPW resonator. 

Note that the equation has a single-valued solution for all values of the flux bias $\Phi_\mathrm{diff}$ if and only if the inductance of the junction $L_\mathrm{J}$ dominates over the inductance $L_\mathrm{CPW}$ of the CPW. This single-valuedness is used below to express the Hamiltonian in terms of the normal modes of the circuit which is linearized around the dc phase. If there are two or more solutions to Eq. \eqref{eq: flux quantization 3}, the treatment becomes more complex. Thus for model~1, we refrain from going into the multivalued regime. However, we observe below that for model~2, no special issues arise even in the multivalued regime.

\begin{figure}[t]
    \centering
    \includegraphics[width=0.85\textwidth]{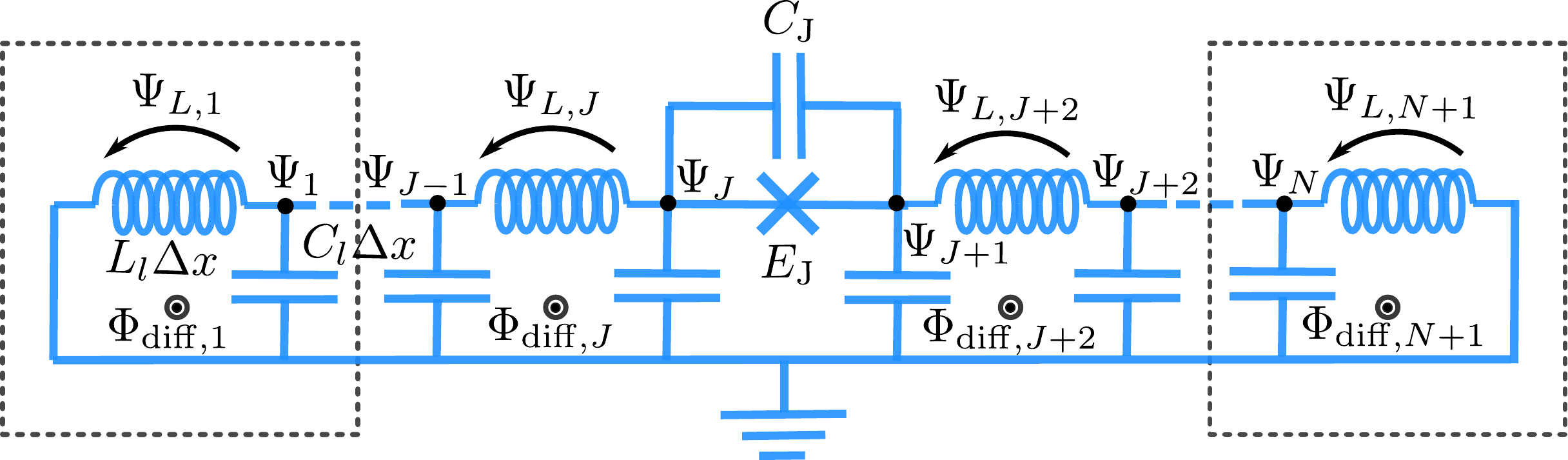}
    \caption{\textbf{Distributed-element circuit model of the unimon.} Discretized model of the unimon circuit in the presence of a flux bias $\Phid$. The circuit model consists of $N$ lumped-element inductors and capacitors such that $J$ inductors and capacitors are located on the left side of the Josephson junction and $N-J$ inductors and capacitors are located on the right side of the Josephson junction. The two boxes with dashed black contours illustrate the repeating elements of the lumped-element circuit on the left and right side of the junction.  The node flux $\Psi_i$ at the $i$th capacitor is defined as the time integral of the corresponding node voltage, whereas $\Psi_{L, i}$ denotes the branch flux across the $i$th inductor.  Note that $\Phi_{\mathrm{diff}, i}$ should be interpreted as the corresponding half difference of the external magnetic fluxes on the two sides of the center conductor of the CPW. A positive applied flux is defined to point out of the page, and hence to give rise to a clockwise rotating dc supercurrent.
    }
    \label{fig: distributed element circuit}
\end{figure}

\noindent \textbf{Derivation of the quantum Hamiltonian of coupled normal modes for the unimon qubit (model 1)} \\
To derive the Hamiltonian for the unimon circuit, we use the discretized circuit model visualized in Fig.~\ref{fig: distributed element circuit}. Our theoretical approach is inspired by  Ref.\cite{bourassa2012josephson}, in which the Hamiltonian is derived for a system consisting of a floating CPW resonator with an embedded Josephson junction in its center conductor. However, we extend the theory to model phase-biased Josephson junctions in order to consider effects arising from an external magnetic field. To summarize our derivation of the Hamiltonian, we first derive classical equations of motion for the distributed-element system. Subsequently, we find the classical normal modes of the system by treating the phase-biased Josephson junction as a linear inductor. After finding the classical normal modes, we make a single-mode approximation, in which we neglect all other modes apart from the mode that is used as a qubit. Finally, the classical single-mode Hamiltonian is quantized by enforcing the canonical commutation relation between phase and charge operators, as a result of which we obtain an accurate approximation for the Hamiltonian of the unimon qubit.

In the discretized circuit diagram shown in Fig.~\ref{fig: distributed element circuit}, we model the CPW of length $2l$ by $N$ lumped-element inductors and capacitors. The Josephson junction at $x_\mathrm{J} \in (-l, l)$ is taken to be located between the capacitors with indices $J$ and $J +1$. Due to the gradiometricity of the circuit, the total external flux in the circuit model corresponds to the half difference of the external fluxes on the two sides of the center conductor in line with the derivation of the previous section.
We use classical circuit theory to derive the equations of motion for the system and to further compute the frequencies and the envelope functions of the normal modes. The equations of motion are obtained from the classical Lagrangian 
\begin{equation}
    \Lag = T - U, \label{eq: Lagrangian ext flux}
\end{equation}
where $T$ and $U$ are the kinetic and the potential energy, respectively.  By choosing the node fluxes $\Psi_i$ as the generalized free coordinates, the total kinetic energy of the circuit can be written as
\begin{gather}
    T = \sum_{i = 1}^N \frac{1}{2} \Cl \Delta x \dot{\Psi}_i^2 + \frac{1}{2} C_\mathrm{J} (\dot{\Psi}_{J+1} - \dot{\Psi}_{J})^2 \label{eq: kinetic energy with flux} \\
    \xrightarrow[N \to \infty]{}  \int_{-l}^{x_\mathrm{J}}\frac{\Cl}{2} \dot{\psi}(x)^2 \mathrm{d}x + \int_{x_\mathrm{J}}^{l}\frac{\Cl}{2} \dot{\psi}(x)^2 \mathrm{d}x  + \frac{1}{2} C_\mathrm{J} (\dot{\psi}(x_\mathrm{J}^+) - \dot{\psi}(x_\mathrm{J}^-))^2, \label{eq: kinetic energy with flux continuum limit}
\end{gather}
where on the first line $\Psi_i = \int_{-\infty}^t V_i(t') \, \mathrm{d}t'$ denotes the node flux related to the $i$th capacitor in terms of the node voltage $V_i$, the index $i$ assumes values from the set $\{1, \dots, J, J+1, \dots, N \}$, $\Delta x = 2l/N$ describes the length scale of the discretization, $\Cl$ denotes the capacitance per unit length of the CPW, and $C_\mathrm{J}$ denotes the junction capacitance that is much smaller than the other relevant capacitances in the circuit. The second line represents the kinetic energy in the continuum limit, $N \to \infty$, where $\Psi_i \rightarrow \psi(x_i, t)$, and $x_\mathrm{J}^-$ and $x_\mathrm{J}^+$ correspond to the locations of the left and right electrode of the Josephson junction, respectively.

Before writing down the potential energy of the circuit, we relate the flux across the $i$th inductor to the external magnetic flux using Faraday's law 
\begin{equation}
    -\Phi_{\mathrm{diff}, i} + \Psi_{i} + \Psi_{L, i} - \Psi_{i-1} = 0 \Rightarrow \Psi_{L, i} = \Psi_{i-1} - \Psi_{i} + \Phi_{\mathrm{diff}, i}, \label{eq: Faradays law ext flux}
\end{equation}
where $\Psi_{L, i}$ denotes the flux across the $i$th inductor, the index $i$ assumes values from the set $\{1, \dots, J, J+2, \dots, N+1 \}$, and $\Phi_{\mathrm{diff}, i}$  corresponds to half of the external magnetic flux difference on the interval $x \in [x_{i-1},x_i]$. Note that there is no inductor between the indices $J$ and $J+1$ due to the presence of the Josephson junction. Using the above equation, we express the potential energy of the system as
\begin{gather}
    U =  \frac{1}{2 \Ll \Delta x} \sum_{i = 1, i \neq J+1}^{N+1} (\Psi_{i-1} - \Psi_i + \Phi_{\mathrm{diff}, i})^2   - \EJ \cos \bigg [ \frac{2\pi}{\Phi_0} (\Psi_{J+1} - \Psi_{J}) \bigg ] \label{eq: potential energy with flux} \\
 \xrightarrow[N \to \infty]{} \int_{-l}^{x_\mathrm{J}}\frac{(\partial_x \psi - s B_\mathrm{diff})^2}{2\Ll}\mathrm{d}x + \int_{x_\mathrm{J}}^{l}\frac{(\partial_x \psi - s B_\mathrm{diff})^2}{2\Ll}\mathrm{d}x \nonumber \\-\EJ \cos \left(\frac{2\pi}{\Phi_0}(\psi(x_\mathrm{J}^+) -\psi(x_\mathrm{J}^-)) \right),  \label{eq: potential energy with flux continuum} 
\end{gather} 
where on the first line we define $\Psi_0 = \Psi_{N+1} = 0$, $\Ll$ denotes the total inductance per unit length, and $\EJ$ denotes the Josephson energy. The second line represents the potential energy in the continuum limit, where $\Phi_\mathrm{diff, i}/\Delta x \rightarrow s B_\mathrm{diff}(x_i)$ with $s$ being the perpendicular distance between the center conductor and the ground plane and $B_\mathrm{diff}(x_i)$ denoting half of the effective magnetic flux density difference on the two sides of the center conductor at the location $x_i$. 

%+  \frac{1}{2 \Ll \Delta x} \sum_{i = J+2}^{N} (\Psi_{i-1} - \Psi_i + \Phi_{\mathrm{ext}, i})^2 

The classical equation of motion for the $i$th node flux $\Psi_i$ ($i \neq J, J+1$)  is obtained from the Euler--Lagrange equation 
\begin{equation}
    \frac{\dint}{\dint t} \frac{\partial \Lag}{\partial \dot{\Psi}_i} - \frac{\partial \Lag}{\partial \Psi_i} = 0.
\end{equation}
Inserting the Lagrangian in equation~\eqref{eq: Lagrangian ext flux} into the Euler--Lagrange equation, we acquire
\begin{equation}
    \Cl \Delta x \ddot{\Psi}_i - \frac{1}{\Ll \Delta x} \left(\Psi_{i + 1} - 2\Psi_i + \Psi_{i - 1} + \Phi_{\mathrm{diff}, i} - \Phi_{\mathrm{diff}, i  +1} \right) = 0, ~~ i  \neq J, J+1.
\end{equation}
By taking the the continuum limit, $\Delta x \rightarrow 0$, of the above equation of motion, we further obtain 
\begin{equation}
    \Cl \ddot{\psi} = \frac{1}{\Ll} \partial_{xx} \psi - \frac{s}{\Ll} \partial_x B_\mathrm{diff}. \label{eq: wave eq. with external flux 3}
\end{equation}
Note that equation~\eqref{eq: wave eq. with external flux 3} corresponds to a wave equation with a source term. In the following calculations, we assume that the applied magnetic field is sufficiently homogeneous in the $x$ direction such that we can  neglect the magnetic-field-dependent term from equation~\eqref{eq: wave eq. with external flux 3} resulting in the wave equation
\begin{equation}
    \ddot{\psi} = v_\mathrm{p}^2\partial_{xx} \psi, \label{eq: wave eq. with homogenous flux}
\end{equation}
where we have defined the phase velocity $v_\mathrm{p} = 1/\sqrt{\Ll \Cl}$.

 Subsequently, we derive the equation of motion for the node flux $\Psi_J$ that corresponds to the left electrode of the junction. According to the Euler--Lagrange equation, we obtain
\begin{equation}
    \Cl \Delta x \ddot{\Psi}_J + C_\mathrm{J} (\ddot{\Psi}_J - \ddot{\Psi}_{J+1})  + I_\mathrm{c} \sin \bigg [ \frac{2\pi}{\Phi_0} (\Psi_J - \Psi_{J+1})  \bigg ] = -\frac{1}{\Ll \Delta x} (\Psi_J - \Psi_{J - 1} -  \Phi_{\mathrm{diff}, J}).
\end{equation}
In the continuum limit, $\Delta x \rightarrow 0$, the above equation of motion can be written as
\begin{equation}
    -C_\mathrm{J} \Delta \ddot{\psi} - I_\mathrm{c} \sin \left(\frac{2\pi}{\Phi_0}\Delta \psi \right) = -\frac{1}{\Ll} \partial_x \psi \bigg |_{x = x_\mathrm{J}^-} + \frac{\Phi_\mathrm{diff}}{2 l \Ll},  \label{eq: boundary cond with external flux 3}
\end{equation} 
where $\Delta \psi = \Psi_{J + 1} - \Psi_J$ denotes the change in the flux across the junction,  $\Phi_\mathrm{diff} = \sum_{i} \Phi_{\mathrm{diff}, i}$ is the total external half flux difference as above, and we have utilized the assumption of a homogenous magnetic field in order to write $\Phi_{\mathrm{diff},J}/\Delta x \rightarrow \Phi_\mathrm{diff}/(2 l)$. Note that our definition of $\Delta \psi$ has a sign convention opposite to $\Psi_{L,i}$.  
Intuitively speaking, the above equation imposes the current continuity condition at the left electrode of the Josephson junction. The term on the left side of equation~\eqref{eq: boundary cond with external flux 3} can be interpreted as the total current across the Josephson junction including the charging of the junction electrodes and a supercurrent across an ideal Josephson element. The term on the right side of equation~\eqref{eq: boundary cond with external flux 3} corresponds to the current in the CPW resonator in the presence of an external magnetic flux as can be inferred from Eq. \eqref{eq: Faradays law ext flux}, according to which the current at location $x_i$ in the center conductor is given by
\begin{equation}
    I = \frac{\Psi_{i-1} - \Psi_i + \Phi_{\mathrm{diff},i}}{\Ll \Delta x} \xrightarrow[N \to \infty]{} -\frac{1}{\Ll} \partial_x \psi \bigg |_{x = x_i} + \frac{\Phi_\mathrm{diff}}{2l \Ll}, \label{eq: current in center conductor}
\end{equation}
where we have again imposed the assumption of a homogeneous flux bias $\Phi_\mathrm{diff}$. Importantly, we can also derive a similar boundary condition at the right electrode of the Josephson junction
\begin{equation}
    -C_\mathrm{J} \Delta \ddot{\psi} - I_\mathrm{c} \sin \left(\frac{2\pi}{\Phi_0}\Delta \psi \right) = -\frac{1}{\Ll} \partial_x \psi \bigg |_{x = x_\mathrm{J}^+} + \frac{\Phi_\mathrm{diff}}{2 l \Ll}.  \label{eq: boundary cond with external flux 3 right}
\end{equation}
%Note that the current flowing in the CPW attains an additional component $\Phi_\mathrm{diff}/(2 l \Ll)$ due to the external flux, and therefore, the current in the center conductor at location $x \neq x_\mathrm{J}$ is given by
%\begin{equation}
%I = -\frac{1}{\Ll} \partial_x\psi + \frac{\Phi_\mathrm{diff}}{2l \Ll}.
%\end{equation}

Classically, the solution to the wave equation in equation~\eqref{eq: wave eq. with homogenous flux}  can be decomposed into a sum of oscillatory normal modes and a dc component corresponding to a spatially and temporally constant current as
\begin{equation}
    \psi(x, t) = \phi_0 u_0(x) + \sum_{n \geq 1} u_n(x) \psi_n(t), \label{eq: flux decomposition with flux}
\end{equation}
where the result applies for $x \neq x_\mathrm{J}$. In the above equation, $\phi_0$ is the coefficient of the dc mode in the units of flux, and $u_0(x)$ is the corresponding dimensionless envelope function. Furthermore, $u_n(x)$ are dimensionless mode envelopes and $\psi_n(t) = c_n\exp(-\mathrm{i} \omega_n t)$ are the corresponding temporally oscillating coefficients in the units of flux with $\omega_n$ being the classical mode frequency. In order for the decomposition to make sense, the mode envelopes must be required to satisfy the wave equation in equation~\eqref{eq: wave eq. with homogenous flux}, and the boundary conditions corresponding to the grounding of the CPW and the current continuity across the Josephson junction [see equations~\eqref{eq: boundary cond with external flux 3} and~\eqref{eq: boundary cond with external flux 3 right}].  

Importantly, the dc component of the current should behave in line with the results of the dc analysis presented in the previous section. To recover equation~\eqref{eq: flux quantization 3} and to comply with the grounding of the CPW, the envelope function of the dc mode must be chosen as a piece-wise linear function
\begin{equation}
	u_0(x) = \begin{cases} (x+l)/(2l), \quad & \textrm{ for } x \in [-l, x_\mathrm{J}), \\(x-l)/(2l), \quad & \textrm{ for } x \in (x_\mathrm{J}, l], \end{cases}
\end{equation}
which corresponds to a spatially constant current of $I = (\Phi_\mathrm{diff} - \phi_0)/(2l \Ll)$ based on Eq. \eqref{eq: current in center conductor} %MM: you could cite to the equation that yields this current. I could not spot it now.
%EH: cited now
and to a dc Josephson phase difference of $\varphi_0 = 2\pi \phi_0/\Phi_0$. Importantly, we  use the decomposition in equation~\eqref{eq: flux decomposition with flux} and rewrite the current continuity condition in equation~\eqref{eq: boundary cond with external flux 3} as
\begin{gather}
    C_\mathrm{J} \sum_{n \geq 1} \omega_n^2 \Delta u_n \psi_n + I_\mathrm{c} \sin \bigg [ \frac{2\pi}{\Phi_0} \bigg( \phi_0 - \sum_{n \geq 1} \Delta u_n \psi_n \bigg) \bigg ] + \overbrace{I_\mathrm{c} \sin\left( \frac{2\pi\phi_0}{\Phi_0}  \right)- I_\mathrm{c} \sin\left( \frac{2\pi\phi_0}{\Phi_0} \right)}^{= 0} \nonumber \\
    =-\frac{1}{\Ll} \left(\frac{\phi_0}{2l } + \sum_{n \geq 1} \psi_n \partial_x u_n \big |_{x = x_\mathrm{J}^-}  \right) + \frac{\Phi_\mathrm{diff}}{2 l \Ll},
\end{gather}
where $\Delta u_n = u_n(x_\mathrm{J}^+) - u_n(x_\mathrm{J}^-)$ denotes the change in the envelope of the $n$th mode across the Josephson junction. The above equation can be grouped into a time-independent part and a time-dependent part that should both vanish 
\begin{gather}
    \bigg \{ I_\mathrm{c} \sin\left( \frac{2\pi\phi_0}{\Phi_0}  \right) + \frac{\phi_0}{2l \Ll} - \frac{\Phi_\mathrm{diff}}{2 l \Ll} \bigg \} + \bigg \{ C_\mathrm{J} \sum_{n \geq 1} \omega_n^2 \Delta u_n \psi_n  \nonumber \\
     +  I_\mathrm{c} \sin \bigg [ \frac{2\pi}{\Phi_0} \bigg( \phi_0 - \sum_{n \geq 1} \Delta u_n \psi_n \bigg) \bigg ] -  I_\mathrm{c} \sin\left( \frac{2\pi\phi_0}{\Phi_0}  \right) + \frac{1}{\Ll}\sum_{n \geq 1}  \psi_n  \partial_x u_n \big |_{x = x_\mathrm{J}^-}  \bigg \} = 0. \label{eq: time independent and dependent bound cond ext flux}
\end{gather}
We simplify the time-dependent part further using the trigonometric identity $\sin(x + y) = \sin(x) \cos(y) + \cos(x) \sin(y) $ as
\begin{gather}
    C_\mathrm{J} \sum_{n \geq 1} \omega_n^2 \Delta u_n \psi_n + I_\mathrm{c} \sin \bigg (\frac{2\pi \phi_0}{\Phi_0} \bigg )\cos \bigg( \frac{2\pi}{\Phi_0} \sum_{n \geq 1} \Delta u_n \psi_n \bigg) \nonumber \\ - I_\mathrm{c} \cos \bigg (\frac{2\pi \phi_0}{\Phi_0} \bigg ) \sin \bigg ( \frac{2\pi}{\Phi_0} \sum_{n \geq 1} \Delta u_n \psi_n \bigg )   -  I_\mathrm{c} \sin\left( \frac{2\pi\phi_0}{\Phi_0}  \right) + \frac{1}{\Ll}\sum_{n \geq 1}\psi_n \partial_x u_n \big |_{x = x_\mathrm{J}^-}   = 0.   
\end{gather}
By invoking the assumption of small ac oscillations $\big( \zeta:=\sum_{n \geq 1} 2 \pi \Delta u_n \psi_n/\Phi_0  \ll 1 \big)$, we write the above equation into a simple form 
\begin{gather}
    \sum_{n \geq 1} \psi_n \bigg [C_\mathrm{J} \omega_n^2 \Delta u_n - \frac{\cos (2\pi \phi_0/\Phi_0)}{L_\mathrm{J}} \Delta u_n  + \frac{1}{\Ll} \partial_x u_n \big |_{x = x_\mathrm{J}^-} \bigg ] \approx 0, \label{eq: time dependent bound cond ext flux}
\end{gather}
where we used the approximations $\sin(\zeta) \approx  \zeta$ and $\cos(\zeta) \approx 1 $ together with the relation $L_\mathrm{J} = \Phi_0/(2 \pi I_\mathrm{c})$. Importantly, each term corresponding to a different $n$ in equation~\eqref{eq: time dependent bound cond ext flux} must vanish independently since all the coefficients $\psi_n \propto \exp(- \mathrm{i} \omega_n t)$ oscillate in time with different frequencies. Note that our assumption of small ac oscillations $\big( \sum_{n \geq 1} 2 \pi \Delta u_n \psi_n/\Phi_0  \ll 1 \big)$ is equivalent to linearizing the circuit around its dc operation point, which also has the consequence of decoupling the different modes of the circuit. In the following equations, we drop the approximation sign for notational simplicity.

Under the assumption of small ac oscillations, it is thus possible to summarize the current continuity conditions across the Josephson junction with the following set of decoupled equations
\begin{gather}
    I_\mathrm{c} \sin \left(\frac{2\pi}{\Phi_0} \phi_0  \right) + \frac{\phi_0}{2l \Ll}  = \frac{\Phi_\mathrm{diff}}{2l \Ll}, \label{eq: boundary cond dc 3} \\
    \omega_m^2 C_\mathrm{J} \Delta u_m - \frac{\cos(2\pi \phi_0/\Phi_0)}{L_\mathrm{J}} \Delta u_m = -\frac{1}{\Ll}\partial_x u_m|_{x = x_\mathrm{J}^-}, \quad m \geq 1, \label{eq: boundary cond ac 3} \\
    -\frac{1}{\Ll}\partial_x u_m|_{x = x_\mathrm{J}^-} = -\frac{1}{\Ll}\partial_x u_m|_{x = x_\mathrm{J}^+}, \quad m \geq 1, \label{eq: boundary cond ac both sides}
\end{gather}
where the final equation essentially requires current continuity on both sides of the junction and it has been obtained by also considering the current continuity condition of the right electrode in equation~\eqref{eq: boundary cond with external flux 3 right}. We make a few important observations based on the above set of boundary conditions. Assuming that the ground-loop inductance is much smaller than the total inductance of the center conductor $2l \Ll$, we have $L_\mathrm{CPW} = 2l \Ll$, in the case of which equation~\eqref{eq: boundary cond dc 3} reduces to the dc flux quantization condition presented in  equation~\eqref{eq: flux quantization 3}. On the other hand, equation~\eqref{eq: boundary cond ac 3} can be interpreted such that the dc supercurrent essentially changes the phase bias of the Josephson junction by $2\pi\phi_0/\Phi_0$ resulting in an effective linear inductance of $L_\mathrm{J}/\cos(2\pi \phi_0/\Phi_0)$ in the limit of small oscillations. Importantly, the effective inductance of the biased Josephson junction $L_\mathrm{J}/\cos(2\pi \phi_0/\Phi_0)$ can be negative for certain values of the dc phase $\varphi_0 = 2\pi \phi_0/\Phi_0$. The operation points corresponding to negative inductances are classically stable with a single energy minimum if the Josephson inductance $L_\mathrm{J}$ exceeds the total inductance of the CPW $L_\mathrm{CPW} = 2l\Ll$. 

Subsequently, we use the derived set of boundary conditions to determine the normal-mode angular frequencies $\omega_m$ and the mode envelope functions $u_m(x)$. To satisfy both the wave equation~\eqref{eq: wave eq. with homogenous flux} and the grounding condition, we use a piece-wise sinusoidal ansatz for each of the envelope functions
\begin{equation}
    u_m(x) = \begin{cases} A_m \sin [ k_m (x + l) ], & \textrm{if } x \in [-l, x_\mathrm{J}) \\  A_m B_m \sin [ k_m (x - l) ], & \textrm{if } x \in (x_\mathrm{J}, l] \end{cases} \label{eq: mode um}
\end{equation}
where $k_m $ is the wavenumber of the $m$th mode obeying $k_m = \omega_m/v_\mathrm{p}$ in order  to satisfy the wave equation, and $A_m$ and $B_m$ are dimensionless constants. We solve $B_m$ by inserting the piece-wise sinusoidal ansatz of the mode envelope function to the boundary condition in equation~\eqref{eq: boundary cond ac both sides}, which yields
\begin{gather}
B_m = \frac{\cos[k_m(x_\mathrm{J} + l)]}{\cos[k_m(x_\mathrm{J} - l)]}. \label{eq: boundary cond Bm}
\end{gather}
 By further inserting the piece-wise sinusoidal mode envelopes to the boundary condition in equation~\eqref{eq: boundary cond ac 3} and using equation~\eqref{eq: boundary cond Bm} to carry out simplifications, we obtain a transcendental equation for the wavenumbers $k_m$ ($m \geq 1$) of the normal modes
\begin{gather}
    k_m l\cos[k_m (x_\mathrm{J} - l)]\cos[k_m (x_\mathrm{J} + l)] - \nonumber \\ \bigg [\frac{C_\mathrm{J}(k_m l)^2 }{\Cl l}
    - \frac{\Ll l}{L_\mathrm{J}} \cos\left(\frac{2\pi\phi_0}{\Phi_0}\right) \bigg ]\sin(2 k_m l)  = 0, \label{eq: transcendental eq for k with flux}
   %-k_m l = \left(\frac{C_\mathrm{J}(k_m l)^2 }{\Cl l} - \frac{\Ll l}{L_\mathrm{J}} \right) \left( \tan(k_m(x_\mathrm{J} - l)) - \tan(k_m(x_\mathrm{J} + l))\right). \label{eq: transcendental eq for k}
\end{gather}
where the value of $\phi_0$ should be obtained by first solving equation~\eqref{eq: boundary cond dc 3}, and the wave numbers $k_m$ are related to the classical mode angular frequencies $\omega_m$ through the relation $\omega_m = v_\mathrm{p} k_m$. Importantly, the mode envelope functions corresponding to solutions of the above equation may or may not have a discontinuity ($\Delta u_m \neq 0$) at the location of the Josephson junction.  The modes with a discontinuity at the junction are the modes of interest since in the following quantum-mechanical treatment, a mode can have an anharmonic energy spectrum only if the non-linear junction couples to it, i.e., a non-zero current flows across the junction.

\begin{figure}[ht]
	\centering	
	\includegraphics[width = 0.9\textwidth]{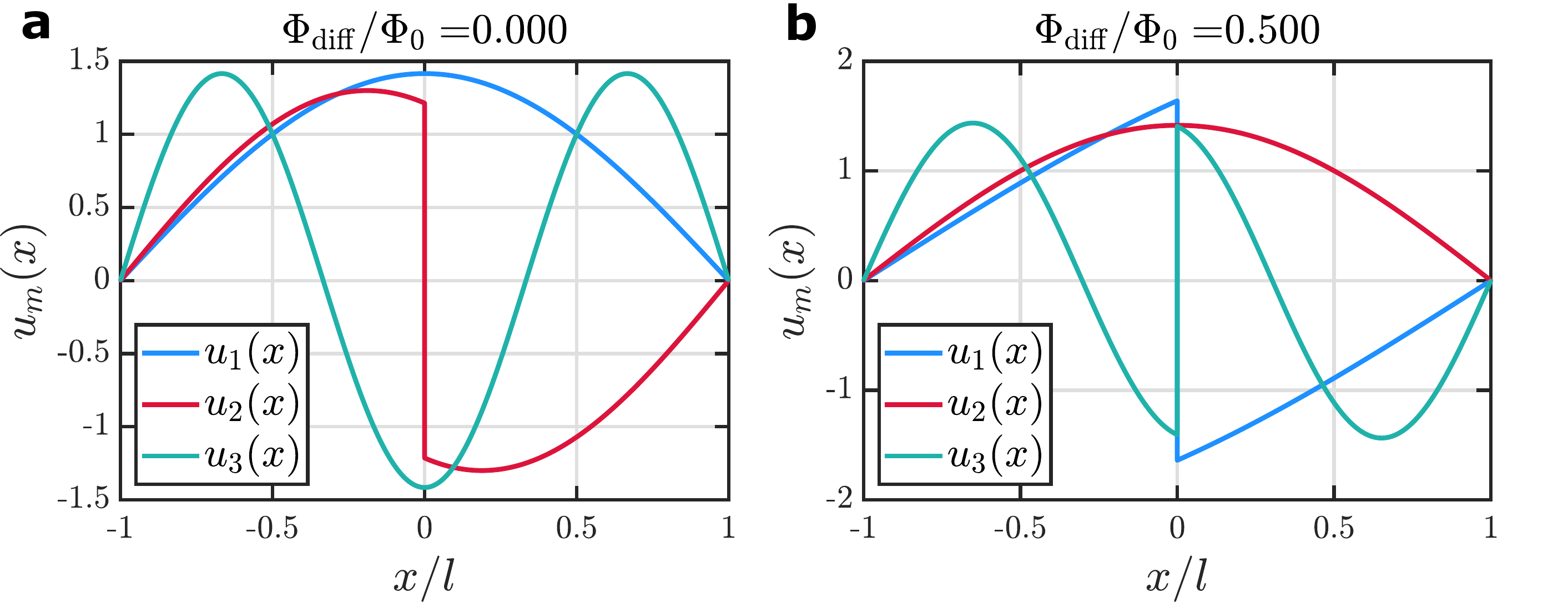}
	%\begin{subfigure}[t]{0.4\textwidth}
    %     \centering
    %     \caption{}
    %    \includegraphics[width=\textwidth]{unimon_mode_envolopes_flux_1.eps}
         %\label{fig:y equals x}
    %\end{subfigure}
    %\begin{subfigure}[t]{0.4\textwidth}
    %     \centering
    %     \caption{}
    %     \includegraphics[width=\textwidth]{unimon_mode_envolopes_flux_2.eps}
         %\label{fig:y equals x}
    %\end{subfigure}
    \\
    \caption{ \label{fig: mode envelopes} \textbf{Mode envelopes of the unimon.} \textbf{a}, \textbf{b}, Flux envelope functions for the three lowest-frequency normal modes at a flux bias of $\Phi_\mathrm{diff}/\Phi_0 = 0.0$ (\textbf{a}) and at $\Phi_\mathrm{diff}/\Phi_0 = 0.5$ (\textbf{b}). The results were obtained using the example parameter set presented in Table~\ref{tab: design parameters}. }
\end{figure}

In Fig.~\ref{fig: mode envelopes}, we illustrate the mode envelope functions of the three normal modes with the lowest frequency at external flux bias  of $\Phi_\mathrm{diff}/\Phi_0 = 0.0$ and $\Phi_\mathrm{diff}/\Phi_0 = 0.5$. The mode envelope functions at the two external fluxes differ due to a different value of the effective Josephson inductance: At $\Phi_\mathrm{diff}/\Phi_0 = 0.0$, the effective Josephson inductance is $L_\mathrm{J}$, whereas at $\Phi_\mathrm{diff}/\Phi_0 = 0.5$, the effective inductance is $-L_\mathrm{J}$. Note that the lowest anharmonic mode corresponds to the second normal mode of the system at $\Phi_\mathrm{diff}/\Phi_0 = 0.0$, whereas the first normal mode is the lowest anharmonic mode at $\Phi_\mathrm{diff}/\Phi_0 = 0.5$.

\begin{table}
\centering
\caption{\label{tab: design parameters} \textbf{Example set of parameter values used in Supplementary Methods I.}  }
\vspace{0.2 cm}
%\resizebox{\textwidth}{!}{
%\begin{tabular}{@{\extracolsep{4pt}} lccc@{}}
\begin{tabular}{cccccccc}
\toprule
$x_\mathrm{J}/l$ & $2l$ (mm) &$\EJ/h$ (GHz)  & $C_\mathrm{J}$ (fF) & $\Cl$ (pF/m) & $\Ll$ ($\mu$H/m) & $Z_0$ ($\Omega$) & $2l\Ll/L_\mathrm{J}$  \\
 \hline
 0.0 & 8.0 & 19.0 & 1.4 & 83 & 0.83 & 100  & 0.772\\
\bottomrule
\end{tabular}%}
\end{table}

Importantly, the classical normal modes satisfy useful orthogonality relations\cite{bourassa2012josephson}
\begin{gather}
\langle u_m, u_n \rangle = \int_{-l}^l  \Cl u_m(x) u_n(x) \, \dint x + C_\mathrm{J} \Delta u_m \Delta u_n = C_\Sigma \delta_{mn}, \label{eq: u orthogonality with flux} \\
 \langle \partial_x u_m, \partial_x u_n \rangle = \int_{-l}^l  \frac{1}{\Ll} \partial_x u_m(x) \partial_x u_n(x) \, \dint x +  \frac{\cos(2\pi \phi_0/\Phi_0)}{L_\mathrm{J}} \Delta u_m \Delta u_n = \frac{\delta_{mn}}{L_m}, \label{eq: u_x orthogonality with flux}
\end{gather}
where  $C_\Sigma = 2 \Cl l + C_\mathrm{J}$ denotes the total capacitance of the circuit, and the effective inductance corresponding to the $m$th mode is given by $L_m =  (C_\Sigma \omega_m^2)^{-1}$. Note that the above orthogonality relations fix the normalization coefficients $\{A_m \}$ introduced in equation~\eqref{eq: mode um}.

Above, we have described how to solve the classical equations of motion for the system and how to determine the frequencies and flux mode envelopes of the normal modes in the limit of small flux oscillations across the junction. In the limit of small oscillations, the classical Hamiltonian decouples to a sum of harmonic oscillators corresponding to the normal modes. Subsequently, we derive an approximation for the classical Hamiltonian of the unimon circuit by invoking a single-mode approximation, in which all other modes apart from the mode used as a qubit are neglected. The single-mode approximation describes the system with a good accuracy since the frequency difference of the anharmonic modes coupled to each other is of the order of 10~GHz in our system and consequently, the coupling between the modes has only a small perturbative effect. 

Note that a more accurate expression for the Hamiltonian may be obtained by keeping more than one of the modes in the following calculations. Such a treatment is needed especially if one considers the case of several solutions to the dc flux in equation~\eqref{eq: flux quantization 3}. Since the mode envelopes provide a complete set of basis functions for the grounded CPW system, an exact expression for the Hamiltonian can, in principle, be obtained by taking into account an infinite number of the modes.  %Eric: Last sentence is only a strong gut-feeling. Hmm..

In the single-mode approximation, we truncate the flux decomposition in equation~\eqref{eq: flux decomposition with flux} as $\psi(x,t) = \phi_0 u_0(x) + \psi_m(t) u_m(x)$, where $m$ is the index of the anharmonic mode ($\Delta u_m \neq 0$) utilized as the qubit. Invoking the single-mode approximation and the continuum limit, the kinetic energy of the circuit in equation~\eqref{eq: kinetic energy with flux} can be expressed as
\begin{equation}
    T = \int_{-l}^l \frac{\Cl}{2} \dot{\psi}_m^2 u_m^2 \dint x + \frac{C_\mathrm{J}}{2} \dot{\psi}_m^2 (\Delta u_m)^2 = \frac{C_\Sigma}{2} \dot{\psi}_m^2,
\end{equation}
where we have used the orthogonality relation in equation~\eqref{eq: u orthogonality with flux} to obtain the final result. Under these approximations, we simplify the potential energy of the circuit in equation~\eqref{eq: potential energy with flux} to obtain 
\begin{align}
    U &= \int_{-l}^l \frac{1}{2 \Ll} \left( \frac{\phi_0}{2l} +  \psi_m \partial_x u_m - \frac{\Phi_\mathrm{diff}}{2l} \right)^2 \dint x - \EJ \cos \bigg [ \frac{2\pi}{\Phi_0}(- \phi_0 + \psi_m \Delta u_m) \bigg ] \nonumber \\ 
    %&=\int_{-l}^l \frac{\psi_m^2 (\partial_x u_m)^2}{2 \Ll} \dint x + \frac{1}{2l \Ll} \int_{-l}^l \psi_m (\partial_x u_m) (\phi_0 - \Phi_\mathrm{diff}) \dint x - \EJ \cos \bigg [ \frac{2\pi}{\Phi_0}(- \phi_0 + \psi_m \Delta u_m) \bigg ] \nonumber \\
    %&= \frac{1}{2}\bigg [ \frac{1}{L_m} - \frac{\cos(2 \pi \phi_0/\Phi_0)}{L_\mathrm{J}} (\Delta u_m)^2  \bigg ] \psi_m^2 + \frac{1}{2l \Ll} \psi_m \Delta u_m ( \Phi_\mathrm{diff} - \phi_0) \nonumber \\
    %& - \EJ \cos \bigg [ \frac{2\pi}{\Phi_0}(- \phi_0 + \psi_m \Delta u_m) \bigg ] \nonumber \\
    &=\frac{\psi_m^2}{2 \tilde{L}_m(\phi_0)}    + \frac{1}{2l \Ll} \psi_m \Delta u_m ( \Phi_\mathrm{diff} - \phi_0) - \EJ \cos \bigg [ \frac{2\pi}{\Phi_0}(- \phi_0 + \psi_m \Delta u_m) \bigg ],
\end{align}
where we have dropped a constant term % on the second line 
and used the orthogonality relation presented in equation~\eqref{eq: u_x orthogonality with flux}. % to obtain  the result on the third line. 
Furthermore, we have defined the effective inductance $\tilde{L}_m(\phi_0)$ in the last step as
\begin{equation}
   \frac{1}{\tilde{L}_m(\phi_0)} = \frac{1}{L_m} - \frac{\cos(2 \pi \phi_0/\Phi_0)}{L_\mathrm{J}} (\Delta u_m)^2.
\end{equation}

Subsequently, we define the flux variable $\phi_m = \psi_m \Delta u_m$ that equals the (time-dependent) flux difference across the Josephson junction. This allows us to write the classical single-mode Lagrangian as
\begin{gather}
    \Lag = \frac{C_m'}{2} \dot{\phi}_m^2 - \frac{\phi_m^2}{2 \tilde{L}_m'}
 - \frac{1}{2l \Ll} \phi_m (\Phi_\mathrm{diff} - \phi_0) + \EJ \cos\bigg [ \frac{2\pi}{\Phi_0}(\phi_m -\phi_0 ) \bigg ], \label{eq: unimon single mode Lagrangian}
 \end{gather}
where $C_m' = C_\Sigma/(\Delta u_m)^2$ and $\tilde{L}_m' = \tilde{L}_m (\Delta u_m)^2$ are the rescaled capacitance and inductance. By taking a Legendre transformation of the Lagrangian, we obtain the corresponding classical Hamiltonian 
\begin{gather}
    \Ham_m = \frac{q_m^2}{2 C_m'} + \frac{\phi_m^2}{2 \tilde{L}_m'}
 +\frac{1}{2l \Ll} \phi_m (\Phi_\mathrm{diff} - \phi_0) - \EJ \cos \bigg [ \frac{2\pi}{\Phi_0}(\phi_m -\phi_0) \bigg ],
\end{gather}
where we have defined the conjugate momentum with units of charge as $q_m = \partial \Lag/ \partial \dot{\phi}_m = C_m' \dot{\psi}_m$.

Finally, we obtain the quantum-mechanical single-mode Hamiltonian by imposing the canonical commutation relation $[\hat{\phi}_m, \hat{q}_m] = \mathrm{i}\hbar$. Furthermore, we define the dimensionless charge operator $\hat{n}_m = \hat{q}_m/(2e)$ and the phase operator $\hat{\varphi}_m = 2 \pi \hat{\phi}_m/\Phi_0$, which allows us to express the quantum Hamiltonian in a form resembling the fluxonium Hamiltonian as
\begin{equation}
    \hat{\Ham}_m = 4\ECm(\varphi_0) \hat{n}_m^2 + \frac{1}{2} E_{L, m}(\varphi_0) \hat{\varphi}_m^2 + E_L \hat{\varphi}_m (\varphi_\mathrm{diff} - \varphi_0) - \EJ \cos(\hat{\varphi}_m  -\varphi_0 ), \label{eq: single mode Hamiltonian with external flux}
\end{equation}
where we have defined $\ECm(\varphi_0) = e^2/[2 C_m'(\varphi_0)]$, $E_{L, m}(\varphi_0) = \Phi_0^2/(2\pi)^2/\tilde{L}_m'(\varphi_0)$, $E_L = \Phi_0^2/(2\pi)^2/(2l \Ll)$, $\varphi_\mathrm{diff} = 2\pi \Phi_\mathrm{diff}/\Phi_0$, and $\varphi_0 = 2\pi \phi_0/\Phi_0$. Note that the conjugate operators $\hat{\varphi}_m$ and $\hat{n}_m$ satisfy the commutation relation $[\hat{\varphi}_m, \hat{n}_m] = \mathrm{i}$.

\begin{figure}[t]
	\centering	
	\includegraphics[width = 0.9\textwidth]{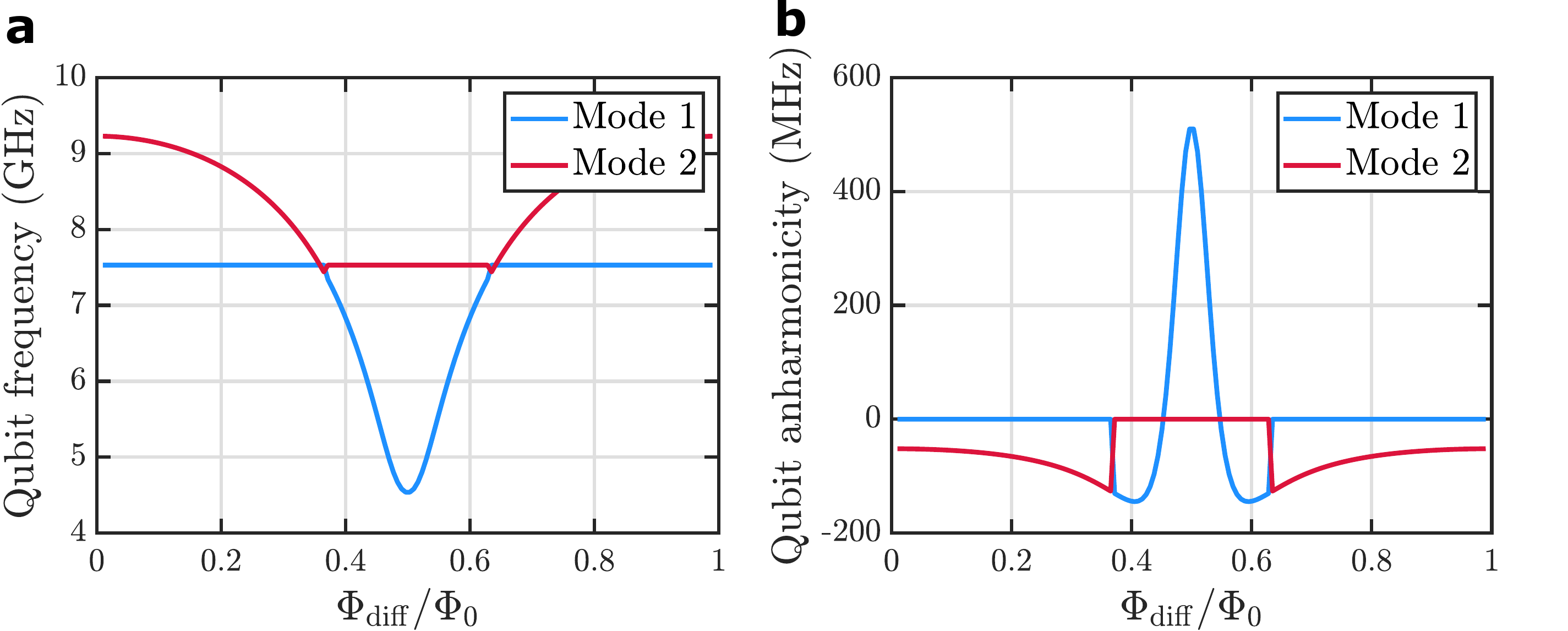}
    \\
    \caption{ \label{fig: theoretical freq and anharmonicity} \textbf{Numerically computed frequency and anharmonicity of the unimon.} \textbf{a}, Qubit frequency $f_{01,m}$ as a function of the flux bias $\Phid$ for the two lowest-frequency modes. \textbf{b}, Anharmonicity $\alpha_m/(2\pi)$ as a function of $\Phid$ for the two lowest-frequency modes. The results were obtained using the example parameter set presented in Table~\ref{tab: design parameters}. Note that the results were computed individually for each of the modes by enforcing the single-mode approximation and diagonalizing the Hamiltonian in equation~\eqref{eq: single mode Hamiltonian with external flux}.} 
\end{figure}

To study the qubit frequency and anharmonicity of the unimon, we numerically diagonalize the Hamiltonian in equation~\eqref{eq: single mode Hamiltonian with external flux} for flux biases in the range $\Phid/\Phi_0 \in [0, 1]$. For each of the modes, the corresponding qubit frequency and anharmonicity are computed as
\begin{gather}
    f_{01, m} = \frac{E_{1,m} - E_{0,m}}{h}, \\ 
    \frac{\alpha_{m}}{2\pi} = \frac{(E_{2,m} - E_{1,m}) - (E_{1,m} - E_{0,m})}{h},
\end{gather}
where  $f_{01, m}$ and $\alpha_m/(2\pi)$ denote the qubit frequency and anharmonicity corresponding to the $m$th mode, whereas $E_{i,m}$ is the $i$th eigenenergy of the $m$th mode. In Fig.~\ref{fig: theoretical freq and anharmonicity}, we illustrate the numerically computed qubit frequency and anharmonicity of the two lowest-frequency modes as functions of the flux bias. Importantly, there is a flux-insensitive sweet spot at $\Phid/\Phi_0 = 0.5$ that corresponds to the smallest frequency and highest anharmonicity of the lowest-frequency mode. For the parameter values presented in Table~\ref{tab: design parameters}, the qubit frequency at $\Phid/\Phi_0 = 0.5$ is approximately 4.5 GHz, whereas the corresponding anharmonicity is +510 MHz. Thus, the sweet spot $\Phid/\Phi_0 = 0.5$ corresponds to the optimal operation point of the unimon.

To gain intuitive understanding of the enhanced anharmonicity at $\Phid/\Phi_0 = 0.5$, we note that the single-mode Hamiltonian simplifies into
\begin{equation}
    \hat{\Ham}_m = 4\ECm(\pi) \hat{n}_m^2 + \frac{1}{2} E_{L, m}(\pi) \hat{\varphi}_m^2  + \EJ \cos(\hat{\varphi}_m ), \label{eq: single mode Hamiltonian at half flux}
\end{equation}
where we have used the fact that $\varphi_\mathrm{diff} = \varphi_0 = \pi$, which is valid if $E_\mathrm{J} \leq E_L$. By expanding the cosine potential as a Taylor series, we obtain 
\begin{equation}
     \hat{\Ham}_m = 4\ECm(\pi) \hat{n}_m^2 + \frac{E_{L, m}(\pi) - E_\mathrm{J}}{2} \hat{\varphi}_m^2  + \frac{\EJ}{24} \hat{\varphi}_m^4 + \mathcal{O}(\hat{\varphi}_m^6). 
\end{equation}
Importantly, the unimon is designed to work in the regime $\EJ \lesssim E_{L, m}(\pi)$. Thus, the enhanced anharmonicity can be attributed to the partial cancellation of the inductive energy $E_{L, m}(\pi)$ and the Josephson energy $E_\mathrm{J}$, as result of which the quadratic potential energy term practically vanishes while the quartic and higher order terms are preserved. 

%\section*{Supplementary Methods II: Path-integral-based model for the unimon}
%\clearpage
\pagebreak
\noindent \textbf{Path-integral-based model for the unimon (model 2)} \\
%\subsection{Path-integral-based model for the unimon}
We start from the classical action for the electromagnetic field of the unimon, which can be expressed with the help of the temporally dependent flux density $\psi(x, t)$ in the unimon circuit based on the continuum-limit Lagrangian given by Eqs. \eqref{eq: Lagrangian ext flux}, \eqref{eq: kinetic energy with flux continuum limit}, \eqref{eq: potential energy with flux continuum} as
\begin{multline}
    S[\psi(x, t)] = \int \left(
        \int\limits_{-l}^{x_\mathrm J} \left\{\frac{\Cl \dot \psi(x, t)^2}{2} -
        \frac{\left[\partial_x \psi(x, t) - s B_\mathrm{diff}(x) \right]^2}{2 \Ll}\right\}\;\mathrm dx \right . \\ \left .
        +
        \int\limits_{x_\mathrm J}^{l} \left\{\frac{\Cl \dot \psi(x, t)^2}{2} -
        \frac{\left[\partial_x \psi(x, t) - s B_\mathrm{diff}(x) \right]^2}{2 \Ll}\right\}\;\mathrm dx
        + \frac{C_\mathrm J \left[\dot \psi\left(x_\mathrm J^-, t\right) - \dot \psi
        \left(x_\mathrm J^+, t\right)\right]^2}{2} \right .  \\ \left . \phantom{\int\limits_{l}^{x_\mathrm J}} + E_\mathrm J \cos\left\{
            \frac{2\pi}{\Phi_0} \left[\psi\left(x_\mathrm J^-, t\right) -
            \psi\left(x_\mathrm J^+, t\right)\right]
            \right\}
        \right)\;\mathrm dt
\end{multline}
The spectrum of the unimon can be extracted from the partition function $Z(\beta)$, where $\beta = 1 / (k_\mathrm B
T)$. The partition function of the unimon can be expressed through a path integral of an exponent of an Euclidean action in the imaginary time $\tau = \textrm{i}t$ as
\begin{equation}
    Z(\beta) = \int \bm D[\psi(x, \tau)] e^{-\frac 1{\hbar} S^\mathrm{E}[\psi(x,
    \tau)]},
\end{equation}
where the path integral is taken over all periodic trajectories in imaginary time and the Euclidean action $S^\mathrm E[\psi(x, \tau)] = S[\psi(x,-i\tau)]$ is given by 
\begin{multline}
    S^\mathrm{E}[\psi(x, \tau)] = \int\limits_0^{\hbar \beta} \left(
        \int\limits_{-l}^{x_\mathrm J} \left\{\frac{\Cl [\partial_\tau \psi(x, \tau)]^2}{2} +
        \frac{\left[\partial_x \psi(x, \tau) - s B_\mathrm{diff}(x) \right]^2}{2 \Ll}\right\}\;\mathrm dx \right . \\ \left.
        +
        \int\limits_{x_\mathrm J}^{l} \left\{\frac{\Cl [\partial_\tau \psi(x, \tau)]^2}{2} +
        \frac{\left[\partial_x \psi(x, \tau) - s B_\mathrm{diff}(x) \right]^2}{2 \Ll}\right\}\;\mathrm dx \right . \\ \left .
        + \frac{C_\mathrm J \left[\partial_\tau \psi(x_\mathrm J^-, \tau) -
        \partial_\tau \psi
        (x_\mathrm J^+, \tau)\right]^2}{2} -  E_\mathrm J \cos\left\{
            \frac{2\pi}{\Phi_0} \left[\psi(x_\mathrm J^-, \tau) -
            \psi(x_\mathrm J^+, \tau)\right]
            \right\}
        \right)\;\mathrm d\tau ,
\end{multline}
where the interval of imaginary time is determined by the inverse temperature $\beta$.
First, we define $\psi'(x, \tau) = \psi(x, \tau) - \Phi_\mathrm{diff}(x)$,  
where
\begin{equation}
    \Phi_\mathrm{diff}(x) = s \left\{
        \begin{split}
            \int\limits_{-l}^x B_\mathrm{diff}(x')\;\mathrm dx',~x < x_\mathrm J  \\
            -\int\limits_{x}^l B_\mathrm{diff}(x')\;\mathrm dx',~x > x_\mathrm J.  \\
        \end{split}
        \right.
\end{equation}
Then the action takes the following form:
\begin{multline}
    S^\mathrm{E}[\psi'(x, \tau)] = 
    \int\limits_0^{\hbar \beta} \left(
        \int\limits_{-l}^{x_\mathrm J} \left\{\frac{\Cl [\partial_\tau \psi'(x, \tau)]^2}{2} +
        \frac{\left[\partial_x \psi'(x, \tau) \right]^2}{2 \Ll}\right\}\;\mathrm dx \right . \\ \left.
        +
        \int\limits_{x_\mathrm J}^{l} \left\{\frac{\Cl [\partial_\tau \psi'(x, \tau)]^2}{2} -
        \frac{\left[\partial_x \psi'(x, \tau) \right]^2}{2 \Ll}\right\}\;\mathrm dx 
        + \frac{C_\mathrm J \left[\partial_\tau \psi'(x_\mathrm J^-, \tau) -
        \partial_\tau \psi'
        (x_\mathrm J^+, \tau)\right]^2}{2} \right . \\ \left .
        \phantom{\int\limits_{x_\mathrm J}^l}-  E_\mathrm J \cos\left\{
            \frac{2\pi}{\Phi_0} \left[\psi'(x_\mathrm J^-, \tau) -
            \psi'(x_\mathrm J^+, \tau) + \Phi_\mathrm{diff}\right]
            \right\}
        \right)\;\mathrm d\tau,
\end{multline}
where $\Phi_\mathrm{diff} = \Phi_\mathrm{diff}(x_\mathrm J^-) - \Phi_\mathrm{diff}(x_\mathrm J^+)$. 

In the following derivation, we eliminate all the linear resonator degrees of freedom and arrive at a model expressed in terms of a single variable, namely phase difference across the junction. To this end, we seek for fixed $\psi'(x_\mathrm J^-, \tau)$ and $\psi'(x_\mathrm J^+,\tau)$ classical trajectories which minimize the Euclidean action of the linear resonator parts of the system. We first focus on the left half of the resonator with $x \in [-l, x_\mathrm J)$ and then adjust the results for the right side of the resonator.
The equation for the classical trajectory $\psi_\mathrm c(x, \tau)$ corresponds to the Euler--Lagrange equation for the Euclidean action that is given as
\begin{equation}
    \partial_\tau^2 \psi_\mathrm c(x, \tau) + v_\mathrm p^2 \partial_x^2
    \psi_\mathrm c(x, \tau) = 0
\end{equation}
where $v_\mathrm{p}^2 = 1 / (\Ll \Cl)$.
Since we integrate only over the periodic in imaginary time trajectories with the fixed value of the flux at the junction, we impose the following boundary conditions:
\begin{gather}\label{eq:48}
    \psi_\mathrm c(-l, \tau) = 0, \textrm{ for } \tau\in[0,\hbar\beta],\\\label{eq:49}
    \psi_\mathrm c(x_\mathrm J, \tau) = \psi'(x_\mathrm J^-, \tau), \textrm{ for } \tau\in[0,\hbar\beta], \\\label{eq:50}
    \psi_\mathrm c(x, 0) = \psi_\mathrm c(x, \hbar \beta), \textrm{ for } x\in[-l,x_\mathrm J].
\end{gather}
We expand the flux at the junction into a Fourier series as
\begin{equation}\label{eq:51}
    \psi'(x_\mathrm J^-, \tau) = \frac{1}{\hbar \beta}
    \sum\limits_{n=-\infty}^{\infty} \psi'_n(x_\mathrm J^-) e^{-\textrm{i} \omega_n
    \tau},
\end{equation}
where $\omega_n = 2\pi n k_\mathrm{B} T / \hbar$ are the bosonic Matsubara frequencies.
Consequently, the classical trajectory is given by the following expression:
\begin{equation}
    \psi_\mathrm c(x, \tau) = \frac 1{\hbar \beta} \sum\limits_{n=-\infty}^\infty
    \psi'_n(x_\mathrm J^-) \frac{\sinh\left[\frac{\omega_n}{v_\mathrm p} (x +
    l)\right]}{\sinh\left[\frac{\omega_n}{v_\mathrm p}(x_\mathrm J + l)\right]} e^{-i\omega_n
    \tau} .
\end{equation}
We make a substitution $\psi'(x, \tau) = \psi_\mathrm c(x, \tau) + \tilde
\psi(x, \tau)$ in the path integral expression for the partition function. Thus the action for the Euclidean left half of the resonator reads as
\begin{multline}
    S^\mathrm E_\mathrm{l} \left[\tilde \psi(x, \tau), \psi'(x_\mathrm J^-, \tau)\right] = \\ \int\limits_0^{\hbar\beta}
    \int\limits_{-l}^{x_\mathrm J} \left\{\frac{\Cl [\partial_\tau
    \psi_\mathrm c(x, \tau)]^2}{2} + \frac{[\partial_x \psi_\mathrm c(x, \tau)]^2}{2\Ll} + \frac{\Cl [\partial_\tau
    \tilde \psi(x, \tau)]^2}{2} + \frac{[\partial_x \tilde
    \psi(x, \tau)]^2}{2\Ll}\right\}\;\mathrm dx\;\mathrm d\tau = \\
    \int\limits_{-l}^{x_\mathrm J} \left\{\frac{\Cl [\partial_\tau
    \tilde \psi(x, \tau)]^2}{2} + \frac{[\partial_x \tilde
    \psi(x, \tau)]^2}{2\Ll}\right\}\;\mathrm dx\;\mathrm d\tau +
    \frac{1}{\hbar \beta}\sum\limits_{n=-\infty}^\infty |\psi'_n(x_\mathrm
    J^-)|^2 \frac{\omega_n}{2 Z \tanh\left(\frac{\omega_n}{v_p} l_\mathrm l\right)} ,
\end{multline}
where $Z = \sqrt{\Ll / \Cl}$ and $l_\mathrm{l} = l + x_\mathrm J$. Repeating the above
procedure for the right half of the resonator, we obtain the following
expression for Euclidean action of the whole unimon with~$\psi'(x_\mathrm J^-, \tau)$ and~$\psi'(x_\mathrm J^+,\tau)$ given:
\begin{multline}
    S^\mathrm{E}\left[\tilde \psi(x, \tau), \psi'(x_\mathrm J^-, \tau), \psi'(x_\mathrm J^+, \tau)\right] = \int\limits_0^{\hbar \beta} \left(\rule{0cm}{1.2cm}\right.
        \int\limits_{-l}^{x_\mathrm J} \left\{\frac{\Cl \left[\partial_\tau \tilde \psi(x, \tau)\right]^2}{2} +
        \frac{\left[\partial_x \tilde \psi(x, \tau)\right]^2}{2 \Ll}\right\}\;\mathrm dx
         \\ +
        \int\limits_{x_\mathrm J}^{l} \left\{\frac{\Cl \left[\partial_\tau \tilde \psi(x, \tau)\right]^2}{2} +
        \frac{\left[\partial_x \tilde \psi(x, \tau)\right]^2}{2 \Ll}\right\}\;\mathrm dx
         +\frac{C_\mathrm J \left[\partial_\tau \psi'(x_\mathrm J^-, \tau) -
        \partial_\tau \psi'
        (x_\mathrm J^+, \tau)\right]^2}{2}   \\ -  E_\mathrm J \cos\left\{
            \frac{2\pi}{\Phi_0} \left[\psi'(x_\mathrm J^-, \tau) -
            \psi'(x_\mathrm J^+, \tau) + \Phi_\mathrm{diff}\right]
            \right\}
        \left.\rule{0cm}{1.2cm}\right)\;\mathrm d\tau  + \frac{1}{\hbar \beta}
    \sum\limits_{n=-\infty}^{+\infty}\frac{\omega_n}{2Z}\left[\frac{\left|\psi'_n(x_\mathrm
    j^-)\right|^2}{\tanh\left(\frac{\omega_n}{v_\mathrm p} l_\mathrm {l}\right)} + \frac{\left|\psi'_n(x_\mathrm
    j^+)\right|^2}{\tanh\left(\frac{\omega_n}{v_\mathrm p} l_\mathrm {r}\right)}\right] ,
\end{multline}
where $l_\mathrm{r} = l - x_\mathrm J$.
The partition function is expressed through this action as
\begin{equation}
    Z(\beta) = \int \bm D\left[\tilde \psi(x, \tau), \psi'(x_\mathrm J^-, \tau),
    \psi'(x_\mathrm J^+, \tau)\right] e^{-\frac{1}{\hbar} S^\mathrm E\left[\tilde \psi(x, \tau), \psi'(x_\mathrm J^-, \tau),
    \psi'(x_\mathrm J^+, \tau)\right]}.
\end{equation}
The field $\tilde \psi(x, \tau)$ is uncoupled from $\psi'(x_\mathrm J^\pm, \tau)$ since the boundary conditions of the total flux $\psi'(x, \tau) = \psi_\mathrm c(x, \tau) + \tilde\psi(x, \tau)$ are obeyed by the classical part according to equations~\eqref{eq:48}--\eqref{eq:50} and consequently the field $\tilde \psi(x, \tau)$ has zero boundary conditions independent of $\psi'(x_\mathrm J^\pm, \tau)$. Thus the field $\tilde \psi(x, \tau)$ can be integrated out. Since it satisfies zero boundary conditions at $x = -l$, $x =
x_\mathrm J$, and $x = l$, it is equal to a product of
partition functions of two grounded $\lambda/2$ resonators with lengths
$l_\mathrm{l}$ and $l_\mathrm{r}$. Consequently, we have
\begin{equation}
    Z(\beta) = Z'(\beta) \int \bm D[\psi'(x_\mathrm J^-, \tau),
    \psi'(x_\mathrm J^+, \tau)] e^{-\frac{1}{\hbar} S^\mathrm E_\mathrm J[\psi'(x_\mathrm J^-, \tau),
    \psi'(x_\mathrm J^+, \tau)]} ,
\end{equation}
where
\begin{equation}
    Z'(\beta) = \prod\limits_{m=1}^{\infty} \frac{1}{\left(1 - e^{-\hbar
    \beta \frac{mv_\mathrm p}{l_\mathrm{l}}}\right)\left(1 - e^{-\hbar
    \beta \frac{mv_\mathrm p}{l_\mathrm{r}}}\right)},
\end{equation}
\begin{multline}
    S^\mathrm{E}_\mathrm J[\psi'(x_\mathrm J^-, \tau), \psi'(x_\mathrm J^+, \tau)] = \int\limits_0^{\hbar \beta} \left(
        \frac{C_\mathrm J \left[\partial_\tau \psi'(x_\mathrm J^-, \tau) -
        \partial_\tau \psi'
        (x_\mathrm J^+, \tau)\right]^2}{2} \right . \\ \left .-  E_\mathrm J \cos\left\{
            \frac{2\pi}{\Phi_0} \left[\psi'(x_\mathrm J^-,\tau) -
            \psi'(x_\mathrm J^+,\tau) + \Phi_\mathrm{diff}\right]
            \right\}
        \right)\;\mathrm d\tau +  \frac{1}{\hbar \beta}
    \sum\limits_{n=-\infty}^{+\infty}\left[\left|\psi'_n(x_\mathrm
    J^-)\right|^2 K_{\mathrm ln}+\left|\psi'_n(x_\mathrm
    J^+)\right|^2 K_{\mathrm rn}\right] ,
\end{multline}
and
\begin{equation}
    K_{\alpha n} = \frac{\omega_n}{2Z\tanh\left(\frac{\omega_n}{v_\mathrm p} l_{\mathrm \alpha}\right)},\textrm{ for }\alpha\in\{\textrm{r}, \textrm{l}\}.
\end{equation}

Here, we introduce $$\psi_+(\tau) = \frac{1}{2}\left[\psi'(x_\mathrm J^-, \tau) +
\psi'(x_\mathrm J^+,\tau)\right],$$ $$\psi_-(\tau) = \psi'(x_\mathrm J^-,\tau) -
\psi'(x_\mathrm J^+,\tau),$$ and $$K_{\pm,n} = K_{\mathrm l n} \pm K_{\mathrm r n},$$ and express the corresponding Euclidean action in terms of these variables as 
\begin{multline}
    S^\mathrm{E}_\mathrm J[\psi_+(\tau), \psi_-(\tau)] =
    \int\limits_0^{\hbar \beta} \left(
        \frac{C_\mathrm J \left[\partial_\tau \psi_-(\tau)\right]^2}{2} -  E_\mathrm J \cos\left\{
            \frac{2\pi}{\Phi_0} \left[\psi_-(\tau) + \Phi_\mathrm{diff}\right]
            \right\}
        \right)\;\mathrm d\tau + \\ \frac{1}{\hbar \beta}
    \sum\limits_{n=-\infty}^{+\infty}\left[\left|\psi_{+,n} +
    \frac{1}{2} K_{+,n}^{-1} K_{-,n} \psi_{-,n}\right|^2 K_{+,n}
    +\frac{1}{4}\left(
    K_{+,n} - K_{-,n}^2 K_{+,n}^{-1}
        \right)|\psi_{-,n}|^2\right],
\end{multline}
where $\psi_{\pm,n}$ are the Fourier components of $\psi_\pm(\tau)$ in the spirit of equation~\eqref{eq:51}.
We proceed by introducing $\psi_{+,n}' = \psi_{+,n} +
\frac{1}{2}K_{+,n}^{-1} K_{-,n} \psi_{-,n}$ and integrate out the $\psi_+'$ 
degree of freedom to obtain   
\begin{equation}
    Z(\beta) = Z'(\beta) Z_+(\beta) \int \bm D[\psi_-(\tau)]
    e^{-\frac{1}{\hbar}S^\mathrm E_-[\psi_-(\tau)]} ,
\end{equation}
where
\begin{equation}
    Z_+(\beta) = \lim_{M\to\infty}\prod\limits_{n=-M}^{M} \frac{\mathcal N_M}{\sqrt{K_{+,n}}}, 
\end{equation}
where $\mathcal N_M$ are normalization constants needed to enforce the convergence of the product. Thus, we obtain 
\begin{equation}\label{eq:66}
    S^\mathrm{E}_-[\psi_-(\tau)] = 
    \int\limits_0^{\hbar \beta} \left\{
        \frac{C_\mathrm J \left[\partial_\tau \psi_-(\tau) \right]^2}{2} -  E_\mathrm J \cos\left[
            \frac{2\pi}{\Phi_0} \left(\psi_-(\tau) + \Phi_\mathrm{diff}\right)
            \right]
        \right\}\;\mathrm d\tau + \\ \frac{1}{\hbar \beta}
    \sum\limits_{n=-\infty}^{+\infty}    \frac{1}{4}K_n|\psi_{-,n}|^2,
\end{equation}
where
\begin{equation}
    K_n = K_{+,n} - K_{-,n}^2 K_{+,n}^{-1} = \frac{2 \omega_n}{Z \left[\tanh\left(\frac{\omega_n}{v_\mathrm p} l_\mathrm l\right) + \tanh\left(\frac{\omega_n}{v_\mathrm p} l_\mathrm
    r\right)\right]}.
    \label{eq:S64}
\end{equation}

Until this point for model 2, we have not applied any approximations, i.e., the expression for the action $S_-^\mathrm E[\psi_-(\tau)]$ is exact. However, this action is challenging to utilize further in analytical calculations since it is both non-Gaussian and non-local in imaginary time. The corresponding low-temperature partition function is determined by the low-Matsubara-frequency contribution of the non-local kernel $K_n$. To proceed, we expand this kernel in the vicinity of $\omega_n = 0$ as
\begin{equation}
    K_n \approx \frac{1}{l\Ll} + \frac{\Cl(l^2 + 3 x_\mathrm J^2)}{3l} \omega_n^2. 
    \label{eq:S65}
\end{equation}
These terms correspond to an effective capacitance and inductance, induced by the
resonator. Consequently, we write a non-Gaussian action that is local in imaginary time as
\begin{equation}
    S^\mathrm{E}_0[\psi_-(\tau)] = \int\limits_0^{\hbar \beta} \left(
        \frac{C_\mathrm{eff} \left[\partial_\tau \psi_-(\tau)\right]^2}{2} +
        \frac{\psi_-^2(\tau)}{2 L_\mathrm{eff}} -  E_\mathrm J \cos\left\{
            \frac{2\pi}{\Phi_0} \left[\psi_-(\tau) + \Phi_\mathrm{diff}\right]
            \right\}
        \right)\;\mathrm d\tau,
\end{equation}
where
\begin{gather}
    C_\mathrm{eff} = C_\mathrm J + \frac{\Cl (l^2 + 3 x_\mathrm J^2)}{6 l}, \\  
    L_\mathrm{eff} = {2l\Ll}.
\end{gather}
Partition function, corresponding to this action, can be evaluated
numerically by diagonalization of the following Hamiltonian:
\begin{equation}
    \hat H_0 = \frac{\hat Q_-^2}{2C_\mathrm{eff}} + \frac{\hat\psi_-^2}{2 L_\mathrm{eff}} - E_\mathrm J\cos\left[
            \frac{2\pi}{\Phi_0} \left(\hat\psi_- + \Phi_\mathrm{diff}\right)
            \right],
%    \hat H^{(0)}(\psi_-) = -\frac{\hbar^2}{2 C_\mathrm{eff}} \frac{\mathrm d^2}{\mathrm d\psi_-^2}  + \frac{\psi_-^2}{2 L_\mathrm{eff}} - E_\mathrm J\cos\left[\frac{2\pi}{\Phi_0} \left(\psi_- + \Phi_\mathrm{diff}\right)\right].
\end{equation}
where $[\hat\psi_-,\hat Q_-]=\textrm{i}\hbar$.
This Hamiltonian has indentical form to that of the lumped-element unimon and can be used to obtain a qualitative spectrum of unimon. However it is too inaccurate for any quantitative analysis so we need to develop a better approximation. 

To proceed, we rewrite the non-local kernel~\eqref{eq:S64} as follows:
\begin{equation}
    K_n(\omega_n) =
    \frac{\textrm{i} \omega_n \left[ \cos \left(\frac{2\textrm{i} \omega_n}{v_\mathrm p} x_\mathrm J\right) + \cos\left(
    \frac{2\textrm{i} \omega_n}{v_\mathrm p} l\right)\right]}{Z \sin \left(\frac{2\textrm{i} \omega_n}{v_\mathrm p} l\right)}. %=
\end{equation}
Importantly, $K_n(\omega_n)$ has poles at frequencies of the $\lambda/2$ resonator modes
$\omega_n = \pm \textrm{i} \pi k v_\mathrm p / (2 l)$, $k \in \mathbb{Z}_+$. % 1, 2, \ldots$. 
The residues of these poles are equal to $\pm \textrm{i} \pi k v_\mathrm p^2 / (4 l^2 Z) \left[(-1)^k \cos \left(\frac{\pi k x_\mathrm J}{l}\right) + 1\right]$. 
We approximate the resonator with a set of $M$ auxiliary modes with frequencies $\Omega_k = \pi k v_\mathrm{p} / (2 l)$, inductively coupled to the non-linear oscillator described by $\psi_-$ degree of freedom. Thus we consider a trial action
of the form
\begin{multline}
    S^\mathrm E_\mathrm{trial} = \int \limits_0^{\hbar \beta} \left(
        \frac{C[\partial_\tau \psi_-(\tau)]^2}{2} +
        \frac{\psi_-^2(\tau)}{2L_\psi} - E_\mathrm J \cos
        \left\{\frac{2\pi}{\Phi_0}\left[\psi_-(\tau) + \Phi_\mathrm{diff}\right]\right\} \right . \\ \left . +
        \sum\limits_{k=1}^M \left\{
            \frac{C[\partial_t \chi_k (\tau)]^2}{2} + \frac{C \Omega_k^2 \chi_k^2(\tau)}{2}
            + \alpha_k{\chi_k(\tau) \psi_-(\tau)}
            \right\}
        \right)\;\mathrm d\tau
     = \\
    \int\limits_0^{\hbar \beta}\left(
        \frac{C [\partial_\tau \psi_-(\tau)]^2}{2} + \frac{\psi_-^2(\tau)}{2
        L_\psi}- E_\mathrm J
        \cos \left\{\frac{2\pi}{\Phi_0} \left[\psi_-(\tau) + \Phi_\mathrm{diff}\right]\right\}
        \right)\;\mathrm d\tau + \\ \frac{1}{\hbar \beta}
        \sum\limits_{k=1}^M
    \sum\limits_{n=-\infty}^\infty \left [
        \frac{C}{2}\left(\omega_n^2 + \Omega_k^2\right)
        \left|\chi_{k,n} + \frac{\alpha_k \psi_{-,n}}{C(\omega_n^2 + \Omega_k^2)
        }\right|^2 - \frac{\alpha_k^2|\psi_{-,n}|^2}{2 C (\omega_n^2 +
        \Omega_k^2)}\right],
\end{multline}
where $\{\chi_k\}_{k=1}^M$ are the flux variables of the auxiliary modes, $\chi_{k,n}$ is the $n$th Fourier coefficient of $\chi_k(\tau)$, and $C$, $L_\psi$, and $\alpha_k$ are coefficients used below to match the trial action with that in equation~\eqref{eq:66}. To this end, we define $\chi_{k,n}' = \chi_{k,n} + \alpha_k / [(C (\omega_n^2 + \Omega_k^2)] \psi_{-,n}$ and integrate out variables $\chi_{k,n}'$.
We obtain an effective action for the variable $\psi_-$ as
\begin{multline}
    S^\mathrm E_M = \int \limits_0^{\hbar \beta} \left(
        \frac{C[\partial_\tau \psi_-(\tau)]^2}{2} +
        \frac{\psi_-^2(\tau)}{2L_\psi} - E_\mathrm J \cos
        \left\{\frac{2\pi}{\Phi_0}\left[\psi_-(\tau) + \Phi_\mathrm{diff}\right]\right\}         \right)\;\mathrm d\tau \\
        -\frac{1}{\hbar\beta} \sum\limits_{n=-\infty}^{+\infty}
        \sum\limits_{k=1}^M \frac{\alpha_k^2|\psi_{-,n}|^2}{2 C (\omega_n^2 +
        \Omega_k^2)}.
\end{multline}
The kernel of non-local part of the action also has poles in the frequency domain at $\omega_n = \pm \textrm{i} \Omega_k$. The residues of the kernel at these poles are equal to $\pm \textrm{i} \alpha_k^2 / (4 C \Omega_k)$. Thus it is natural to equate the residues of the exact kernel $K_n$ to the residues of this trial kernel as
\begin{equation}
    \frac{\alpha_k^2}{4 C \Omega_k} = \frac{\Omega_k^2}{4 \pi k Z} \left[(-1)^k
    \cos \left(\frac{\pi k x_\mathrm J}{l}\right) + 1\right].
\end{equation}
We have two more free parameters, namely, $C$ and $L_\psi$ which we choose for the second-order Taylor expansion of the trial kernel to match with that of the exact kernel given in equation~\eqref{eq:S65}. Consequently, we have
\begin{gather}
    \frac{1}{2 L_\psi} - \sum\limits_{k=1}^M \frac{\alpha_k^2}{2 C
    \Omega_k^2} = \frac{1}{4 l \Ll}, \\
    \frac{C}{2} + \sum\limits_{k=1}^M \frac{\alpha_k^2}{2 C \Omega_k^4} =
    \frac{C_\mathrm J}{2} + \frac{\Cl (l^2 + 3 x_\mathrm J^2)}{12 l}.
\end{gather}
Finally, the spectrum of the unimon is given by the Hamiltonian
\begin{equation}
    \hat H_M = \frac{\hat Q_-^2}{2C} + \frac{\hat\psi_-^2}{2 L_\psi} - E_\mathrm J \cos\left[\frac{2\pi}{\Phi_0}\left(\hat\psi_- + \Phi_\mathrm{diff}\right)\right] + \sum\limits_{k=1}^M\left( \frac{\hat q_k^2}{2C} + \frac{C \Omega_k^2 \hat\chi_k^2}{2}+ \alpha_k \hat\chi_k \hat\psi_-\right),
\end{equation}
where $[\hat \chi_k,\hat q_m]=\textrm{i}\hbar\delta_{km}$, $[\hat\psi_-,\hat Q_-]=\textrm{i}\hbar$, and all other single-operator commutators are zero.

In model~2 of this work, we choose $M=2$, i.e., we consider two lowest auxiliary modes. Thus in general, we need to solve a 3D Schr\"odinger equation which is feasible with the modern computational techniques.
Note that if the unimon is symmetric ($x_\mathrm J = 0$), the first mode is decoupled from the junction, i.e., $\alpha_1 = 0$. However, the second mode is strongly coupled to the junction, and hence it cannot be neglected or treated perturbatively. In this case, the problem is reduced to solving a 2D Schr\"odinger equation which is computationally convenient.

\pagebreak
\section*{Supplementary Methods II: Derivation of the theoretical model for a coupled unimon-resonator system}

\begin{figure}[t!]
	\centering	
	\includegraphics[width = 0.95\textwidth]{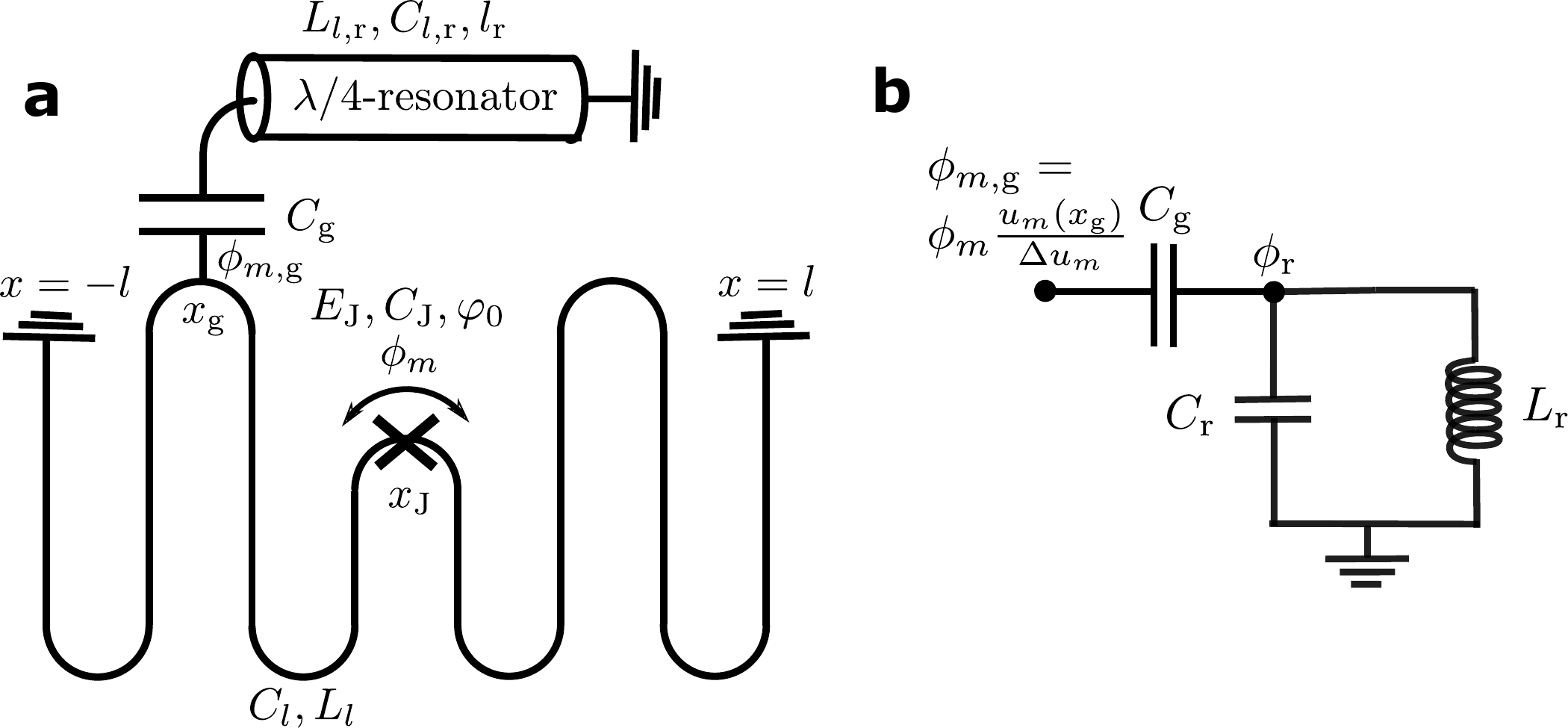}
    \\
    \caption{ \label{fig: unimon-resonator} \textbf{Schematic diagram of a coupled unimon-resonator system.} \textbf{a}, Schematic illustration of a coupled system consisting of a unimon qubit and $\lambda/4$ readout resonator together with symbols corresponding to the relevant circuit parameters. Here, $2l$ is the total length of the unimon circuit, $x_\mathrm{J}$ is the junction location, $\EJ$ is the Josephson energy, $C_\mathrm{J}$ is the capacitance of the junction, $\varphi_0$ is the dc Josephson phase, $\phi_m$ is the flux difference across the junction owing to the qubit mode $m$, $x_\mathrm{g}$ is the location of the coupling capacitance, $C_\mathrm{g}$ is the coupling capacitance between the unimon and its readout resonator, $\phi_{m,\mathrm{g}}$ is the flux of the unimon center conductor at the coupling location, $L_{l, \mathrm{r}}$ is the inductance per unit length of the readout resonator, $C_{l, \mathrm{r}}$ is the capacitance per unit length of the readout resonator, and $l_\mathrm{r}$ is the length of the readout resonator.   
    \textbf{b}, Circuit diagram corresponding to the coupling capacitance and the lumped-element approximation of the readout resonator in a coupled unimon-resonator system. Here, $\phi_\mathrm{r}$ is the flux at the end of the readout resonator, and $C_\mathrm{r}$ and $L_\mathrm{r}$ are the effective capacitance and inductance of the readout resonator. }
\end{figure}

\noindent \textbf{Hamiltonian for the coupled unimon-resonator system}\\
%%Eric needs to read Thu 24.3.
Here, we derive the Hamiltonian for a coupled system consisting of a unimon qubit and a $\lambda/4$  readout resonator using an approach extended from the theoretical model 1 presented in Supplementary Methods I. In the derivation, we take into account the qubit mode of the unimon and the lowest-frequency mode of the readout resonator. See Fig.~\ref{fig: unimon-resonator} for a schematic illustration of the coupled system and a lumped-element circuit model of the coupling. 
 
 We begin by writing the Lagrangians of the different parts of the coupled system. Referring to equation~\eqref{eq: unimon single mode Lagrangian}, we write the classical Lagrangian owing to the bare qubit mode $m$  of the unimon as
\begin{equation}
\Lag_\mathrm{u} = \frac{C_m'}{2} \dot{\phi}_m^2 - \frac{\phi_m^2}{2 \tilde{L}_m'} - \frac{1}{2l\Ll}\phi_m(\Phi_\mathrm{diff} - \phi_0) + \EJ \cos[2\pi (\phi_m - \phi_0)/\Phi_0],
\end{equation}
where as above $C_m'$ is the effective capacitance of the qubit mode $m$, $\phi_m$ is the flux across the Josephson junction owing to the $m$th mode, $ \tilde{L}_m'$ is the effective inductance of the qubit mode, $2l$ is the length of the unimon, $\EJ$ is the Josephson energy, $\phi_0$ is the flux across the Josephson junction owing to the dc component, $\Ll$ is the inductance per unit length of the unimon, and  $\Phi_\mathrm{diff}$ is the half of the external flux difference. %Below in Sec. \ref{sec: recipe for dispersive shift}, we will give an explicit recipe for computing the dispersive shift starting from the device parameters of the unimon and there, we will explicitly define the effective inductances and capacitances in a (hopefully) understandable way. 
In the following, we assume that the coupling to the readout resonator does not affect the mode envelope functions of the unimon. %However, we observe that the coupling capacitance modifies the total capacitance of the unimon. 

To write the Lagrangian related to the capacitive coupling, we note that the flux at the coupling location $x_\mathrm{g}$ is related to the flux across the Josephson junction as 
\begin{equation}
    \phi_{m, \mathrm{g}} = \phi_m u_m(x_\mathrm{g})/(\Delta u_m), 
\end{equation}
where $u_m(x_\mathrm{g})$ is the envelope function of the $m$th mode evaluated at the coupling location $x_\mathrm{g}$ and $\Delta u_m$ is the discontinuity in the envelope function $\Delta u_m = u_m(x_\mathrm{J}^+) - u_m(x_\mathrm{J}^-)$ across the junction. Thus, the Lagrangian corresponding to the capacitive coupling between the unimon and the resonator (see Fig.~\ref{fig: unimon-resonator}) can be written as
\begin{equation}
\Lag_\mathrm{g} = \frac{1}{2} C_\mathrm{g} \left(\dot{\phi}_m \frac{u_m(x_\mathrm{g})}{\Delta u_m} - \dot{\phi}_\mathrm{r} \right)^2,
\end{equation}
where $\phi_\mathrm{r}$ is the node flux of the readout resonator at the coupling location. Next, the Lagrangian of the resonator is expressed as
\begin{equation}
\Lag_\mathrm{r} = \frac{1}{2} C_\mathrm{r} \dot{\phi}_\mathrm{r}^2 - \frac{\phi_\mathrm{r}^2}{2L_\mathrm{r}},
\end{equation}
where $C_\mathrm{r}$ and $L_\mathrm{r}$ are the effective capacitance and inductance of the  readout resonator. For a  $\lambda/4$  readout resonator, we have $C_\mathrm{r} = C_{l, \mathrm{r}}l_\mathrm{r}/2$, and $L_\mathrm{r} = 8L_{l,\mathrm{r}}l_\mathrm{r}/\pi^2$, where $C_{l,\mathrm{r}}$ and $L_{l,\mathrm{r}}$ are the capacitance and inductance per unit length of the readout resonator and $l_\mathrm{r}$ is the length of the readout resonator. Note also that the characteristic impedance of the lumped-element resonator mode is $Z_\mathrm{r} = \sqrt{L_\mathrm{r}/C_\mathrm{r}} = 4Z_\mathrm{tr}/\pi$, where $Z_\mathrm{tr} = \sqrt{L_{l,\mathrm{r}}/C_{l,\mathrm{r}}}$.  Thus, the total Lagrangian of the qubit-resonator system is given by 
\begin{equation}
\Lag = \Lag_\mathrm{u} + \Lag_\mathrm{g} + \Lag_\mathrm{r}.    
\end{equation}

To derive the Hamiltonian of the circuit, we first write the kinetic part of the total Lagrangian in a matrix form as
\begin{equation}
T = \frac{1}{2} \begin{pmatrix} \dot{\phi}_m &  \dot{\phi}_\mathrm{r} \end{pmatrix} \begin{pmatrix}  C_m' + C_\mathrm{g} \frac{u_m(x_\mathrm{g})^2}{(\Delta u_m)^2} & -C_\mathrm{g} \frac{u_m(x_\mathrm{g})}{\Delta u_m} \\ -C_\mathrm{g} \frac{u_m(x_\mathrm{g})}{\Delta u_m} & C_\mathrm{r} + C_\mathrm{g} \end{pmatrix}\begin{pmatrix} \dot{\phi}_m \\  \dot{\phi}_\mathrm{r} \end{pmatrix} = \frac{1}{2} \dot{\bm{\phi}}^T \bm{C}  \dot{\bm{\phi}},
\end{equation}
  where ${\bm{\phi}} = ({\phi}_m,  {\phi}_\mathrm{r})^T$ is a node flux vector and $\bm{C}$ is the corresponding capacitance matrix. With the help of the matrix notation, we can express the conjugate charges in terms of the node fluxes as
\begin{equation}
\bm{q} = \frac{\partial \Lag}{\partial  \dot{\bm{\phi}}} = \bm{C} \dot{\bm{\phi}},
\end{equation}
where $\bm{q} =  ( q_m, q_\mathrm{r})^T$ 
is a vector containing the conjugate charges. Using the matrix notation, the classical Hamiltonian can  be written as
\begin{equation}
\Ham= \dot{\bm{\phi}}^T \frac{\partial \Lag}{\partial  \dot{\bm{\phi}}} - \Lag = \frac{1}{2} \bm{q}^T \bm{C}^{-1} \bm{q} + U(\bm{\phi}). \label{eq: Hamiltonian formula}
\end{equation}
We can approximate the matrix elements of $\bm{C}^{-1}$ as
\begin{align}
\bm{C}^{-1}(1, 1) &= \frac{C_\mathrm{g} + C_\mathrm{r}}{\left[C_m' + C_\mathrm{g} \frac{u_m(x_\mathrm{g})^2}{(\Delta u_m)^2}\right]( C_\mathrm{r} + C_\mathrm{g}) -C_\mathrm{g}^2 \frac{u_m(x_\mathrm{g})^2}{(\Delta u_m)^2} } \approx \frac{1}{C_m' + C_\mathrm{g} \frac{u_m(x_\mathrm{g})^2}{(\Delta u_m)^2}}, \\
\bm{C}^{-1}(1, 2) &= \frac{C_\mathrm{g} \frac{u_m(x_\mathrm{g})}{\Delta u_m} }{\left[C_m' + C_\mathrm{g} \frac{u_m(x_\mathrm{g})^2}{(\Delta u_m)^2}\right]( C_\mathrm{r} + C_\mathrm{g}) -C_\mathrm{g}^2 \frac{u_m(x_\mathrm{g})^2}{(\Delta u_m)^2}} \approx  \frac{C_\mathrm{g} \frac{u_m(x_\mathrm{g})}{\Delta u_m} }{\left[C_m' + C_\mathrm{g} \frac{u_m(x_\mathrm{g})^2}{(\Delta u_m)^2}\right]( C_\mathrm{r} + C_\mathrm{g}) }, \\
\bm{C}^{-1}(2, 2) &= \frac{ C_m' + C_\mathrm{g} \frac{u_m(x_\mathrm{g})^2}{(\Delta u_m)^2}}{\left[C_m' + C_\mathrm{g} \frac{u_m(x_\mathrm{g})^2}{(\Delta u_m)^2}\right]( C_\mathrm{r} + C_\mathrm{g}) -C_\mathrm{g}^2 \frac{u_m(x_\mathrm{g})^2}{(\Delta u_m)^2}} \approx \frac{1}{C_\mathrm{r} + C_\mathrm{g}},
\end{align}
where we have used the valid assumption that the coupling capacitance $C_\mathrm{g}$ is much smaller than the effective unimon capacitance $C_m'$ and the resonator capacitance $C_\mathrm{r}$. Inserting these approximations to the Hamiltonian in equation~\eqref{eq: Hamiltonian formula}, we obtain 
\begin{align}
\Ham &= \bigg \{ \frac{q_m^2 (\Delta u_m)^2}{2 C_{\mathrm{u}, \mathrm{tot}}} + \frac{\phi_m^2}{2 \tilde{L}_m'} +  \frac{1}{2l\Ll}\phi_m(\Phi_\mathrm{diff} - \phi_0) - \EJ \cos[2\pi (\phi_m - \phi_0)/\Phi_0] \bigg \} \nonumber \\
         &+ \bigg \{ \frac{q_m q_\mathrm{r} C_\mathrm{g} \frac{u_m(x_\mathrm{g})}{\Delta u_m}}{\frac{C_{\mathrm{u}, \mathrm{tot}}}{(\Delta u_m)^2} C_{\mathrm{r}, \mathrm{tot}}} \bigg \} +  \bigg \{ \frac{q_\mathrm{r}^2}{2 C_{\mathrm{r}, \mathrm{tot}} }+ \frac{\phi_\mathrm{r}^2}{2L_\mathrm{r}} \bigg \},
\end{align}
where we have defined the total capacitance of the unimon as $C_{\mathrm{u}, \mathrm{tot}} = C_m'(\Delta u_m)^2  + C_\mathrm{g} u_m(x_\mathrm{g})^2 = 2 \Cl l + C_\mathrm{J} +   C_\mathrm{g} u_m(x_\mathrm{g})^2 $ and the total capacitance of the readout resonator as $C_{\mathrm{r}, \mathrm{tot}} = C_\mathrm{r} + C_\mathrm{g}$. In the above equation, the first curly brackets correspond to the unimon, the second curly brackets correspond to the coupling, and the third curly brackets correspond to the readout resonator.  Finally, we quantize the above Hamiltonian to obtain 
\begin{align}
\hat{\Ham} &= \bigg \{ 4 \ECm^\mathrm{tot} \hat{n}_m^2 + \frac{1}{2} E_{L, m} \hat{\varphi}_m^2 + E_L \hat{\varphi}_m(\varphi_\mathrm{diff} - \varphi_0)  - \EJ \cos(\hat{\varphi}_m - \varphi_0) \bigg \} \nonumber  \\
&+\bigg \{  \frac{4e^2  C_\mathrm{g} u_m(x_\mathrm{g})\Delta u_m}{C_{u,\mathrm{tot}} C_{\mathrm{r}, \mathrm{tot}}} \hat{n}_m \hat{n}_\mathrm{r} \bigg \} + \bigg \{ 4E_{C, \mathrm{r}}^\mathrm{tot} \hat{n}_\mathrm{r}^2 + \frac{1}{2} E_{L, \mathrm{r}} \hat{\varphi}_\mathrm{r}^2 \bigg \} , \label{eq: coupled hamiltonian charge and phase}
\end{align}
where we have defined the charge operators as $\hat{n}_i = \hat{q}_i/(2e)$ and the phase operators as $\hat{\varphi}_i = 2\pi \hat{\phi}_i/\Phi_0$ for $i \in \{m, \textrm{r}\}$. The charge and phase operators satisfy the usual commutation relation $[\hat{\varphi}_i, \hat{n}_j] = \mathrm{i} \delta_{ij}$. Furthermore, we have defined the following energy scales for the $m$th mode of the unimon $\ECm^\mathrm{tot} = e^2 \Delta u_m^2/(2 C_{\mathrm{u}, \mathrm{tot}})$, $E_{L, m} = \Phi_0^2/(2\pi)^2/\tilde{L}_m'$, and $E_L =  \Phi_0^2/(2\pi)^2/(2l\Ll)$. For the readout resonator, we have defined the energy scales $E_{C, \mathrm{r}}^\mathrm{tot} = e^2/(2 C_{\mathrm{r}, \mathrm{tot}})$, and $E_{L, \mathrm{r}} = \Phi_0^2/(2\pi)^2/L_\mathrm{r}$. 

It is instructive to express the Hamiltonian in equation~\eqref{eq: coupled hamiltonian charge and phase} using the eigenbasis of the unimon, and the annihilation and creation operators of the resonator. The unimon part of the Hamiltonian can be diagonalized, e.g., in the phase basis to obtain the eigenenergies $\{ \hbar \omega_j \}$ and eigenstates $\{ |j \rangle \}$ of the bare unimon. When it comes to the resonator, the annihilation and creation operators are related to the charge and phase operators $\hat{n}_\mathrm{r}$ and $\hat{\varphi}_\mathrm{r}$ as
\begin{gather}
 \hat{n}_\mathrm{r} = -\frac{\mathrm{i}}{2} \left(\frac{E_{L,\mathrm{r}}}{2 E_{C, \mathrm{r}}^\mathrm{tot}} \right)^{1/4} (\ah_\mathrm{r} - \ad_\mathrm{r}), \label{eq: nr}\\
\hat{\varphi}_\mathrm{r} = \left(\frac{2 E_{C, \mathrm{r}}^\mathrm{tot}}{E_{L,\mathrm{r}} } \right)^{1/4} (\ah_\mathrm{r} + \ad_\mathrm{r}), \label{eq: phir}
\end{gather} 
which allows us to express the resonator Hamiltonian as
\begin{equation}
\hat{\Ham}_\mathrm{r} = \hbar \omega_\mathrm{r} \ad_\mathrm{r} \ah_\mathrm{r}, \label{eq: resonator hamiltonian}
\end{equation}
where $\omega_\mathrm{r} = \sqrt{8 E_{C, \mathrm{r}}^\mathrm{tot}E_{L,\mathrm{r}} }/\hbar$.  

Using equations~\eqref{eq: coupled hamiltonian charge and phase}--\eqref{eq: resonator hamiltonian}, the Hamiltonian of the coupled unimon-resonator system can be written as
\begin{equation}
\hat{\Ham} = \hbar \omega_\mathrm{r} \ad_\mathrm{r} \ah_\mathrm{r} + \sum_j \hbar \omega_j |j\rangle \langle j| - 2\mathrm{i} e^2\frac{C_\mathrm{g} u_m(x_\mathrm{g})\Delta u_m}{C_{\mathrm{u}, \mathrm{tot}} C_{\mathrm{r}, \mathrm{tot}}} \left ( \frac{E_{L,\mathrm{r}}}{2 E_{C, \mathrm{r}}^\mathrm{tot}} \right )^{1/4} \hat{n}_m (\ah_\mathrm{r} - \ad_\mathrm{r}).  \label{eq: coupled Hamiltonian} 
\end{equation} 
In the eigenbasis of the unimon, we can express the charge operator $\hat{n}_m$ further as
\begin{equation}
\hat{n}_m = -\mathrm{i} \sum_{i, j} \langle i |\mathrm{i} \hat{n}_m | j\rangle |i \rangle \langle j |  = -\mathrm{i} \sum_{i, j} n_{ij} |i \rangle \langle j |, \label{eq: charge operator}
\end{equation} 
where we have defined $n_{ij}= \langle i |\mathrm{i} \hat{n}_m | j\rangle$.  If the eigenenergies and states of the unimon have been solved in the phase basis, the matrix elements can be evaluated as
\begin{equation}
n_{ij} = \int_{-\infty}^\infty  \Psi_i^*(\varphi) \frac{\partial}{\partial \varphi} \Psi_{j}(\varphi) \mathrm{d} \varphi,
\end{equation}
where $\Psi_j(\varphi) =\langle \varphi | j \rangle$ is the $j$th eigenstate of the unimon in the phase basis. Note also that $n_{ij}^* =  \langle i |\mathrm{i} \hat{n}_m | j\rangle^\dagger = -\langle j |\mathrm{i} \hat{n}_m | i\rangle = -n_{ji}$. 

The qubit--resonator coupling term in the Hamiltonian can be simplified as
\begin{align}
\hat{\Ham}_\mathrm{c} &= -2\mathrm{i} e^2\frac{C_\mathrm{g} u_m(x_\mathrm{g})\Delta u_m}{C_{\mathrm{u}, \mathrm{tot}} C_{\mathrm{r}, \mathrm{tot}}} \left ( \frac{E_{L,\mathrm{r}}}{2 E_{C, \mathrm{r}}^\mathrm{tot}} \right )^{1/4} \left(-\mathrm{i} \sum_{i, j} n_{ij} |i \rangle \langle j | \right) (\ah_\mathrm{r} - \ad_\mathrm{r}) \nonumber \\
&= 2 e^2\frac{C_\mathrm{g} u_m(x_\mathrm{g})\Delta u_m}{C_{\mathrm{u}, \mathrm{tot}} C_{\mathrm{r}, \mathrm{tot}}} \left ( \frac{E_{L,\mathrm{r}}}{2 E_{C, \mathrm{r}}^\mathrm{tot}} \right )^{1/4} \sum_{i, j}\left( n_{ij} |i \rangle \langle j | \ad_\mathrm{r} + n_{ij}^* |j \rangle \langle i | \ah_\mathrm{r} \right)  \nonumber \\
&= \hbar \sum_{i,j} \left( g_{ij} |i \rangle \langle j | \ad_\mathrm{r} +  g_{ij}^* |j \rangle \langle i | \ah_\mathrm{r} \right), \label{eq: coupling Hamiltonian using g}
\end{align} 
where the coupling strength $g_{ij}$ is given by 
\begin{align}
g_{ij} &= \frac{2 e^2}{\hbar}\frac{C_\mathrm{g} u_m(x_\mathrm{g})\Delta u_m}{C_{\mathrm{u}, \mathrm{tot}} C_{\mathrm{r}, \mathrm{tot}}} \left ( \frac{E_{L,\mathrm{r}}}{2 E_{C, \mathrm{r}}^\mathrm{tot}} \right )^{1/4} n_{ij} \nonumber \\
%&=\frac{2 e^2}{\hbar}\frac{C_\mathrm{g} u_m(x_\mathrm{g})\Delta u_m}{C_{\mathrm{u}, \mathrm{tot}}} \frac{1}{2\sqrt{\pi}} \sqrt{R_K Z_{\mathrm{r}, \mathrm{tot}}} \omega_\mathrm{r} n_{ij} \\
&=2 \omega_\mathrm{r} \frac{C_\mathrm{g} u_m(x_\mathrm{g})\Delta u_m}{C_{\mathrm{u}, \mathrm{tot}}} \sqrt{\frac{Z_{\mathrm{r}, \mathrm{tot}} \pi}{R_K}} n_{ij} \label{eq: exact gij} \\
&\approx 2 \omega_\mathrm{r} \frac{C_\mathrm{g} u_m(x_\mathrm{g})\Delta u_m}{C_{\mathrm{u}, \mathrm{tot}}} \sqrt{\frac{4 Z_\mathrm{tr}}{R_K}} n_{ij} \label{eq: approx gij}, 
\end{align}
where $R_K = h/e^2$ is the von Klitzing constant and  $Z_{\mathrm{r},\mathrm{tot}}  = \sqrt{L_\mathrm{r} /C_{\mathrm{r}, \mathrm{tot}}}$ is the effective impedance of the resonator. In the simplifications, we have used the equation
\begin{equation}
\left ( \frac{E_{L,\mathrm{r}}}{2 E_{C, \mathrm{r}}^\mathrm{tot}} \right )^{1/4}\frac{1}{ C_{\mathrm{r}, \mathrm{tot}}} = \frac{1}{2\sqrt{\pi}}\sqrt{R_K Z_{\mathrm{r}, \mathrm{tot}}} \omega_\mathrm{r},
\end{equation}
and noted that $Z_{\mathrm{r}, \mathrm{tot}} \approx Z_\mathrm{r} = 4Z_\mathrm{tr}/\pi$ for a $\lambda/4$-readout resonator. 

After all these simplifications, the Hamiltonian of the coupled system in equation~\eqref{eq: coupled Hamiltonian} can be expressed as
\begin{equation}
\hat{\Ham} = \hbar \omega_\mathrm{r} \ad_\mathrm{r} \ah_\mathrm{r} + \sum_j \hbar \omega_j |j\rangle \langle j| +  \hbar \sum_{i,j} \left( g_{ij} |i \rangle \langle j | \ad_\mathrm{r} +  g_{ij}^* |j \rangle \langle i | \ah_\mathrm{r} \right).  \label{eq: coupled Hamiltonian final}
\end{equation}

\noindent \textbf{Approximation of the Hamiltonian in the dispersive limit} \\
In this section, we derive an approximation for the coupled Hamiltonian in the dispersive regime ($|\omega_1 - \omega_0 - \omega_\mathrm{r}| \gg g_{01}$), which allows us to obtain an equation for the dispersive shift of the readout resonator coupled to a unimon. Importantly, the Hamiltonian of the coupled system in equation~\eqref{eq: coupled Hamiltonian final} is of identical form to the Hamiltonian used in Appendix~B of Ref.\cite{blais2021circuit} in the derivation of the dispersive approximation for a general qubit-resonator Hamiltonian. Thus, we apply the results of Ref.\cite{blais2021circuit}, and obtain the following dispersive approximation of the coupled Hamiltonian
\begin{equation}
\hat{\Ham}_\mathrm{disp}  \approx \hbar \omega_\mathrm{r} \ad_\mathrm{r} \ah_\mathrm{r} + \sum_{j} \hbar (\omega_j + \Lambda_j) |j\rangle \langle j| + \sum_{j} \hbar \chi_j \ad_\mathrm{r} \ah_\mathrm{r}  |j \rangle \langle j|, \label{eq: Blais disp Hamiltonian}
\end{equation}
where 
\begin{align}
\Lambda_j &= \sum_{i=0}^{\infty} \chi_{ij}, \label{eq: lambda j} \\
\chi_j &= \sum_{i=0}^\infty (\chi_{ij} - \chi_{ji}), \label{eq: chi j}
\end{align}
with 
\begin{equation}
\chi_{ij} = \frac{|g_{ij}|^2}{\omega_j - \omega_i - \omega_\mathrm{r}}.
\end{equation}

If the dispersive Hamiltonian in equation~\eqref{eq: Blais disp Hamiltonian} is projected to the lowest two levels of the qubit, the Hamiltonian can be further simplified as
\begin{equation}
\hat{\Ham}_\mathrm{disp} \approx  \hbar \omega_\mathrm{r}' \ad_\mathrm{r} \ah_\mathrm{r} - \frac{\hbar \omega_{01}'}{2} \sz - \hbar \chi  \ad_\mathrm{r} \ah_\mathrm{r}  \sz, \label{eq: Blais disp Hamiltonian projected}
\end{equation}
where we have defined $\omega_\mathrm{r}'  = \omega_\mathrm{r} + (\chi_0 + \chi_1)/2$, $\sz = |0\rangle \langle 0 | -  |1\rangle \langle 1|$, 
$\omega_{01}' = \omega_1 - \omega_0 + \Lambda_1 - \Lambda_0$, and the dispersive shift is given by 
\begin{equation}
\chi = (\chi_1 - \chi_0) / 2. \label{eq: Dispersive shift}
\end{equation}
% Note that we use the convention $\sz = |1\rangle \langle 1 | -  |0\rangle \langle 0|$ in equation~\eqref{eq: Blais disp Hamiltonian projected}. 
According to the dispersive Hamiltonian in equation~\eqref{eq: Blais disp Hamiltonian projected}, the frequency of the readout resonator attains a frequency shift of $\pm \chi$ depending on whether the unimon is in the ground state or in the excited state. Thus, we can use dispersive readout similarly to the conventional transmon systems in order to measure the state of the qubit.

Let us then derive a useful approximate expression for the dispersive shift at $\Phid/\Phi_0 = 0.5$. At this sweet spot, the symmetry of the wave functions ensures that $g_{i, i + 2k} = 0 = n_{i, i + 2k}$, where $k \in \mathbb{Z}$. Furthermore, the parameter regime of the unimon ensures that  $|g_{i, i + 1}|  \gg |g_{i, i + 2k + 1}|$, where $k \geq 1$.  Thus, it is a reasonably accurate approximation to only take into account the coupling terms of the form $g_{i, i+1}$ or $g_{i+1, i}$ and neglect the rest of the terms. Within this approximation, we can simplify the equation of the dispersive shift by noting that 
\begin{align}
\chi_0 &= \sum_{i=0}^\infty (\chi_{i0} - \chi_{0i}) \approx \chi_{10} - \chi_{01},
\end{align}
and 
\begin{align}
\chi_1 &= \sum_{i=0}^\infty (\chi_{i1} - \chi_{1i}) \approx \chi_{01} - \chi_{10} + \chi_{21} - \chi_{12}. 
\end{align}
By inserting these results to equation~\eqref{eq: Dispersive shift}, we obtain 
\begin{align}
\chi &\approx \frac{\chi_{01} - \chi_{10} + \chi_{21} - \chi_{12}- (\chi_{10} - \chi_{01})}{2} \nonumber \\
%&= \chi_{10} - \chi_{01} + \frac{ \chi_{21} - \chi_{12}}{2} \nonumber \\
&= |g_{01}|^2 \left(\frac{1}{\omega_{01} - \omega_\mathrm{r}} - \frac{-1}{\omega_{01} + \omega_\mathrm{r}}  \right) + \frac{|g_{12}|^2}{2}  \left(\frac{-1}{\omega_{12} + \omega_\mathrm{r}} - \frac{1}{\omega_{12} - \omega_\mathrm{r}}  \right)  \nonumber \\
&\approx \frac{ |g_{01}|^2}{\omega_{01} - \omega_\mathrm{r}} - \frac{1}{2} \frac{|g_{12}|^2}{\omega_{12} - \omega_\mathrm{r}} =\chi_{01} - \frac{1}{2} \chi_{12} \label{eq: dispersive shift approximation}, 
\end{align}
where we have discarded small terms inversely proportional to the sum of two angular frequencies. % $\chi_{10}$ and $\chi_{21}$. 
Within this approximation, the equation for the dispersive shift is similar to that typically used for transmons\cite{blais2021circuit} apart from the fact that the equation for $g_{ij}$ is different for the unimon. 

\begin{figure}[ht!]
    \centering
    \includegraphics[width = 0.5\textwidth]{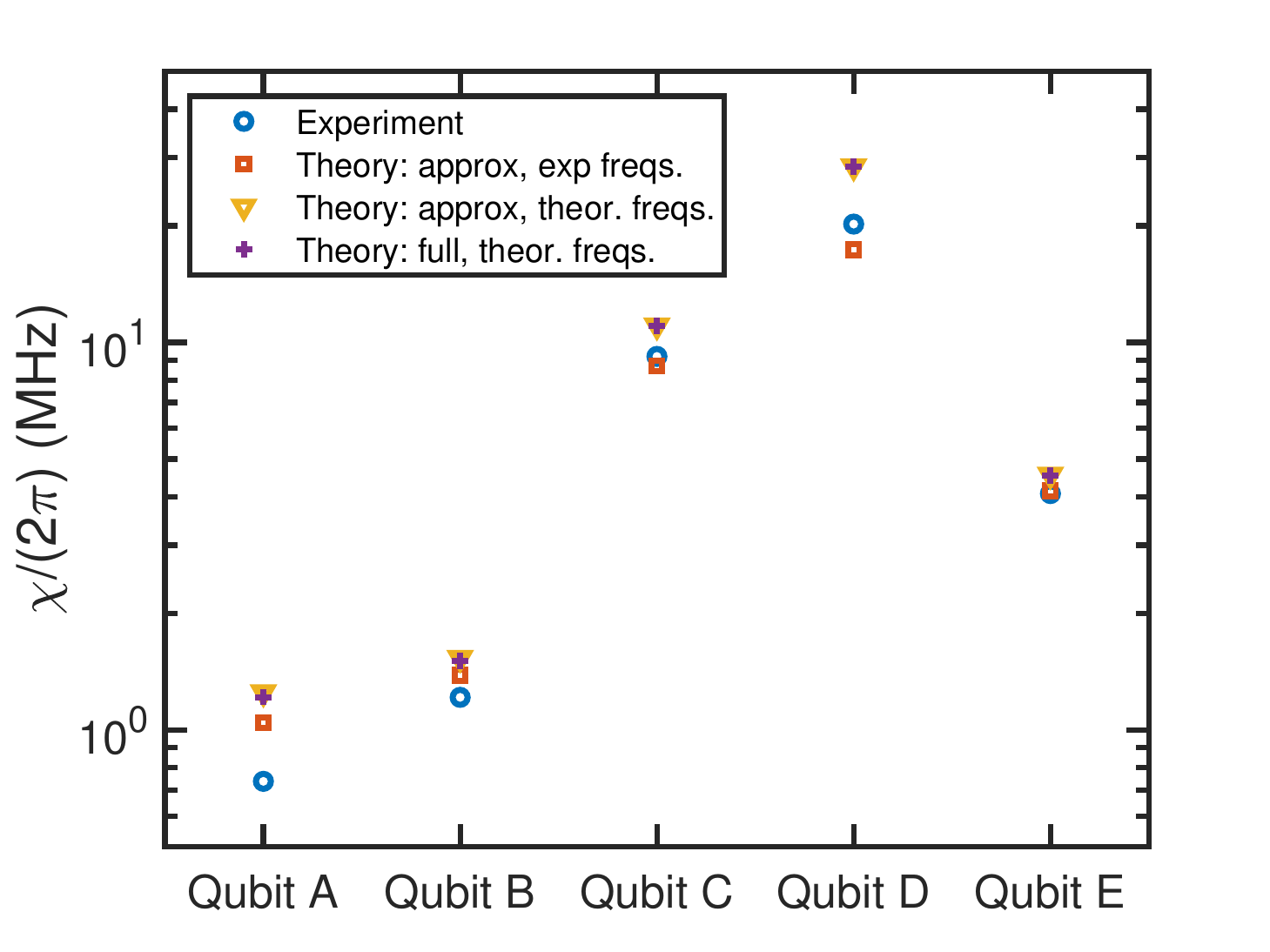}
    \caption{\label{fig: disp shifts} \textbf{Measured and theoretical dispersive shifts.} Measured dispersive shifts (blue circles) for the five qubits characterized in this work together with theoretical predictions. The orange squares show the theoretical prediction based on the approximation in equation~\eqref{eq: dispersive shift approximation} and experimentally measured frequencies $\omega_{01}$ and $\omega_{12}$. The yellow triangles show the theoretical prediction corresponding to the approximation in equation~\eqref{eq: dispersive shift approximation} and numerically computed frequencies $\omega_{01}$ and $\omega_{12}$ based on the Hamiltonian in equation~\eqref{eq: single mode Hamiltonian with external flux} and the fitted circuit parameters presented in Extended Data Table~1. The violet crosses present the theoretical prediction corresponding to  equation~\eqref{eq: Dispersive shift} and numerically computed frequencies $\omega_{01}$ and $\omega_{12}$ based on the Hamiltonian in equation~\eqref{eq: single mode Hamiltonian with external flux} and the fitted circuit parameters presented in Extended Data Table~1. For all of the theoretical predictions, we use a coupling capacitance $C_\mathrm{g}$ estimated from the avoided crossing of the unimon and the resonator.   }
\end{figure}

As illustrated in Fig.~\ref{fig: disp shifts}, we compare the experimentally measured dispersive shifts $\chi/(2\pi)$ to theoretical predictions based on equations~\eqref{eq: Dispersive shift}  and \eqref{eq: dispersive shift approximation}. This allows us to make the following observations: The best agreement with the experimental results is obtained with the approximation in equation~\eqref{eq: dispersive shift approximation} and the experimentally measured transition frequencies $\omega_{01}/(2\pi)$ and $\omega_{12}/(2\pi)$, in the case of which the theoretical predictions and experimental measurements agree within 0--15\% for qubits B--E and within 32\% for qubit A. Secondly, we observe that the theoretically predicted dispersive shift is overestimated if we use numerically computed transition frequencies $\omega_{01}/(2\pi)$ and $\omega_{12}/(2\pi)$ based on the fitted circuit parameters presented in Extended Data Table~1. This arises from the fact that the theoretically predicted anharmonicity is slightly higher than the experimentally measured anharmonicity as can be seen from Fig.~3(a) of the main text. Furthermore, we observe that the theoretical predictions based on equation~\eqref{eq: Dispersive shift} and the approximation in equation~\eqref{eq: dispersive shift approximation} agree within a few percent, thus validating the approximation in equation~\eqref{eq: dispersive shift approximation}.

\pagebreak
\section*{Supplementary Methods III: Theoretical models for the relaxation rate of the unimon}
In this section, we provide models for estimating the relaxation rate of a unimon qubit due to different noise sources. If a qubit is only susceptible to a noise source $\lambda$, the $T_1$ decay rate of the qubit can be described with an expression that is of the Fermi golden rule type\cite{schoelkopf2003qubits}
\begin{equation}
    \Gamma_1 = \frac{1}{T_1} = \frac{|\langle 0 | \partial \hat{\Ham}_m/ \partial \lambda |1\rangle |^2}{\hbar^2} S_{\lambda}(\omega_{01}), \label{eq: general loss rate}
\end{equation}
where  $S_{\lambda}(\omega_{01})$ is the symmetrized noise power spectral density of the variable $\lambda$ at the qubit frequency $\omega_{01}$. The symmetrized noise power spectral density is defined as\cite{schoelkopf2003qubits}
\begin{equation}
    S_{\lambda}(\omega) = \int_{-\infty}^\infty \mathrm{d} t (\mathrm{e}^{+\mathrm{i} \omega t} + \mathrm{e}^{-\mathrm{i} \omega t}) \langle \lambda (t) \lambda(0) \rangle, 
\end{equation}
where $\langle \lambda (t) \lambda(0) \rangle$ denotes an ensemble average over the noise realizations. In the calculations below, we use the Hamiltonian $\hat{\Ham}_m$ in Eq.~\eqref{eq: single mode Hamiltonian with external flux} or its extensions, which means that that we employ the theoretical model 1 derived in Supplementary Methods I. Below, we consider relaxation due to flux noise (ohmic and $1/f$), dielectric losses, inductive losses, radiative losses, and Purcell decay through the readout resonator. In the case where many uncorrelated noise sources are relevant for the qubit, the $T_1$ decay rates of each noise source simply add such that the $T_1$ of the qubit is the inverse of the sum of all different decay rates.

\noindent \textbf{Flux noise}\\
In the case of the unimon, flux noise is caused by fluctuations of the external flux difference. Fluctuations of the external flux difference can be caused by current fluctuations in the flux line or environmental flux noise. Current fluctuations in the resistive flux line can be described with an ohmic power spectral density, whereas the environmental flux noise is often well-described by $1/f$-type noise in superconducting circuits\cite{bylander2011noise}. 

Using equation~\eqref{eq: single mode Hamiltonian with external flux}, we evaluate the matrix element $\langle 0 | \partial \hat{\Ham}_m/ \partial \Phid |1\rangle$ needed to compute the relaxation rate due to flux noise as
\begin{equation}
\langle 0 | \partial \hat{\Ham}_m/ \partial \Phid |1\rangle = \frac{2\pi E_L}{\Phi_0}\langle 0 | \phim | 1\rangle. \label{eq: flux noise overlap}
\end{equation}
Note that the relaxation is caused by fluctuations at the qubit frequency and therefore, the dc Josephson phase $\varphi_0$ can be taken to be a constant. 

To estimate the relaxation rate due to ohmic flux noise, we note that the current in the flux line is related to the external flux bias according to $\Phid = MI$, where $M$ is the mutual inductance of the flux coupling. In this case, the noise power spectral density of $\Phid$ can be written as
\begin{equation}
    S_{\Phid} (\omega) = M^2 S_I (\omega) = \frac{2M^2 \hbar \omega}{R} \coth\left( \frac{\hbar \omega}{2 k_\mathrm{B} T} \right), \label{eq: ohmic flux noise}
\end{equation}
where $R$ is the resistance of the flux line, $k_\mathrm{B}$ is the Boltzmann constant, $T$ is temperature of the resistance, and we have utilized the noise power spectral density for the current in a resistor based on the derivation of Ref.\cite{schoelkopf2003qubits}. By combining equations~\eqref{eq: flux noise overlap} and \eqref{eq: ohmic flux noise}, we obtain the following equation for the relaxation rate due to Ohmic flux noise
\begin{equation}
    \Gamma_1^M = \frac{8\pi^2 E_L^2M^2\omega_{01}}{\Phi_0^2 \hbar R} |\langle 0 | \phim |1\rangle |^2 \coth \left(\frac{\hbar\omega_{01}}{2 k_\mathrm{B} T} \right).
\end{equation}

For the $1/f$ noise, the noise power spectral density is given by 
\begin{equation}
    S_\Phi(\omega) = \frac{2\pi A_{\Phid}^2}{\omega}, \label{eq: 1 per f flux noise}
\end{equation}
where the constant $A_{\Phid}$ yields the noise power spectral density at 1~Hz. By combining equations~\eqref{eq: flux noise overlap} and \eqref{eq: 1 per f flux noise}, we obtain the following equation for the relaxation rate due to $1/f$ flux noise
\begin{equation}
    \Gamma_1^{1/f} = 8\pi^3\frac{E_L^2}{\hbar^2} \frac{A_{\Phid}^2}{\Phi_0^2} \frac{|\langle 0 | \phim |1\rangle |^2}{\omega_{01}} .
\end{equation}

\begin{figure}[t]
	\centering	
	\includegraphics[width = \textwidth]{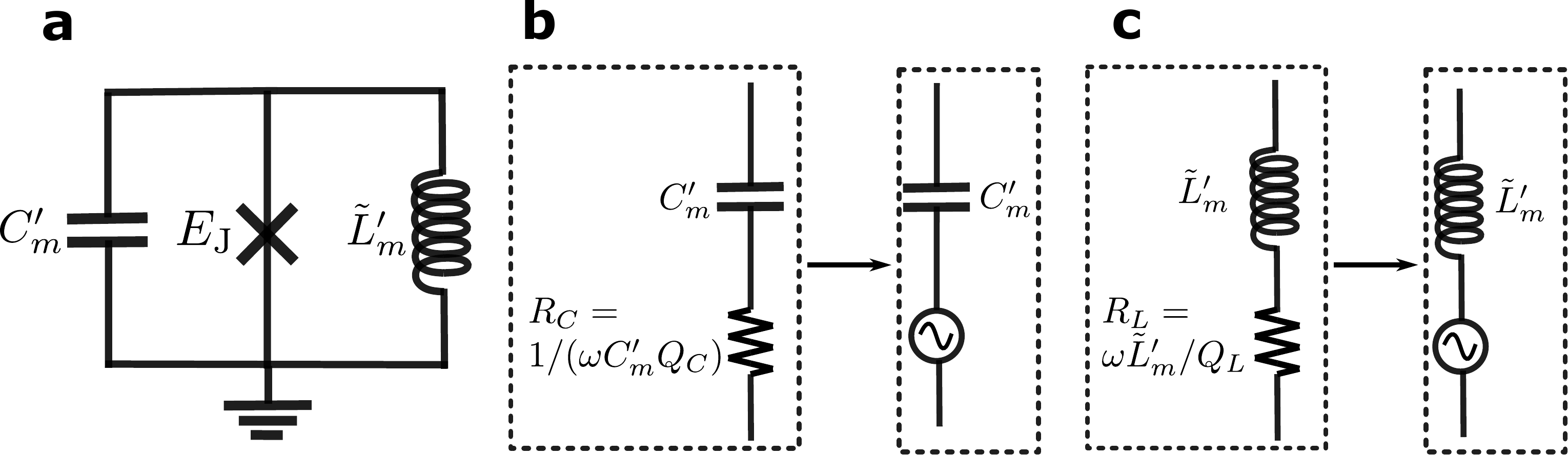}
    \\
    \caption{ \label{fig: unimon T1} \textbf{Circuit models for relaxation rate models.} \textbf{a}, Lumped-element circuit diagram of the unimon. \textbf{b}, Lossy capacitor can be modeled with an ideal capacitor in series with an effective resistance. \textbf{c}, Lossy inductor can be modeled with an inductor in series with a resistance. In (\textbf{b}) and (\textbf{c}), fluctuations in the voltage across the effective resistance can result in the relaxation of the qubit.}
\end{figure}

\noindent \textbf{Dielectric losses} \\
To estimate the relaxation rate due to dielectric losses, we assume that the capacitance $C_m'$ in the lumped-element circuit model of the unimon [see Fig.~\ref{fig: unimon T1}(a)] suffers from losses. These losses can be modeled with a series resistance $R_C = 1/(\omega C_m' Q_C)$ corresponding to a fixed quality factor of $Q_C$ as illustrated in  Fig.~\ref{fig: unimon T1}(b). Due to voltage fluctuations of the effective resistance $R_C$, an additional term is introduced to the Hamiltonian of the unimon such that the Hamiltonian becomes
\begin{equation}
    \hat{\Ham}_m' = \hat{\Ham}_m + 2eV_{R_C}\hat{n}_m,
\end{equation}
where $V_{R_C}$ denotes the voltage across the effective resistance. We  estimate the matrix element $\langle 0 | \partial \hat{H}_m' / \partial V_{R_C} | 1\rangle$ and obtain
\begin{equation}
    \langle 0 | \partial \hat{H}_m' / \partial V_{R_C} | 1\rangle = 2e \langle 0 | \hat{n}_m | 1 \rangle. \label{eq: overlap diel losses}
\end{equation}
By inserting equation~\eqref{eq: overlap diel losses} into equation~\eqref{eq: general loss rate}, we acquire that the relaxation rate owing to dielectric losses is given as
\begin{equation}
    \Gamma_1^\mathrm{cap} =  \frac{16 \ECm}{\hbar Q_C} |\langle 0 | \hat{n}_m | 1 \rangle|^2 \coth\left( \frac{\hbar \omega_{01}}{2 k_\mathrm{B} T} \right), \label{eq: relaxation rate diel losses}
\end{equation}
where we have used the fact that the noise power spectral density of the voltage across the effective resistor can be  written as
\begin{equation}
    S_{V_R}(\omega) = \frac{2 \hbar}{Q_C C_m'} \coth\left( \frac{\hbar \omega}{2 k_\mathrm{B} T} \right).
\end{equation}
Note that equation~\eqref{eq: relaxation rate diel losses} resembles the dielectric loss rate of fluxonium qubits \cite{hazard2019nanowire, zhang2021universal}.

\noindent \textbf{Inductive losses} \\
To model inductive losses, we assume that the inductor $\tilde{L}_m'$ in the lumped-element circuit model of the unimon is connected in series with a resistance $R_C =\omega \tilde{L}_m'/ Q_L$. Here, the inductor is assumed to yield a fixed quality factor of $Q_L$. Due to the presence of the series resistance, the quadratic potential energy associated with $\tilde{L}_m'$ is modified into the form 
$1/(2\tilde{L}_m')(\hat{\phi}_m - \phi_{R})^2$, where $\phi_{R}$ denotes the fluctuating flux across the series resistance. As a result, we need to evaluate the matrix element $\langle 0 | \partial \hat{\Ham}_m / \partial \phi_{R_L} | 1 \rangle $, which gives us
\begin{equation}
    \langle 0 | \partial \hat{\Ham}_m / \partial \phi_{R} | 1 \rangle = \frac{\Phi_0}{2\pi \tilde{L}_m'} \langle 0 | \hat{\varphi}_m| 1\rangle. \label{eq: overlap inductive loss}
\end{equation}
The noise power spectral density for fluctuations of $\phi_{R}$ can be evaluated as\cite{vool2017introduction}
\begin{equation}
    S_{\phi_{R}}(\omega) = \frac{1}{\omega^2}S_{V_{R}}(\omega) =  \frac{2 \hbar \tilde{L}_m'}{Q_L} \coth\left( \frac{\hbar \omega}{2 k_\mathrm{B} T} \right), \label{eq: noise inductive loss}
\end{equation}
where $V_R$ denotes the voltage across the effective resistor. 
By combining equations~\eqref{eq: overlap inductive loss} and \eqref{eq: noise inductive loss}, we obtain the following equation for the relaxation rate due to inductive losses
\begin{equation}
    \Gamma_1^\mathrm{ind} = \frac{2 E_{L, m}}{\hbar Q_L} |\langle 0 | \hat{\varphi}_m| 1\rangle|^2\coth\left( \frac{\hbar \omega_{01}}{2 k_\mathrm{B} T} \right), \label{eq: relaxation rate inductive loss}
\end{equation}
which resembles the inductive relaxation rate of fluxonium qubits \cite{hazard2019nanowire, zhang2021universal}.

\noindent \textbf{Radiative losses} \\
The capacitive coupling to the drive line can also give rise to relaxation due to radiative losses. We can modify an expression of the radiative relaxation rate for a fluxonium\cite{zhang2021universal} to obtain the following result for the unimon
\begin{equation}
    \Gamma_1^\mathrm{rad} = \frac{\omega_{01}}{Q_\mathrm{rad}}\coth\left(\frac{\hbar \omega_{01}}{2 k_\mathrm{B}T}\right) |\langle 0 | \hat{n}_m |1\rangle |^2, \label{eq: radiative losses}
\end{equation}
where $Q_\mathrm{rad}$ is an effective quality factor associated with the relaxation to the drive line. 

\noindent \textbf{Purcell decay to the readout resonator} \\
To estimate the Purcell decay to the readout resonator, we note that the Hamiltonian of a coupled unimon-resonator system in equation~\eqref{eq: coupled Hamiltonian final} is similar to the Hamiltonian of a coupled transmon-resonator system apart from different $\{\omega_j \}$ and $\{g_{ij} \}$. Thus, we can approximate the Purcell decay rate at a low temperature using a result similar to transmon qubits\cite{koch2007charge}
\begin{equation}
    \Gamma_1^\mathrm{P} = \kappa \frac{|g_{01}|^2}{(\omega_{01} - \omega_\mathrm{r})^2}, \label{eq: Purcell decay}
\end{equation}
where $\kappa/(2\pi)$ is the linewidth of the readout resonator in Hz.

\begin{figure}[ht!]
    \centering
    \includegraphics[width=0.5\textwidth]{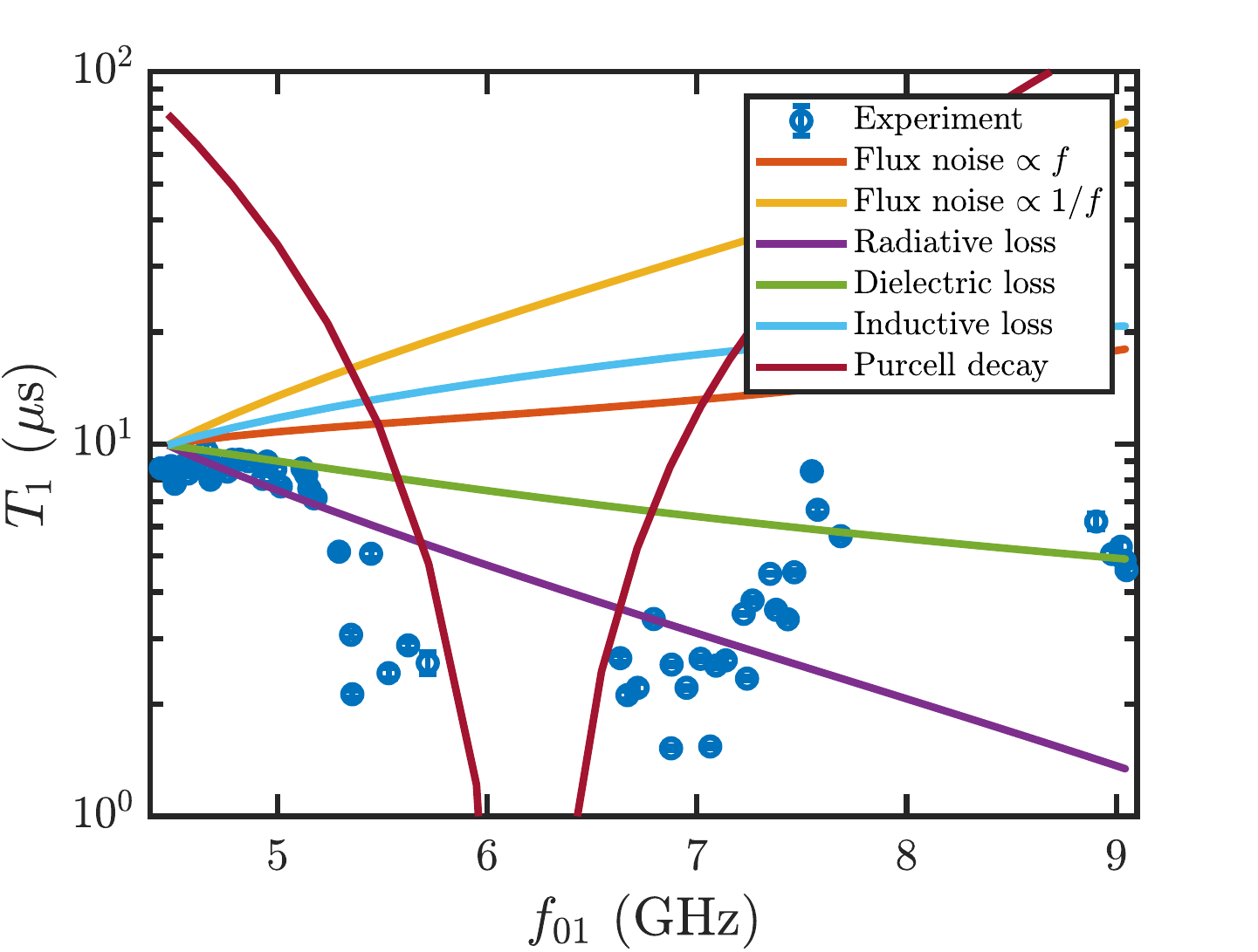}
    \caption{\label{fig: T1} \textbf{Measured $T_1$ together with theoretical models.} Measured $T_1$ (blue circles) for qubit B as a function of qubit frequency together with theoretical predictions based on taking into account a single decay channel at a time as indicated. The theoretical predictions are based on the following results: ohmic flux noise in equation~\eqref{eq: ohmic flux noise}, $1/f$ flux noise in equation~\eqref{eq: 1 per f flux noise}, dielectric losses in equation~\eqref{eq: relaxation rate diel losses}, inductive losses in equation~\eqref{eq: relaxation rate inductive loss}, radiative losses in equation~\eqref{eq: radiative losses} and Purcell decay in equation~\eqref{eq: Purcell decay}. The theoretically predicted relaxation times, apart from the Purcell decay,  were scaled to coincide with the experimental data at $\Phid/\Phi_0 = 0.5$ corresponding to $f_{01} \approx 4.5$ GHz.   }
\end{figure}

\noindent \textbf{Comparing frequency dependence of measured $T_1$ to theoretical models} \\
In Fig.~\ref{fig: T1}, we compare the measured $T_1$ of the qubit B to theoretical predictions based on equations~\eqref{eq: ohmic flux noise}, \eqref{eq: 1 per f flux noise}, \eqref{eq: relaxation rate diel losses}, \eqref{eq: relaxation rate inductive loss}, \eqref{eq: radiative losses}, and \eqref{eq: Purcell decay}. We have scaled the theoretically predicted relaxation times, apart from the Purcell decay, to coincide with the experimental data at $\Phid/\Phi_0 = 0.5$ in order to compare, which relaxation mechanism could explain the measured data. Based on our analysis, the dielectric losses and Purcell decay appear to be the dominating loss mechanisms for the qubit B. When considering the dielectric loss as the dominating loss mechanism, a good fit to the data is obtained with a quality factor of $Q_C = 1.7 \times 10^5$.  

Note that our analysis assumes that the quality factors $Q_L$, $Q_C$, and $Q_\mathrm{rad}$ are independent of the qubit frequency, which may be violated in practice. Namely, the flux mode envelope functions of the unimons are frequency-dependent, and thus, the electric field strengths in different parts of the circuit vary with the qubit frequency.

\clearpage
\section*{Supplementary References}
\bibliographystyle{naturemag}
\bibliography{supplementary}